\documentclass[aps,prb,twocolumn,amsmath,amssymb,10pt,aps,longbibliography,superscriptaddress,citeautoscript,bibnotes,floatfix,nobibnotes]{revtex4-2}

\usepackage[utf8]{inputenc}
\usepackage[T1]{fontenc}
\usepackage{textcomp} 

\usepackage{amsmath}   
\usepackage{amssymb}   
\usepackage{physics}   
\usepackage{bm}        
\usepackage{mathrsfs}  
\usepackage{mathtools} 

\usepackage{caption}  
\usepackage{graphicx} 
\graphicspath{ {./Figures/} }

\usepackage{subfigure} 

\usepackage{tabularx} 
\usepackage{booktabs} 
\usepackage{multirow} 
\usepackage{dcolumn}  

\usepackage[colorlinks=true,linkcolor=blue,citecolor=blue]{hyperref}
\usepackage{xr-hyper}
\usepackage[normalem]{ulem} 
\usepackage{soul} 
\usepackage{amsthm}  
\usepackage{color,xcolor} 
\usepackage{lipsum}

\usepackage{comment}
\usepackage{soul}

\newcommand{\cyan}[1]{{\color{cyan} #1}}

\usepackage{soul}

\begin{document}
\title{Theory of High-Temperature Fractional Chern Phases in Monolayer van der Waals Kagome Materials A$_2$T$_3$X$_4$}

\author{Soomin Oh}
\affiliation{Department of Physics, Korea Advanced Institute of Science and Technology, Daejeon 34141, Republic of Korea}

\author{Jiyong Han}
\affiliation{Department of Physics and Astronomy, Seoul National University, Seoul 08826, Republic of Korea}

\author{Jinhyeok Choi}
\affiliation{Department of Physics and Astronomy, Seoul National University, Seoul 08826, Republic of Korea}

\author{Donghae Seo}
\affiliation{Natural Science Research Institute, Korea Advanced Institute of Science and Technology, Daejeon 34141, Republic of Korea}

\author{Mingu Kang}
\thanks{iordia@snu.ac.kr}
\affiliation{Department of Physics and Astronomy, Seoul National University, Seoul 08826, Republic of Korea}

\author{Gil Young Cho}
\thanks{gilyoungcho@kaist.ac.kr}
\affiliation{Department of Physics, Korea Advanced Institute of Science and Technology, Daejeon 34141, Republic of Korea}
\affiliation{Center for Artificial Low Dimensional Electronic Systems, Institute for Basic Science, Pohang 37673, Korea}

\begin{abstract}
Fractional Chern insulators bring the fractional quantum Hall effect, with its anyonic excitations, to zero magnetic field. So far they have been realized in moiré materials, where the long superlattice period weakens electronic interactions and confines these phases to cryogenic temperatures. Early proposals instead envisioned topological flat bands on atomic-scale lattices, where fractionalization could persist to far higher temperatures, but a suitable material has been lacking. Here we theoretically identify monolayers of the van der Waals kagome family $\mathrm{A}_2\mathrm{T}_3\mathrm{X}_4$ as such a platform, combining intrinsic topological flat bands and strong Coulomb interactions with gate-tunable carrier density and layer assembly. Exact diagonalization of models derived from density functional theory reveals ferromagnetism over broad filling ranges, supporting diverse fractional Chern insulators, anomalous composite Fermi liquids, and charge-ordered phases. In particular, in $\mathrm{Cs}_2\mathrm{Pd}_3\mathrm{S}_4$, the fractional Chern insulator exhibits a many-body gap of approximately 26 meV in the largest system accessible with band mixing allowing at most one interband electron-hole excitation. Layer assembly further provides a natural route to fractional quantum spin Hall insulators: bilayers with antiparallel magnetizations support this phase at realistic separations. These results identify an experimentally accessible material family in which chemical composition, carrier density, and stacking offer complementary routes to fractionalization at enhanced energy scales and control over topological order.
\end{abstract}

\maketitle

\section{Introduction}
Fractional quantum Hall phases are among the most remarkable forms of quantum matter, hosting quasiparticles that carry fractional charge and obey anyonic statistics~\cite{PRL1982TsuiTwoDimensional, PRL1983LaughlinAnomalous, PRL1984HalperinStatistics, PRL1984ArovasFractional}. These exotic excitations lie at the heart of topological quantum computation~\cite{AnnPhys2003KitaevFaulttolerant, RMP2008NayakNonAbelian}, connecting fundamental many-body physics with the prospect of protected quantum information. Recent experimental breakthroughs in twisted bilayer MoTe$_2$ and rhombohedral multilayer graphene have realized fractional Chern insulators~\cite{Nature2023CaiSignatures, Nature2023ZengThermodynamic, Nature2023ParkObservation, PRX2023XuObservation, Nature2024LuFractional, Nature2024JiLocal, NatMater2025XieTunable, NatPhys2026ParkObservation, PRL2026PanOptical}, fulfilling a long-standing theoretical vision of fractional quantum Hall physics without an external magnetic field~\cite{PRL2011NeupertFractional, PRL2011SunNearly, PRX2011RegnaultFractional}. Yet the prospect of room-temperature fractionalization envisioned in early proposals~\cite{PRL2011TangHighTemperature} remains distant, as the long superlattice length scales constrain their characteristic interaction energies. Larger many-body gaps could bring fractionalization within reach of a broader range of experimental probes like photoemission spectroscopy~\cite{PRB2025PichlerSingleparticle} and potentially improve the robustness of prospective quantum technologies. Beyond these advantages, alternative materials can exhibit distinct phase diagrams, offering opportunities to investigate how fractionalized phases compete with other electronic order.

\begin{figure*}
    \centering
    \includegraphics[width=1\linewidth]{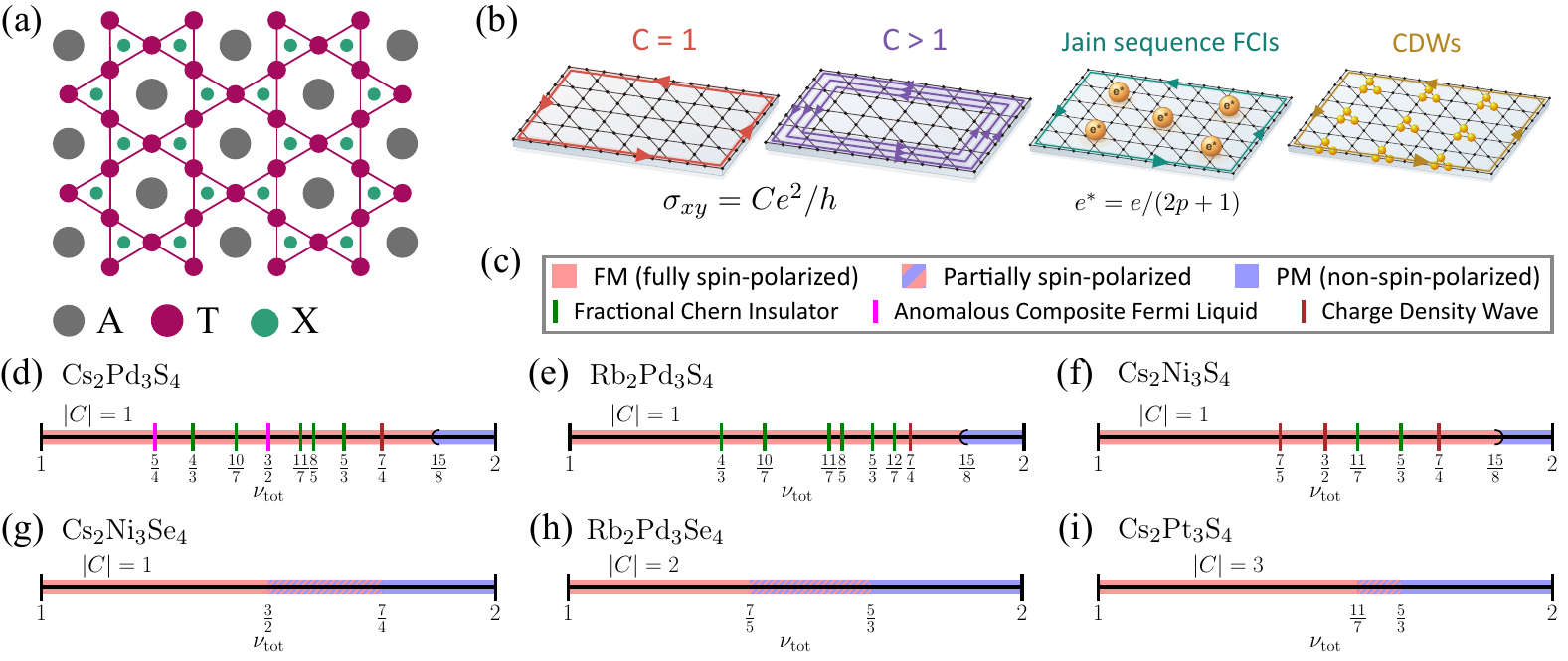}
    \caption{\textbf{Fractional Chern States and Phase Diagrams of Kagome Monolayers.} \textbf{(a)} Schematic compound structure for monolayer A$_2$T$_3$X$_4$. \textbf{(b)} Different types of Chern insulators and symmetry-broken states exhibited by A$_2$T$_3$X$_4$ monolayers. The higher Chern numbers $2$ and $3$ appear in Rb\textsubscript{2}Pd\textsubscript{3}Se\textsubscript{4}, and Cs\textsubscript{2}Pt\textsubscript{3}S\textsubscript{4}.  \textbf{(c-i)} \textbf{(c)} Phases that we have identified for \textbf{(d)} Cs\textsubscript{2}Pd\textsubscript{3}S\textsubscript{4}, \textbf{(e)} Rb\textsubscript{2}Pd\textsubscript{3}S\textsubscript{4}, \textbf{(f)} Cs\textsubscript{2}Ni\textsubscript{3}S\textsubscript{4}, \textbf{(g)} Cs\textsubscript{2}Ni\textsubscript{3}Se\textsubscript{4}, \textbf{(h)} Rb\textsubscript{2}Pd\textsubscript{3}Se\textsubscript{4}, and \textbf{(i)} Cs\textsubscript{2}Pt\textsubscript{3}S\textsubscript{4}. FM and PM stand for ferromagnet and paramagnet. }
    \label{Fig1}
\end{figure*}

In this work, we identify monolayers of the van der Waals kagome family $\mathrm{A}_2\mathrm{T}_3\mathrm{X}_4$ as a platform for diverse fractionalized phases, higher Chern bands, and symmetry breaking [Fig.~\ref{Fig1}]. {Bulk members of this family were synthesized and structurally characterized decades ago~\cite{ZeitschriftfueranorganischeundallgemeineChemie1970BrongerUeber, JournaloftheLessCommonMetals1971BrongerCs2Pd3S4, ZeitschriftfueranorganischeundallgemeineChemie1991BrongerSchichtstrukturen}, and their electronic properties have been explored by subsequent studies~\cite{JPSJ1999FukamachiMagnetization, JournalofPhysicsandChemistryofSolids2003HondouCharacterization, PRB2004NawaiElectronic, PRB2022LiSuperconductivity, AppliedSurfaceScience2023ZhuFamily, JPCM2024DuCrystal, SciAdv2024VillalpandoAccessing}. Their recent exfoliation into monolayers enables electrostatic control of the carrier density~\cite{NatPhys2026BongMonolayer}, opening new opportunities to explore correlated phases in these materials.} Band structure calculations reveal an isolated kagome band manifold near the Fermi level, with spin--orbit coupling endowing its nearly flat upper band with nontrivial topology [Fig.~\ref{Fig2_DFT_TB_comparison}]. The atomic lattice spacing also supports substantially larger Coulomb energy scales than those typical of moir\'e superlattices, opening a route to fractionalization at elevated temperatures. Together, these features and the tunability of spin--orbit coupling, Coulomb interactions, and band dispersion of the $\mathrm{A}_2\mathrm{T}_3\mathrm{X}_4$ family provide a broad, experimentally accessible parameter space for exploring fractional Chern insulators. To explore these possibilities systematically, we study monolayers of six experimentally exfoliated members~\cite{NatPhys2026BongMonolayer}: $\mathrm{Cs}_2\mathrm{Pd}_3\mathrm{S}_4$, $\mathrm{Rb}_2\mathrm{Pd}_3\mathrm{S}_4$, $\mathrm{Cs}_2\mathrm{Ni}_3\mathrm{S}_4$, $\mathrm{Cs}_2\mathrm{Ni}_3\mathrm{Se}_4$, $\mathrm{Rb}_2\mathrm{Pd}_3\mathrm{Se}_4$, and $\mathrm{Cs}_2\mathrm{Pt}_3\mathrm{S}_4$.
We perform exact diagonalization of tight-binding models fitted to density functional theory calculations, using the Ohno interaction to interpolate between onsite Hubbard repulsion and the long-range Coulomb interaction~\cite{Theoreticachimicaacta1964OhnoRemarks}. The Ohno interaction provides a minimal phenomenological description of interactions in the low-energy $d_{xy}$ bands, capturing the finite onsite repulsion associated with the orbital extent while retaining nonlocal Coulomb tail~\cite{PRB2011GundraTheory, PRL2013SchulerOptimal, PRB2016GoliCorrelated}.

Our key findings are as follows. We find robust ferromagnetism over broad filling ranges, supporting a remarkably rich variety of correlated phases encompassing high Chern bands, Jain-sequence fractional Chern insulators, anomalous composite Fermi liquids, and charge-density waves [Fig.~\ref{Fig1}(d)-(i)]. Notably, the fractional Chern insulator in {$\mathrm{Cs}_2\mathrm{Pd}_3\mathrm{S}_4$} at {$\nu_{\rm tot}=4/3$} retains a finite-size many-body gap of approximately $26\,\mathrm{meV}$ with band mixing between the two upper kagome bands in the largest accessible system size [Fig.~\ref{Fig3_EDspectra}(a)], exceeding typical theoretical gaps in moir\'e systems~\cite{PRResearch2020RepellinChern, PRL2021LiuGateTunable, PRB2023ReddyFractional, PRB2023ReddyGlobal, PRResearch2023Morales-DuranPressureenhanced, PRB2024AbouelkomsanBand, PRB2024AhnNonAbelian, PRL2024DongAnomalous, PNAS2024XuMaximally, PRL2024WangFractional, PRL2024DongTheory, NatCommun2026AhnSuperconductivity}. 
We also consider bilayers made of kagome materials with negligible interlayer tunneling across a hexagonal boron nitride (hBN) spacer. Antiparallel layer magnetizations select opposite Chern-band chiralities [Fig.~\ref{Fig4_bilayers}(a)], providing a natural route to fractional quantum spin Hall insulators, whose stabilization in real materials remains a major challenge~\cite{Nature2024KangEvidence, PRResearch2025AbouelkomsanNonAbelian, CommunPhys2026KwanRegarding}. These results establish van der Waals kagome materials as a versatile platform for fractional quantum matter, combining enhanced energy scales with tunable competition among correlated phases. 

\section{Results}  
\begin{figure}
    \centering
    \includegraphics[width=1\linewidth]{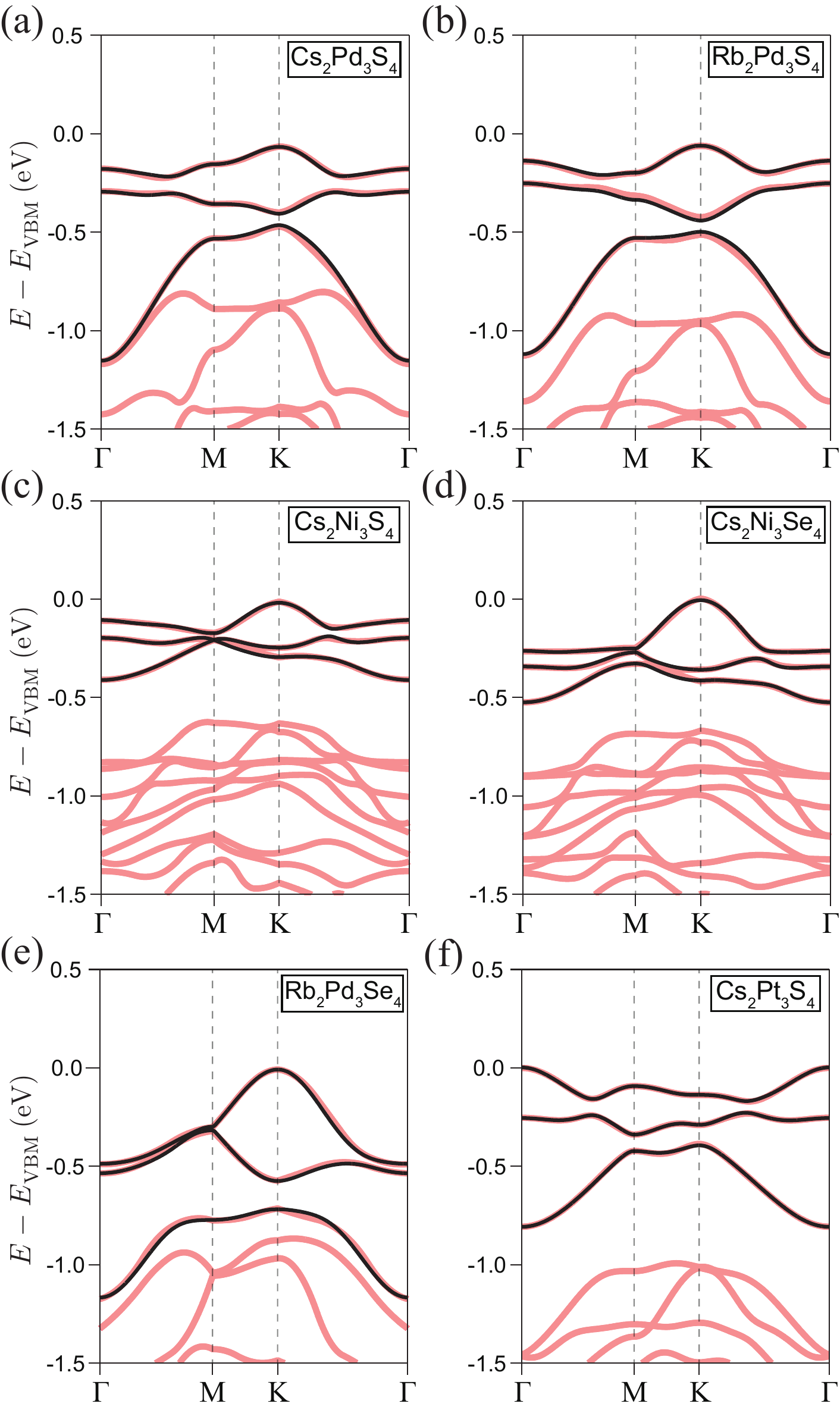}
    \caption{\textbf{Band Structures for Kagome Monolayers.} Black curves show tight-binding dispersions fitted to the DFT bands shown in red for \textbf{(a)} Cs\textsubscript{2}Pd\textsubscript{3}S\textsubscript{4}, \textbf{(b)} Rb\textsubscript{2}Pd\textsubscript{3}S\textsubscript{4}, \textbf{(c)} Cs\textsubscript{2}Ni\textsubscript{3}S\textsubscript{4}, \textbf{(d)} Cs\textsubscript{2}Ni\textsubscript{3}Se\textsubscript{4}, \textbf{(e)} Rb\textsubscript{2}Pd\textsubscript{3}Se\textsubscript{4}, and \textbf{(f)} Cs\textsubscript{2}Pt\textsubscript{3}S\textsubscript{4}.}
    \label{Fig2_DFT_TB_comparison}
\end{figure}

\paragraph{Band Structures.}
We begin by examining the band structures obtained from density functional theory (DFT) and their tight-binding representations 
\begin{equation}
    H_0 = \sum_{\bm{k},\alpha,\beta,\sigma}
    h_{\alpha\beta\sigma}(\bm{k})
    c^\dagger_{\bm{k}\alpha\sigma}
    c_{\bm{k}\beta\sigma},
    \label{SEHamiltonian}
\end{equation}
where $c^\dagger_{\bm{k}\alpha\sigma}$ ($c_{\bm{k}\alpha\sigma}$) creates (annihilates) a $d_{xy}$-orbital electron with momentum $\bm{k}$ and spin $\sigma$ on sublattice $\alpha$. $h_{\alpha\beta\sigma}(\bm{k})$ incorporates symmetry-allowed hopping up to ninth-nearest neighbors, including intrinsic spin--orbit coupling. Diagonalization yields three spin-degenerate kagome bands. The tight-binding and spin-orbit coupling parameters are fitted to the DFT calculations [Fig.~\ref{Fig2_DFT_TB_comparison}] and listed in Tables~\cyan{S1} and \cyan{S2}, while the band characteristics are in Table~\cyan{S3}~\cite{SI}. 

Let us comment on the distinct roles of $\mathrm{A}$, $\mathrm{T}$, and $\mathrm{X}$ sites in shaping the kagome bands. The alkali-metal $\mathrm{A}$ sites lie outside the kagome plane and contribute little orbital weight to the low-energy $d_{xy}$ bands, consistent with the nearly identical band structures of $\mathrm{Cs}_2\mathrm{Pd}_3\mathrm{S}_4$ and $\mathrm{Rb}_2\mathrm{Pd}_3\mathrm{S}_4$. The kagome-forming $\mathrm{T}$ atoms directly control hopping and spin-orbit coupling. For example, in the Cs-based sulfides, replacing Ni with Pd increases the full kagome bandwidth from $0.4\,\mathrm{eV}$ to $1.1\,\mathrm{eV}$ [Fig.~\ref{Fig2_DFT_TB_comparison}(a),~(c)], consistent with the greater spatial extent of Pd $4d$ orbitals relative to Ni $3d$ orbitals. Pt $5d$ orbitals produce a larger spin-orbit coupling gap of $0.1\,\mathrm{eV}$ [Fig.~\ref{Fig2_DFT_TB_comparison}(f)], approximately twice that of the Ni and Pd compounds, while the Pt compound's full kagome bandwidth lies between those of its Ni and Pd counterparts. The chalcogen $\mathrm{X}$ sites mediate hopping between $\mathrm{T}$ sites, influencing the destructive interference underlying flat bands~\cite{PRB2008BergmanBand, PRB2023KimRealization}. Replacing S with Se broadens the topmost flat bands in both the Ni and Pd series, consistent with the more extended Se $4p$ orbitals.

$\mathrm{Cs}_2\mathrm{Pd}_3\mathrm{S}_4$ and $\mathrm{Rb}_2\mathrm{Pd}_3\mathrm{S}_4$ offer particularly favorable conditions for fractionalizations. Both combine the narrowest upper flat bands with overall kagome bandwidths among the largest in the family, closely approaching the ideal kagome limit~\cite{PRB2008BergmanBand, PRB2009GuoTopological, NatCommun2020KangTopological}. This proximity is accompanied by small values of the trace-condition violation~\cite{PRB2014RoyBand, SI}
\begin{equation*}
    \mathcal{T} = \frac{1}{2\pi}\int_{\mathrm{BZ}} d^2\bm{k}\,
    \Big[\operatorname{Tr}g(\bm{k})-|\Omega(\bm{k})|\Big] \approx 0.7,
\end{equation*}
where $g(\bm{k})$ and $\Omega(\bm{k})$ are the quantum metric and Berry curvature of the flat bands. Smaller $\mathcal{T}$ indicates a closer approach to the ideal geometry, a feature favorable to fractional Chern insulators~\cite{NatCommun2015JacksonGeometric, PRB2016BauerQuantum, PRResearch2020LedwithFractional, PRL2024ShavitQuantum}. Further, large dispersions of the lower bands can suppress the effect of virtual interband excitations, helping stabilize fractionalization against band mixing.

\paragraph{Ferromagnetism.}
Having characterized the single-particle bands, we now investigate electron--electron correlations. We adopt the Ohno interaction, motivated by first-principles studies of transition-metal compounds showing that this form captures the spatial dependence of $d$-orbital Coulomb matrix elements under local background screening~\cite{PRResearch2023DiSanteElectronic}
\begin{equation}
    V(\mathbf{r})=\frac{U_0}{\sqrt{1+U_0^2r^2/U_C^2}}, \label{realOhnoInteraction}
\end{equation}
where $r=|\mathbf{r}|$, $U_0$ is the onsite Hubbard repulsion, and $U_C=e^2/4\pi\epsilon$ is the Coulomb prefactor, with $\epsilon$ the effective background permittivity. This form recovers $V(0)=U_0$ and approaches the Coulomb tail $V(r)\to U_C/r$ for $r\gg U_C/U_0$. We take $\epsilon=5\epsilon_0$ and a lattice constant $a=6.3\,\text{\AA}$~\cite{NatPhys2026BongMonolayer}, corresponding to a characteristic Coulomb energy $e^2/4\pi\epsilon a\simeq0.457\,\mathrm{eV}$. We set $U_0=3\,\mathrm{eV}$ as a representative value referring to reported screened interactions in transition-metal $d$ orbitals~\cite{PRL2001LichtensteinFiniteTemperature, PRB2011SasiogluEffective, NJP2014PandaCharge, PRMater2022TasInitio, PRB2022TeschHubbard}. 

We use exact diagonalization to investigate spontaneous symmetry breaking in the partially filled flat bands, focusing on ferromagnetism [Fig.~\ref{Fig1}(d)-(i)]. We project onto the two spin components of the uppermost kagome band, treating the four lower spin-resolved bands as fully occupied and inert. Conservation of the total $S_z$ allows us to diagonalize the Hamiltonian independently in each $S_z$ sector and compare the resulting many-body spectra at fixed total filling $\nu_{\text{tot}} =\nu_\uparrow+\nu_\downarrow$, with $0<\nu_{\mathrm{tot}}<2$. In the undoped compounds, the Fermi level lies above the kagome bands and therefore we focus on $1<\nu_{\mathrm{tot}}<2$, corresponding to less than one hole per unit cell relative to the fully occupied flat bands.

The stability of ferromagnetism correlates with the bandwidth of the flat bands~\cite{SI}. In $\mathrm{Cs}_2\mathrm{Pd}_3\mathrm{S}_4$, $\mathrm{Rb}_2\mathrm{Pd}_3\mathrm{S}_4$, and $\mathrm{Cs}_2\mathrm{Ni}_3\mathrm{S}_4$, which have the narrowest upper bands, the ground state exhibits maximal spin polarization and a substantial spin gap of approximately $30\,\mathrm{meV}$ or larger over broad filling ranges (red regions in Fig.~\ref{Fig1}(d)-(i)), except near full filling, $\nu_{\mathrm{tot}}\to2$~\cite{SI}. The energy splittings between different spin sectors decrease toward $\nu_{\mathrm{tot}}=15/8$ and collapse, indicating the emergence of paramagnetism near full filling~\cite{SI}. By contrast, $\mathrm{Cs}_2\mathrm{Ni}_3\mathrm{Se}_4$, $\mathrm{Rb}_2\mathrm{Pd}_3\mathrm{Se}_4$, and $\mathrm{Cs}_2\mathrm{Pt}_3\mathrm{S}_4$, with broader upper bands, exhibit substantially narrower ferromagnetic regimes. Extended filling intervals host partially spin-polarized states with $|S_z|<(N_\uparrow - N_\downarrow)/2$ (where $N_\sigma$ is the total spin-$\sigma$ electron number) and small energy splittings between different spin sectors~\cite{SI} (dashed regions in Fig.~\ref{Fig1}(d)-(i)). At higher fillings, this gives way to a paramagnetic regime (blue regions in Fig.~\ref{Fig1}(d)-(i)). In passing, we note that the qualitative trends in symmetry breaking remain largely unchanged when the Ohno interaction is replaced by Hubbard repulsion~\cite{SI}.

In these compounds, ferromagnetism is accompanied by Chern bands. Despite their weaker ferromagnetism, $\mathrm{Rb}_2\mathrm{Pd}_3\mathrm{Se}_4$ and $\mathrm{Cs}_2\mathrm{Pt}_3\mathrm{S}_4$ host bands with higher Chern numbers, $|C|=2$ and $|C|=3$, whereas the remaining four compounds have $|C|=1$~\cite{SI}.

\paragraph{Fractionalizations.}
We now investigate intrinsic topological order and competing phases in regimes of robust ferromagnetism, focusing on total fillings $\nu_{\mathrm{tot}}=5/4$, $4/3$, $7/5$, $10/7$, $3/2$, $11/7$, $8/5$, $5/3$, and $7/4$. We perform exact diagonalization with the minority-spin upper flat band partially filled and all other bands fully occupied. To capture the leading effects of band mixing, we retain configurations with at most one interband electron--hole excitation~\cite{PRB2025YuMoire, PRL2026GoncalvesSpinless}. We distinguish fractional Chern insulators, charge-density waves, and anomalous composite Fermi liquids using ground-state degeneracy patterns, spectral flow under flux insertion, many-body Chern numbers, structure factors, momentum-space occupations, and the finite-size scaling of low-energy splittings~\cite{PRX2011RegnaultFractional, JPSJ2005FukuiChern, NatCommun2011ShengFractional, PRL2023DongComposite, NatCommun2025Perea-CausinQuantum}. Further details are provided in~\cite{SI}. 

\begin{figure}
    \centering
    \includegraphics[width=1\linewidth]{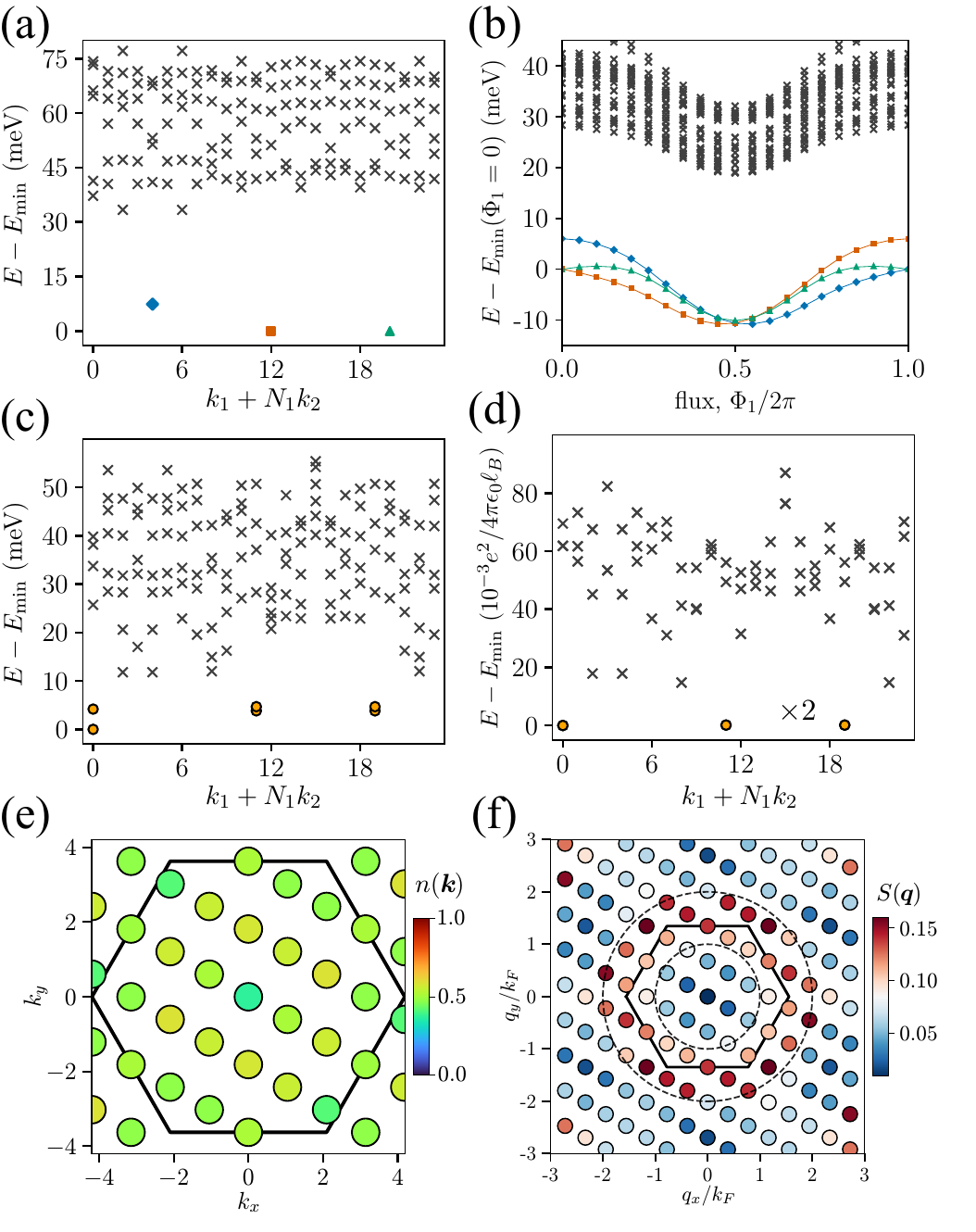}
    \caption{\textbf{Exact Diagonalization of Cs\textsubscript{2}Pd\textsubscript{3}S\textsubscript{4}.} \textbf{(a)} Many-body spectra at $\nu_{\rm tot}=4/3$ of system size $N=24$. The three degenerate ground states (blue diamond, orange square, and green triangle) are separated from the excited states (gray crosses) by a gap of $26\,\mathrm{meV}$. \textbf{(b)} Spectral flow under flux insertion at $\nu_{\rm tot}=4/3$ of system size $N=18$. \textbf{(c)} Many-body spectra at $\nu_{\rm tot}=3/2$ of system size $N=24$. The low-lying states (marked in orange dots) exhibit twofold quasi-degeneracies. \textbf{(d)} Many-body spectra of the half-filled lowest Landau level at $N=24$. \textbf{(e)} Momentum-space occupation number $n(\bm{k})$ and \textbf{(f)} structure factor $S(\bm{q})$ averaged over the six low-lying states in (c). The two dashed circles in (f) are $|\bm q|=k_F$ and $|\bm q|=2k_F$, respectively, where $k_F=\sqrt{A_{\rm BZ}/2\pi}$.}
    \label{Fig3_EDspectra}
\end{figure}

Fig.~\ref{Fig3_EDspectra} summarizes the filling-dependent phases identified by our exact diagonalization calculations. Consistent with their favorable band geometry, $\mathrm{Cs}_2\mathrm{Pd}_3\mathrm{S}_4$ and $\mathrm{Rb}_2\mathrm{Pd}_3\mathrm{S}_4$ host the richest sequences of fractional Chern phases (green solid bars in Fig.~\ref{Fig1}(d)-(i)). In particular, within the one-excitation truncation and at the largest accessible system size, $N=24$, $\mathrm{Cs}_2\mathrm{Pd}_3\mathrm{S}_4$ at $\nu_{\mathrm{tot}}=4/3$ exhibits a threefold ground-state manifold separated from excited states by a many-body gap of approximately $26\,\mathrm{meV}$ [Fig.~\ref{Fig3_EDspectra}(a)]. Its spectral flow under flux insertion [Fig.~\ref{Fig3_EDspectra}(b)] and many-body Chern number $C_{\mathrm{MB}}=1/3$ support its identification as a Laughlin-type fractional Chern insulator~\cite{SI}. At the Jain sequence fillings $\nu_{\mathrm{tot}}=10/7$, $11/7$, $8/5$, and $5/3$, analogous degeneracy patterns, spectral flow, and many-body Chern numbers indicate additional fractional Chern phases~\cite{SI}. Notably, the state at $\nu_{\mathrm{tot}}=5/3$ also exhibits a substantial finite-size gap of approximately $19\,\mathrm{meV}$. Similar cascades of fractional Chern phases occur in $\mathrm{Rb}_2\mathrm{Pd}_3\mathrm{S}_4$ [Fig.~\ref{Fig1}(e)]. Table~\cyan{S5} summarizes the finite-size many-body gaps of the identified fractional Chern states.

At half filling $\nu_{\mathrm{tot}}=3/2$ of $\mathrm{Cs}_2\mathrm{Pd}_3\mathrm{S}_4$, we find evidence for a gapless anomalous composite Fermi liquid (magenta solid bar in Fig.~\ref{Fig1}(d)). The low-energy many-body spectra [Fig.~\ref{Fig3_EDspectra}(c)] closely resemble those of the half-filled lowest Landau level [Fig.~\ref{Fig3_EDspectra}(d)], including the momentum sectors of the lowest-lying states and twofold degeneracies. Furthermore, the occupation number averaged over the ground states [Fig.~\ref{Fig3_EDspectra}(e)]
\begin{equation*}
    n(\bm{k})=\frac{1}{N_{\mathrm{GS}}}\sum_{i\in\mathrm{GS}}
    \langle\Psi_i|c^\dagger_{\bm{k}}c_{\bm{k}}|\Psi_i\rangle,
\end{equation*}
remains nearly uniform, satisfying $|n(\bm{k})-1/2|<0.13$ at the largest accessible system size, $N=24$, with no discernible electronic Fermi surface. This bound is comparable to the corresponding value of $0.17$ reported for the anomalous composite Fermi liquid in twisted bilayer $\mathrm{MoTe}_2$~\cite{PRL2023DongComposite}. Further evidence comes from the projected static structure factor [Fig.~\ref{Fig3_EDspectra}(f)]
\begin{equation*}
    S(\bm{q})=\frac{1}{N}\left[
    \langle\bar{\rho}_{\bm{q}}\bar{\rho}_{-\bm{q}}\rangle
    -\delta_{\bm{q},0}|\langle\bar{\rho}_{\bm{q}}\rangle|^2
    \right],
\end{equation*}
where $\bar{\rho}_{\bm{q}}$ is the density operator projected onto the upper flat band. We observe pronounced peaks near $|\bm{q}|=2k_F$ and nearly vanishing weight for $|\bm{q}|>2k_F$~\cite{SI}. These density correlations, together with the absence of a discernible electronic Fermi surface, support an emergent composite-fermion Fermi surface~\cite{PRB2025PichlerSingleparticle, PRB2023ReddyGlobal, PRL2023DongComposite, PRL2023GoldmanZeroField, PRL2025ShiFractional}. We also find signatures consistent with an anomalous composite Fermi liquid in $\mathrm{Cs}_2\mathrm{Pd}_3\mathrm{S}_4$ at quarter filling of the flat band ($\nu_{\mathrm{tot}}=5/4$)~\cite{SI, PRB2023ReddyGlobal, PRL2023DongComposite, PRL2023GoldmanZeroField}.

At several filling fractions, $\mathrm{Cs}_2\mathrm{Pd}_3\mathrm{S}_4$, $\mathrm{Rb}_2\mathrm{Pd}_3\mathrm{S}_4$, and $\mathrm{Cs}_2\mathrm{Ni}_3\mathrm{S}_4$ exhibit translational symmetry breaking and form charge-density waves, characterized by multifold ground states and Bragg peaks in $S(\bm{q})$ (brown solid bars in Fig.~\ref{Fig1}(d)-(f)). Detailed characterizations are in~\cite{SI}.

\begin{figure}
    \centering
    \includegraphics[width=1\linewidth]{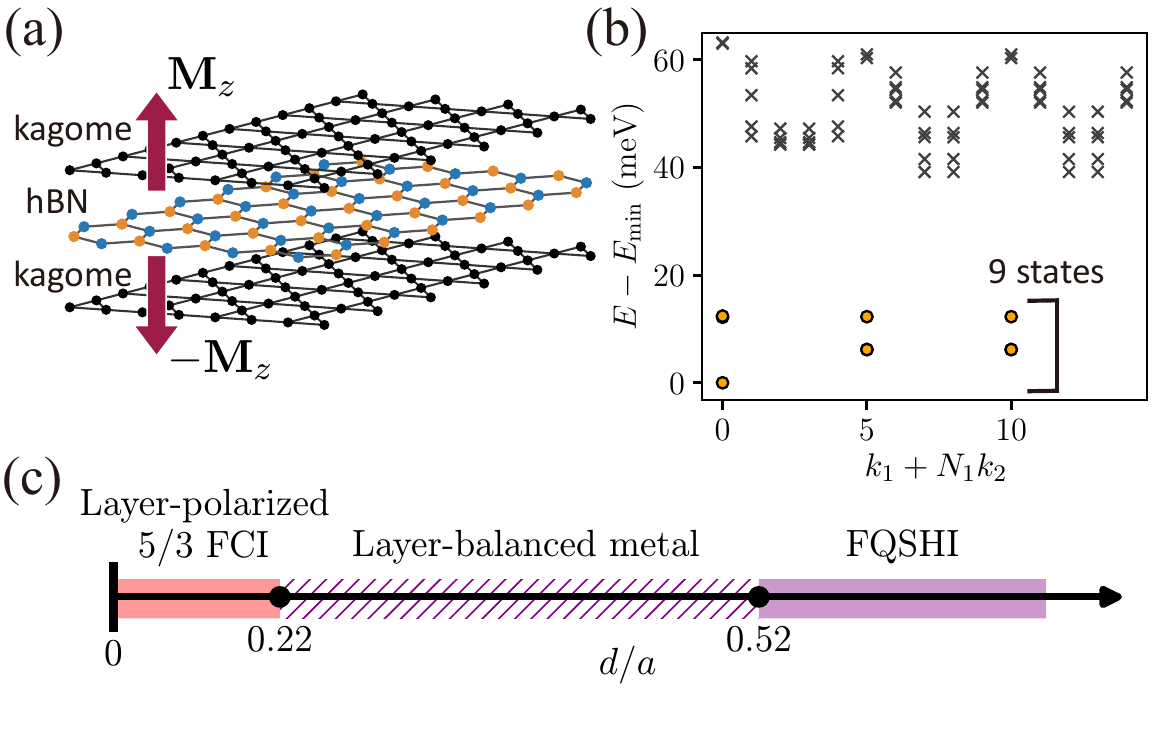}
    \caption{\textbf{Fractional Spin Hall Insulators in Bilayers.} \textbf{(a)} Schematic structure of bilayer kagome separated by hBN with antiparallel magnetizations. \textbf{(b)} Many-body spectra of two $\mathrm{Cs}_2\mathrm{Pd}_3\mathrm{S}_4$ monolayers at $\nu_{\rm tot}=4/3$ per layer, $d/a=1$ for system size $N=15$. The ninefold quasi-degenerate ground states signal a fractional quantum spin Hall phase. \textbf{(c)} Phase diagram for bilayer $\mathrm{Cs}_2\mathrm{Pd}_3\mathrm{S}_4$ as a function of $d/a$. FQSHI stands for fractional quantum spin Hall insulator.}
    \label{Fig4_bilayers}
\end{figure}

\paragraph{Bilayers.}\noindent
The fractional Chern phases in these kagome monolayers naturally motivate bilayer heterostructures, which can be readily assembled using established van der Waals techniques~\cite{NatPhys2026BongMonolayer}. Antiparallel layer magnetizations produce opposite Chern-band chiralities, providing a route to fractional quantum spin Hall insulators. We consider two $\mathrm{Cs}_2\mathrm{Pd}_3\mathrm{S}_4$ monolayers at  $\nu_{\mathrm{tot}}=4/3$ per layer, separated by an hBN spacer that suppresses interlayer tunneling [Fig.~\ref{Fig4_bilayers}(a)]. We assume that ferromagnetism remains robust in each layer, and use the Ohno and Coulomb potentials for intralayer and interlayer interactions.

We use exact diagonalization to map the phase diagram as a function of $d/a$ [Fig.~\ref{Fig4_bilayers}(c)], where $d$ is the interlayer separation and $a=6.3~\text{\AA}$ is the intralayer lattice constant. At small $d/a$, the ground state is fully layer polarized and effectively realizes the monolayer fractional Chern insulator at $\nu_{\mathrm{tot}}=5/3$. As $d/a$ increases, an intermediate layer-balanced metallic phase emerges at $d/a\simeq0.22$ and gives way to a fractional quantum spin Hall insulator at $d/a\simeq0.52$. This insulator effectively realizes two decoupled fractional Chern insulators with opposite chiralities [Fig.~\ref{Fig4_bilayers}(b)]. The measured interlayer spacing of approximately $7\,\text{\AA}$~\cite{NatPhys2026BongMonolayer} corresponds to $d/a\simeq1.1$, well within the fractional quantum spin Hall regime of our model [Fig.~\ref{Fig4_bilayers}(b),(c)], suggesting an experimentally accessible route to realizing this phase.

\section{Discussions}\noindent   
We identify the van der Waals kagome family $\mathrm{A}_2\mathrm{T}_3\mathrm{X}_4$ as a platform for diverse correlated topological phases at enhanced energy scales. Within the ferromagnetic regimes, we find spin-polarized bands with Chern numbers up to $|C|=3$, alongside Jain-sequence fractional Chern insulators, anomalous composite Fermi liquids, and competing charge order. The Pd-based sulfides are particularly favorable: $\mathrm{Cs}_2\mathrm{Pd}_3\mathrm{S}_4$ exhibits a finite-size many-body gap of approximately $26\,\mathrm{meV}$ in the largest accessible system with band mixing that retains at most one interband electron--hole excitation. Bilayers with antiparallel layer magnetizations further support fractional quantum spin Hall insulators at realistic interlayer separations. Chemical diversity, gate-tunable carrier density, and van der Waals layer assembly together make these kagome materials versatile platforms for realizing and controlling fractionalized quantum matter. 

These results motivate several directions for future investigation. The most immediate experimental priority is systematic transport characterization, building on preliminary observations of insulating behavior at partial filling of the flat band~\cite{NatPhys2026BongMonolayer}. Gate-dependent measurements could test for fractional Hall responses and map the competition among the predicted phases. {Notably, experiments on chemically hole-doped bulk CsNi$_3$S$_4$, corresponding to $\nu_{\mathrm{tot}}=1$ of Cs$_2$Ni$_3$S$_4$, have reported an enhanced magnetic moment and an insulating behavior below approximately $68~\mathrm{K}$~\cite{SciAdv2024VillalpandoAccessing}. These observations provide a good reference point to our study, motivating further exploration of magnetic and insulating states through controlled hole doping in this material family.} Secondly, calculations on larger systems using methods such as density matrix renormalization group could clarify the phases at additional fillings, where our finite-size spectra and ground-state diagnostics do not yet establish a definite symmetry-breaking pattern or topological order. Finally, flat bands may also promote other collective instabilities, including superconductivity driven by interaction channels beyond those retained in our minimal model. Together, these efforts could reveal how fractionalization, symmetry breaking, and superconductivity compete and intertwine within a single material family.

\section{Acknowledgements}\noindent
S.~O., D.~S., and G.~Y.~C. acknowledge the financial support from Samsung Science and Technology Foundation under Project Number SSTF-BA2401-03, the NRF of Korea (Grants No. RS-2026-25479545, RS-2024-00410027, RS-2026-25467936, RS-2024-00444725, RS-2023-00256050, RS-2025-25453111, RS-2025-08542968) funded by the Korean Government (MSIT), the Air Force Office of Scientific Research under Award No. FA23862514026, and Institute for Basic Science under project code IBS-R014-D1. S.~O. acknowledges the support from the Agency for Defense Development through a grant funded by the Defense Acquisition Program Administration (DAPA) (UI257011TE).
J.~H., J.~C., and M.~K. are supported by the National Research Foundation (NRF) of Korea through grant nos. RS-2025-23525391 and RS-2026-25496507. M.~K. was supported by the Creative-Pioneering Researchers Program through Seoul National University. M.~K. acknowledges POSCO Science Fellowship from Cheongam Foundation. This work is in part supported by the Korea Max Planck Quantum Materials Centre (KOMQUEST), the Top-Tier Premier International Research Consortium through Grant No. RS-2026-25585727, and the core center program by the Ministry of Education through Grant No. 2021R1A6C101B418.

\bibliography{kagome_A2T3X4_theory}

@misc{SI,
    note = {See Supplementary Information for additional details on DFT calculation, model parameters, band geometry, and exact diagonalization results.}
}

@article{AppliedSurfaceScience2023ZhuFamily,
  title = {A Family of Two-Dimensional Semiconductors with Transition Metal {{Kagome}} Lattice, Large Power Factor and Ultralow Lattice Thermal Conductivity},
  author = {Zhu, Ying and Yuan, Jun-Hui and Fang, Wen-Yu and Sun, Zhi-Gang and Wang, Jiafu},
  year = 2023,
  month = nov,
  journal = {Applied Surface Science},
  volume = {636},
  pages = {157817},
  issn = {0169-4332},
  doi = {10.1016/j.apsusc.2023.157817},
  urldate = {2026-10-22}
}

@article{JournalofPhysicsandChemistryofSolids2003HondouCharacterization,
  title = {Characterization of Magnetism and Electronic Properties of a Ternary Transition Metal Chalcogenide {{Rb}}{$_{2}$}{{Ni}}{$_{3}$}{{S}}{$_{4}$}},
  author = {Hondou, K. and Fujiwara, Y. and Kato, T. and Iio, K. and Saiki, A. and Usuda, M. and Hamada, N.},
  year = 2003,
  month = sep,
  journal = {Journal of Physics and Chemistry of Solids},
  series = {13th {{International Conference}} on {{Ternary}} and {{Multinary Compounds}}},
  volume = {64},
  number = {9},
  pages = {1721--1724},
  issn = {0022-3697},
  doi = {10.1016/S0022-3697(03)00139-2},
  urldate = {2026-10-22}
}

@article{JournaloftheLessCommonMetals1971BrongerCs2Pd3S4,
  title = {{{Cs}}{$_{2}$}{{Pd}}{$_{3}$}{{S}}{$_4$}, Ein Neuer Schichtenstrukturtyp},
  author = {Bronger, W. and Huster, J.},
  year = 1971,
  month = jan,
  journal = {Journal of the Less Common Metals},
  volume = {23},
  number = {1},
  pages = {67--72},
  issn = {0022-5088},
  doi = {10.1016/0022-5088(71)90009-9},
  urldate = {2026-10-22}
}

@article{JPCM2024DuCrystal,
  title = {Crystal Structure, Properties and Pressure-Induced Insulator-Metal Transition in Layered Kagome Chalcogenides},
  author = {Du, Hong and Zheng, Yu and Pei, Cuiying and Yim, Chi-Ming and Qi, Yanpeng and Zhong, Ruidan},
  year = 2024,
  month = jun,
  journal = {J. Phys.: Condens. Matter},
  volume = {36},
  number = {36},
  pages = {365702},
  publisher = {IOP Publishing},
  issn = {0953-8984},
  doi = {10.1088/1361-648X/ad52e0},
  urldate = {2026-10-22},
  langid = {english}
}

@article{JPSJ1999FukamachiMagnetization,
  title = {Magnetization {{Measurements}} and {$^{87}$}{{Rb-NMR}} on {{Rb}}{$_{2}$}{{M}}{$_{3}$}{{S}}{$_4$} ({{M}}={{Ni}} and {{Pd}}) with the {{Kagome Lattice}}},
  author = {Fukamachi, Toshihiko and Kobayashi, Yoshiaki and Nakamura, Akira and Harashina, Hiroshi and Sato, Masatoshi},
  year = 1999,
  month = nov,
  journal = {J. Phys. Soc. Jpn.},
  volume = {68},
  number = {11},
  pages = {3668--3672},
  publisher = {The Physical Society of Japan},
  issn = {0031-9015},
  doi = {10.1143/JPSJ.68.3668},
  urldate = {2026-10-22}
}

@article{PRB2004NawaiElectronic,
  title = {Electronic Structure of the {{Kagom\'e}} Lattice Compound {{Rb}}{$_{2}$}{{Ni}}{$_{3}$}{{S}}{$_{4}$}},
  author = {Nawai, S. and Okazaki, K. and Mizokawa, T. and Fujimori, A. and Hondou, K. and Fujiwara, Y. and Iio, K. and Usuda, M. and Hamada, N.},
  year = 2004,
  month = jan,
  journal = {Phys. Rev. B},
  volume = {69},
  number = {4},
  pages = {045103},
  publisher = {American Physical Society},
  doi = {10.1103/PhysRevB.69.045103},
  urldate = {2026-10-22}
}

@article{PRB2022LiSuperconductivity,
  title = {Superconductivity Arising from Pressure-Induced Emergence of a {{Fermi}} Surface in the Kagome-Lattice Chalcogenide {{Rb}}{$_{2}$}{{Pd}}{$_{3}$}{{Se}}{$_{4}$}},
  author = {Li, Qing and Wu, Yuxiang and Fan, Xinwei and Zhang, Ying-Jie and Zhu, Xiyu and Zhu, Zhengyan and Li, Yiwen and Wen, Hai-Hu},
  year = 2022,
  month = dec,
  journal = {Phys. Rev. B},
  volume = {106},
  number = {21},
  pages = {214501},
  publisher = {American Physical Society},
  doi = {10.1103/PhysRevB.106.214501},
  urldate = {2026-10-22}
}

@article{SciAdv2024VillalpandoAccessing,
  title = {Accessing Bands with Extended Quantum Metric in Kagome {{Cs}}{$_{2}$}{{Ni}}{$_{3}$}{{S}}{$_4$} through Soft Chemical Processing},
  author = {Villalpando, Graciela and Jovanovic, Milena and Hoff, Brianna and Jiang, Yi and Singha, Ratnadwip and Yuan, Fang and Hu, Haoyu and C{\u a}lug{\u a}ru, Dumitru and Mathur, Nitish and Khoury, Jason F. and Dulovic, Stephanie and Singh, Birender and Plisson, Vincent M. and Pollak, Connor J. and Moya, Jaime M. and Burch, Kenneth S. and Bernevig, B. Andrei and Schoop, Leslie M.},
  year = 2024,
  month = sep,
  journal = {Science Advances},
  volume = {10},
  number = {38},
  pages = {eadl1103},
  publisher = {American Association for the Advancement of Science},
  doi = {10.1126/sciadv.adl1103},
  urldate = {2026-10-22}
}

@article{ZeitschriftfueranorganischeundallgemeineChemie1970BrongerUeber,
  title = {{\"Uber Thio- und Selenoniccolate und -palladate der schweren Alkalimetalle}},
  author = {Bronger, W. and Eyck, J. and R{\"u}dorff, W. and St{\"o}ussel, A.},
  year = 1970,
  journal = {Zeitschrift f\"ur anorganische und allgemeine Chemie},
  volume = {375},
  number = {1},
  pages = {1--7},
  issn = {1521-3749},
  doi = {10.1002/zaac.19703750102},
  urldate = {2026-10-22},
  langid = {ngerman}
}

@article{ZeitschriftfueranorganischeundallgemeineChemie1991BrongerSchichtstrukturen,
  title = {{Schichtstrukturen tern\"arer Chalkogenide A{$_2$}M{$_3$}X{$_4$} (A = K, Rb, Cs; M = Ni, Pd, Pt; X = S, Se)}},
  author = {Bronger, W. and Rennau, R. and Schmitz, D.},
  year = 1991,
  journal = {Zeitschrift f\"ur anorganische und allgemeine Chemie},
  volume = {597},
  number = {1},
  pages = {27--32},
  issn = {1521-3749},
  doi = {10.1002/zaac.19915970105},
  urldate = {2026-10-22},
  copyright = {Copyright \copyright{} 1991 Verlag GmbH \& Co. KGaA, Weinheim},
  langid = {ngerman}
}

@article{AnnPhys2003KitaevFaulttolerant,
  title = {Fault-Tolerant Quantum Computation by Anyons},
  author = {Kitaev, A. {\relax Yu}.},
  year = 2003,
  month = jan,
  journal = {Annals of Physics},
  volume = {303},
  number = {1},
  pages = {2--30},
  issn = {0003-4916},
  doi = {10.1016/S0003-4916(02)00018-0},
  urldate = {2026-09-23}
}

@article{CommunPhys2026KwanRegarding,
  title = {Regarding the Existence of Abelian Fractional Topological Insulators in Twisted {{MoTe}}{$_2$} and Related Systems},
  author = {Kwan, Yves H. and Wagner, Glenn and Yu, Jiabin and Dagnino, Andrea Kouta and Jiang, Yi and Xu, Xiaodong and Bernevig, B. Andrei and Neupert, Titus and Regnault, Nicolas},
  year = 2026,
  month = jan,
  journal = {Commun Phys},
  volume = {9},
  number = {1},
  pages = {52},
  publisher = {Nature Publishing Group},
  issn = {2399-3650},
  doi = {10.1038/s42005-025-02483-6},
  urldate = {2026-10-08},
  copyright = {2026 The Author(s)},
  langid = {english}
}

@article{ComputationalMaterialsScience1996KresseEfficiency,
  title = {Efficiency of Ab-Initio Total Energy Calculations for Metals and Semiconductors Using a Plane-Wave Basis Set},
  author = {Kresse, G. and Furthm{\"u}ller, J.},
  year = 1996,
  month = jul,
  journal = {Computational Materials Science},
  volume = {6},
  number = {1},
  pages = {15--50},
  issn = {0927-0256},
  doi = {10.1016/0927-0256(96)00008-0},
  urldate = {2026-10-15}
}

@article{ComputerPhysicsCommunications2014MostofiUpdated,
  title = {An Updated Version of Wannier90: {{A}} Tool for Obtaining Maximally-Localised {{Wannier}} Functions},
  shorttitle = {An Updated Version of Wannier90},
  author = {Mostofi, Arash A. and Yates, Jonathan R. and Pizzi, Giovanni and Lee, Young-Su and Souza, Ivo and Vanderbilt, David and Marzari, Nicola},
  year = 2014,
  month = aug,
  journal = {Computer Physics Communications},
  volume = {185},
  number = {8},
  pages = {2309--2310},
  issn = {0010-4655},
  doi = {10.1016/j.cpc.2014.05.003},
  urldate = {2026-09-23}
}

@article{ComputerPhysicsCommunications2018vanSettenPseudoDojo,
  title = {The {{PseudoDojo}}: {{Training}} and Grading a 85 Element Optimized Norm-Conserving Pseudopotential Table},
  shorttitle = {The {{PseudoDojo}}},
  author = {{van Setten}, M. J. and Giantomassi, M. and Bousquet, E. and Verstraete, M. J. and Hamann, D. R. and Gonze, X. and Rignanese, G. -M.},
  year = 2018,
  month = may,
  journal = {Computer Physics Communications},
  volume = {226},
  pages = {39--54},
  issn = {0010-4655},
  doi = {10.1016/j.cpc.2018.01.012},
  urldate = {2026-09-23}
}

@article{JPCM2009GiannozziQUANTUM,
  title = {{{QUANTUM ESPRESSO}}: A Modular and Open-Source Software Project for Quantum Simulations of Materials},
  shorttitle = {{{QUANTUM ESPRESSO}}},
  author = {Giannozzi, Paolo and Baroni, Stefano and Bonini, Nicola and Calandra, Matteo and Car, Roberto and Cavazzoni, Carlo and Ceresoli, Davide and Chiarotti, Guido L and Cococcioni, Matteo and Dabo, Ismaila and Dal Corso, Andrea and {de Gironcoli}, Stefano and Fabris, Stefano and Fratesi, Guido and Gebauer, Ralph and Gerstmann, Uwe and Gougoussis, Christos and Kokalj, Anton and Lazzeri, Michele and {Martin-Samos}, Layla and Marzari, Nicola and Mauri, Francesco and Mazzarello, Riccardo and Paolini, Stefano and Pasquarello, Alfredo and Paulatto, Lorenzo and Sbraccia, Carlo and Scandolo, Sandro and Sclauzero, Gabriele and Seitsonen, Ari P and Smogunov, Alexander and Umari, Paolo and Wentzcovitch, Renata M},
  year = 2009,
  month = sep,
  journal = {J. Phys.: Condens. Matter},
  volume = {21},
  number = {39},
  pages = {395502},
  issn = {0953-8984},
  doi = {10.1088/0953-8984/21/39/395502},
  urldate = {2026-09-23},
  langid = {english}
}

@article{JPCM2017GiannozziAdvanced,
  title = {Advanced Capabilities for Materials Modelling with {{Quantum ESPRESSO}}},
  author = {Giannozzi, P and Andreussi, O and Brumme, T and Bunau, O and Buongiorno Nardelli, M and Calandra, M and Car, R and Cavazzoni, C and Ceresoli, D and Cococcioni, M and Colonna, N and Carnimeo, I and Dal Corso, A and {de Gironcoli}, S and Delugas, P and DiStasio, R A and Ferretti, A and Floris, A and Fratesi, G and Fugallo, G and Gebauer, R and Gerstmann, U and Giustino, F and Gorni, T and Jia, J and Kawamura, M and Ko, H-Y and Kokalj, A and K{\"u}{\c c}{\"u}kbenli, E and Lazzeri, M and Marsili, M and Marzari, N and Mauri, F and Nguyen, N L and Nguyen, H-V and {Otero-de-la-Roza}, A and Paulatto, L and Ponc{\'e}, S and Rocca, D and Sabatini, R and Santra, B and Schlipf, M and Seitsonen, A P and Smogunov, A and Timrov, I and Thonhauser, T and Umari, P and Vast, N and Wu, X and Baroni, S},
  year = 2017,
  month = oct,
  journal = {J. Phys.: Condens. Matter},
  volume = {29},
  number = {46},
  pages = {465901},
  publisher = {IOP Publishing},
  issn = {0953-8984},
  doi = {10.1088/1361-648X/aa8f79},
  urldate = {2026-09-23},
  langid = {english}
}

@article{JPSJ2005FukuiChern,
  title = {Chern {{Numbers}} in {{Discretized Brillouin Zone}}: {{Efficient Method}} of {{Computing}} ({{Spin}}) {{Hall Conductances}}},
  shorttitle = {Chern {{Numbers}} in {{Discretized Brillouin Zone}}},
  author = {Fukui, Takahiro and Hatsugai, Yasuhiro and Suzuki, Hiroshi},
  year = 2005,
  month = jun,
  journal = {J. Phys. Soc. Jpn.},
  volume = {74},
  number = {6},
  pages = {1674--1677},
  publisher = {The Physical Society of Japan},
  issn = {0031-9015},
  doi = {10.1143/JPSJ.74.1674},
  urldate = {2026-08-18}
}

@article{NatCommun2011ShengFractional,
  title = {Fractional Quantum {{Hall}} Effect in the Absence of {{Landau}} Levels},
  author = {Sheng, D. N. and Gu, Zheng-Cheng and Sun, Kai and Sheng, L.},
  year = 2011,
  month = jul,
  journal = {Nat Commun},
  volume = {2},
  number = {1},
  pages = {389},
  publisher = {Nature Publishing Group},
  issn = {2041-1723},
  doi = {10.1038/ncomms1380},
  urldate = {2026-10-09},
  copyright = {2011 The Author(s)},
  langid = {english}
}

@article{NatCommun2015JacksonGeometric,
  title = {Geometric Stability of Topological Lattice Phases},
  author = {Jackson, T. S. and M{\"o}ller, Gunnar and Roy, Rahul},
  year = 2015,
  month = nov,
  journal = {Nat Commun},
  volume = {6},
  number = {1},
  pages = {8629},
  publisher = {Nature Publishing Group},
  issn = {2041-1723},
  doi = {10.1038/ncomms9629},
  urldate = {2026-10-09},
  copyright = {2015 The Author(s)},
  langid = {english}
}

@article{NatCommun2020KangTopological,
  title = {Topological Flat Bands in Frustrated Kagome Lattice {{CoSn}}},
  author = {Kang, Mingu and Fang, Shiang and Ye, Linda and Po, Hoi Chun and Denlinger, Jonathan and Jozwiak, Chris and Bostwick, Aaron and Rotenberg, Eli and Kaxiras, Efthimios and Checkelsky, Joseph G. and Comin, Riccardo},
  year = 2020,
  month = aug,
  journal = {Nat Commun},
  volume = {11},
  number = {1},
  pages = {4004},
  publisher = {Nature Publishing Group},
  issn = {2041-1723},
  doi = {10.1038/s41467-020-17465-1},
  urldate = {2026-10-09},
  copyright = {2020 The Author(s)},
  langid = {english}
}

@article{NatCommun2025Perea-CausinQuantum,
  title = {Quantum Anomalous {{Hall}} Crystals in Moir\'e Bands with Higher {{Chern}} Number},
  author = {{Perea-Causin}, Raul and Liu, Hui and Bergholtz, Emil J.},
  year = 2025,
  month = jul,
  journal = {Nat Commun},
  volume = {16},
  number = {1},
  pages = {6875},
  publisher = {Nature Publishing Group},
  issn = {2041-1723},
  doi = {10.1038/s41467-025-62224-9},
  urldate = {2026-10-09},
  copyright = {2025 The Author(s)},
  langid = {english}
}

@article{NatCommun2026AhnSuperconductivity,
  title = {Superconductivity Proximate to Non-{{Abelian}} Fractional Spin {{Hall}} Insulator in Twisted Bilayer {{MoTe}}{$_{2}$}},
  author = {Ahn, Cheong-Eung and Seo, Donghae and Lee, Gyeoul and Kim, Youngwook and Cho, Gil Young},
  year = 2026,
  month = may,
  journal = {Nat Commun},
  volume = {17},
  number = {1},
  pages = {7005},
  publisher = {Nature Publishing Group},
  issn = {2041-1723},
  doi = {10.1038/s41467-026-73777-8},
  urldate = {2026-10-16},
  copyright = {2026 The Author(s)},
  langid = {english}
}

@article{NatMater2025XieTunable,
  title = {Tunable Fractional {{Chern}} Insulators in Rhombohedral Graphene Superlattices},
  author = {Xie, Jian and Huo, Zihao and Lu, Xin and Feng, Zuo and Zhang, Zaizhe and Wang, Wenxuan and Yang, Qiu and Watanabe, Kenji and Taniguchi, Takashi and Liu, Kaihui and Song, Zhida and Xie, X. C. and Liu, Jianpeng and Lu, Xiaobo},
  year = 2025,
  month = jul,
  journal = {Nat. Mater.},
  volume = {24},
  number = {7},
  pages = {1042--1048},
  publisher = {Nature Publishing Group},
  issn = {1476-4660},
  doi = {10.1038/s41563-025-02225-7},
  urldate = {2026-10-08},
  copyright = {2025 The Author(s), under exclusive licence to Springer Nature Limited}
}

@article{NatPhys2026BongMonolayer,
  title = {Monolayer van Der {{Waals}} Kagome Lattice for High-Temperature Flat-Band Physics},
  author = {Bong, Hyungkun and Won, Choongjae and Lee, Seongyong and Oh, Soomin and Nguyen, Phuong X. and Chaturvedi, Raghav and Oh, Sehyun and Han, Jiyong and Malinowski, Paul and Mowers, Chad and Kim, Jimin and Ryu, Sae Hee and Lee, Hyunjong and Choi, Jinhyeok and Yer, Sangsu and Jozwiak, Chris and Bostwick, Aaron and Rotenberg, Eli and Polley, Craig M. and Watson, Matthew D. and Kim, Timur K. and Yang, Heejun and Lee, Jieun and Mak, Kin Fai and Shan, Jie and Park, Jae-Hoon and Cho, Gil Young and Shen, Kyle M. and Kang, Mingu},
  year = 2026,
  month = oct,
  journal = {Nat. Phys.},
  publisher = {Nature Publishing Group},
  issn = {1745-2481},
  doi = {10.1038/s41567-026-03492-0},
  eid = {},
  urldate = {2026-10-16},
  copyright = {2026 The Author(s), under exclusive licence to Springer Nature Limited},
  langid = {english}
}

@article{NatPhys2026ParkObservation,
  title = {Observation of Dissipationless Fractional {{Chern}} Insulator},
  author = {Park, Heonjoon and Li, Weijie and Hu, Chaowei and Beach, Christiano and Gon{\c c}alves, Miguel and {Mendez-Valderrama}, Juan Felipe and {Herzog-Arbeitman}, Jonah and Taniguchi, Takashi and Watanabe, Kenji and Cobden, David and Fu, Liang and Bernevig, B. Andrei and Regnault, Nicolas and Chu, Jiun-Haw and Xiao, Di and Xu, Xiaodong},
  year = 2026,
  month = mar,
  journal = {Nat. Phys.},
  volume = {22},
  number = {3},
  pages = {389--395},
  publisher = {Nature Publishing Group},
  issn = {1745-2481},
  doi = {10.1038/s41567-025-03167-2},
  urldate = {2026-08-19},
  copyright = {2026 The Author(s), under exclusive licence to Springer Nature Limited},
  langid = {english}
}

@article{Nature2023CaiSignatures,
  title = {Signatures of Fractional Quantum Anomalous {{Hall}} States in Twisted {{MoTe}}{$_{2}$}},
  author = {Cai, Jiaqi and Anderson, Eric and Wang, Chong and Zhang, Xiaowei and Liu, Xiaoyu and Holtzmann, William and Zhang, Yinong and Fan, Fengren and Taniguchi, Takashi and Watanabe, Kenji and Ran, Ying and Cao, Ting and Fu, Liang and Xiao, Di and Yao, Wang and Xu, Xiaodong},
  year = 2023,
  month = oct,
  journal = {Nature},
  volume = {622},
  number = {7981},
  pages = {63--68},
  publisher = {Nature Publishing Group},
  issn = {1476-4687},
  doi = {10.1038/s41586-023-06289-w},
  urldate = {2026-08-18},
  copyright = {2023 The Author(s), under exclusive licence to Springer Nature Limited},
  langid = {english}
}

@article{Nature2023ParkObservation,
  title = {Observation of Fractionally Quantized Anomalous {{Hall}} Effect},
  author = {Park, Heonjoon and Cai, Jiaqi and Anderson, Eric and Zhang, Yinong and Zhu, Jiayi and Liu, Xiaoyu and Wang, Chong and Holtzmann, William and Hu, Chaowei and Liu, Zhaoyu and Taniguchi, Takashi and Watanabe, Kenji and Chu, Jiun-Haw and Cao, Ting and Fu, Liang and Yao, Wang and Chang, Cui-Zu and Cobden, David and Xiao, Di and Xu, Xiaodong},
  year = 2023,
  month = oct,
  journal = {Nature},
  volume = {622},
  number = {7981},
  pages = {74--79},
  publisher = {Nature Publishing Group},
  issn = {1476-4687},
  doi = {10.1038/s41586-023-06536-0},
  urldate = {2026-08-18},
  copyright = {2023 The Author(s), under exclusive licence to Springer Nature Limited},
  langid = {english}
}

@article{Nature2023ZengThermodynamic,
  title = {Thermodynamic Evidence of Fractional {{Chern}} Insulator in Moir\'e {{MoTe}}{$_{2}$}},
  author = {Zeng, Yihang and Xia, Zhengchao and Kang, Kaifei and Zhu, Jiacheng and Kn{\"u}ppel, Patrick and Vaswani, Chirag and Watanabe, Kenji and Taniguchi, Takashi and Mak, Kin Fai and Shan, Jie},
  year = 2023,
  month = oct,
  journal = {Nature},
  volume = {622},
  number = {7981},
  pages = {69--73},
  publisher = {Nature Publishing Group},
  issn = {1476-4687},
  doi = {10.1038/s41586-023-06452-3},
  urldate = {2026-08-21},
  copyright = {2023 The Author(s), under exclusive licence to Springer Nature Limited},
  langid = {english}
}

@article{Nature2024JiLocal,
  title = {Local Probe of Bulk and Edge States in a Fractional {{Chern}} Insulator},
  author = {Ji, Zhurun and Park, Heonjoon and Barber, Mark E. and Hu, Chaowei and Watanabe, Kenji and Taniguchi, Takashi and Chu, Jiun-Haw and Xu, Xiaodong and Shen, Zhi-Xun},
  year = 2024,
  month = nov,
  journal = {Nature},
  volume = {635},
  number = {8039},
  pages = {578--583},
  publisher = {Nature Publishing Group},
  issn = {1476-4687},
  doi = {10.1038/s41586-024-08092-7},
  urldate = {2026-10-08},
  copyright = {2024 The Author(s), under exclusive licence to Springer Nature Limited}
}

@article{Nature2024KangEvidence,
  title = {Evidence of the Fractional Quantum Spin {{Hall}} Effect in Moir\'e {{MoTe}}{$_{2}$}},
  author = {Kang, Kaifei and Shen, Bowen and Qiu, Yichen and Zeng, Yihang and Xia, Zhengchao and Watanabe, Kenji and Taniguchi, Takashi and Shan, Jie and Mak, Kin Fai},
  year = 2024,
  month = apr,
  journal = {Nature},
  volume = {628},
  number = {8008},
  pages = {522--526},
  publisher = {Nature Publishing Group},
  issn = {1476-4687},
  doi = {10.1038/s41586-024-07214-5},
  urldate = {2026-08-21},
  copyright = {2024 The Author(s), under exclusive licence to Springer Nature Limited},
  langid = {english}
}

@article{Nature2024LuFractional,
  title = {Fractional Quantum Anomalous {{Hall}} Effect in Multilayer Graphene},
  author = {Lu, Zhengguang and Han, Tonghang and Yao, Yuxuan and Reddy, Aidan P. and Yang, Jixiang and Seo, Junseok and Watanabe, Kenji and Taniguchi, Takashi and Fu, Liang and Ju, Long},
  year = 2024,
  month = feb,
  journal = {Nature},
  volume = {626},
  number = {8000},
  pages = {759--764},
  publisher = {Nature Publishing Group},
  issn = {1476-4687},
  doi = {10.1038/s41586-023-07010-7},
  urldate = {2026-08-21},
  copyright = {2024 The Author(s), under exclusive licence to Springer Nature Limited},
  langid = {english}
}

@article{NJP2014PandaCharge,
  title = {A Charge Self-Consistent {{LDA}}+{{DMFT}} Study of the Spectral Properties of Hexagonal {{NiS}}},
  author = {Panda, S K and Thunstr{\"o}m, P and Di Marco, I and Sch{\"o}tt, J and Delin, A and Dasgupta, I and Eriksson, O and Sarma, D D},
  year = 2014,
  month = sep,
  journal = {New J. Phys.},
  volume = {16},
  number = {9},
  pages = {093049},
  publisher = {IOP Publishing},
  issn = {1367-2630},
  doi = {10.1088/1367-2630/16/9/093049},
  urldate = {2026-10-09},
  langid = {english}
}

@article{PNAS2024XuMaximally,
  title = {Maximally Localized {{Wannier}} Functions, Interaction Models, and Fractional Quantum Anomalous {{Hall}} Effect in Twisted Bilayer {{MoTe}}{$_{2}$}},
  author = {Xu, Cheng and Li, Jiangxu and Xu, Yong and Bi, Zhen and Zhang, Yang},
  year = 2024,
  month = feb,
  journal = {Proceedings of the National Academy of Sciences},
  volume = {121},
  number = {8},
  pages = {e2316749121},
  publisher = {Proceedings of the National Academy of Sciences},
  doi = {10.1073/pnas.2316749121},
  urldate = {2026-10-08}
}

@article{PRB1994BlochlProjector,
  title = {Projector Augmented-Wave Method},
  author = {Bl{\"o}chl, P. E.},
  year = 1994,
  month = dec,
  journal = {Phys. Rev. B},
  volume = {50},
  number = {24},
  pages = {17953--17979},
  publisher = {American Physical Society},
  doi = {10.1103/PhysRevB.50.17953},
  urldate = {2026-10-15}
}

@article{PRB1996KresseEfficient,
  title = {Efficient Iterative Schemes for Ab Initio Total-Energy Calculations Using a Plane-Wave Basis Set},
  author = {Kresse, G. and Furthm{\"u}ller, J.},
  year = 1996,
  month = oct,
  journal = {Phys. Rev. B},
  volume = {54},
  number = {16},
  pages = {11169--11186},
  publisher = {American Physical Society},
  doi = {10.1103/PhysRevB.54.11169},
  urldate = {2026-10-15}
}

@article{PRB2008BergmanBand,
  title = {Band Touching from Real-Space Topology in Frustrated Hopping Models},
  author = {Bergman, Doron L. and Wu, Congjun and Balents, Leon},
  year = 2008,
  month = sep,
  journal = {Phys. Rev. B},
  volume = {78},
  number = {12},
  pages = {125104},
  publisher = {American Physical Society},
  doi = {10.1103/PhysRevB.78.125104},
  urldate = {2026-10-09}
}

@article{PRB2009GuoTopological,
  title = {Topological Insulator on the Kagome Lattice},
  author = {Guo, H.-M. and Franz, M.},
  year = 2009,
  month = sep,
  journal = {Phys. Rev. B},
  volume = {80},
  number = {11},
  pages = {113102},
  publisher = {American Physical Society},
  doi = {10.1103/PhysRevB.80.113102},
  urldate = {2026-10-09}
}

@article{PRB2011GundraTheory,
  title = {Theory of the Electro-Optical Properties of Graphene Nanoribbons},
  author = {Gundra, Kondayya and Shukla, Alok},
  year = 2011,
  month = feb,
  journal = {Phys. Rev. B},
  volume = {83},
  number = {7},
  pages = {075413},
  publisher = {American Physical Society},
  doi = {10.1103/PhysRevB.83.075413},
  urldate = {2026-10-08}
}

@article{PRB2011SasiogluEffective,
  title = {Effective {{Coulomb}} Interaction in Transition Metals from Constrained Random-Phase Approximation},
  author = {{\c S}a{\c s}{\i}o{\u g}lu, Ersoy and Friedrich, Christoph and Bl{\"u}gel, Stefan},
  year = 2011,
  month = mar,
  journal = {Phys. Rev. B},
  volume = {83},
  number = {12},
  pages = {121101},
  publisher = {American Physical Society},
  doi = {10.1103/PhysRevB.83.121101},
  urldate = {2026-10-09}
}

@article{PRB2014RoyBand,
  title = {Band Geometry of Fractional Topological Insulators},
  author = {Roy, Rahul},
  year = 2014,
  month = oct,
  journal = {Phys. Rev. B},
  volume = {90},
  number = {16},
  pages = {165139},
  publisher = {American Physical Society},
  doi = {10.1103/PhysRevB.90.165139},
  urldate = {2026-08-21}
}

@article{PRB2016BauerQuantum,
  title = {Quantum Geometry and Stability of the Fractional Quantum {{Hall}} Effect in the {{Hofstadter}} Model},
  author = {Bauer, David and Jackson, T. S. and Roy, Rahul},
  year = 2016,
  month = jun,
  journal = {Phys. Rev. B},
  volume = {93},
  number = {23},
  pages = {235133},
  publisher = {American Physical Society},
  doi = {10.1103/PhysRevB.93.235133},
  urldate = {2026-10-09}
}

@article{PRB2016GoliCorrelated,
  title = {Correlated Electronic Properties of Some Graphene Nanoribbons: {{A DMRG}} Study},
  shorttitle = {Correlated Electronic Properties of Some Graphene Nanoribbons},
  author = {Goli, V. M. L. Durga Prasad and Prodhan, Suryoday and Mazumdar, Sumit and Ramasesha, S.},
  year = 2016,
  month = jul,
  journal = {Phys. Rev. B},
  volume = {94},
  number = {3},
  pages = {035139},
  publisher = {American Physical Society},
  doi = {10.1103/PhysRevB.94.035139},
  urldate = {2026-10-08}
}

@article{PRB2022TeschHubbard,
  title = {Hubbard {{$U$}} Parameters for Transition Metals from First Principles},
  author = {Tesch, Rebekka and Kowalski, Piotr M.},
  year = 2022,
  month = may,
  journal = {Phys. Rev. B},
  volume = {105},
  number = {19},
  pages = {195153},
  publisher = {American Physical Society},
  doi = {10.1103/PhysRevB.105.195153},
  urldate = {2026-10-09}
}

@article{PRB2023KimRealization,
  title = {Realization of Flat Bands by Lattice Intercalation in Kagome Metals},
  author = {Kim, Dongwook and Liu, Feng},
  year = 2023,
  month = may,
  journal = {Phys. Rev. B},
  volume = {107},
  number = {20},
  pages = {205130},
  publisher = {American Physical Society},
  doi = {10.1103/PhysRevB.107.205130},
  urldate = {2026-10-09}
}

@article{PRB2023ReddyFractional,
  title = {Fractional Quantum Anomalous {{Hall}} States in Twisted Bilayer {{MoTe}}{$_2$} and {{WSe}}{$_{2}$}},
  author = {Reddy, Aidan P. and Alsallom, Faisal and Zhang, Yang and Devakul, Trithep and Fu, Liang},
  year = 2023,
  month = aug,
  journal = {Phys. Rev. B},
  volume = {108},
  number = {8},
  pages = {085117},
  publisher = {American Physical Society},
  doi = {10.1103/PhysRevB.108.085117},
  urldate = {2026-08-21}
}

@article{PRB2023ReddyGlobal,
  title = {Toward a Global Phase Diagram of the Fractional Quantum Anomalous {{Hall}} Effect},
  author = {Reddy, Aidan P. and Fu, Liang},
  year = 2023,
  month = dec,
  journal = {Phys. Rev. B},
  volume = {108},
  number = {24},
  pages = {245159},
  publisher = {American Physical Society},
  doi = {10.1103/PhysRevB.108.245159},
  urldate = {2026-08-18}
}

@article{PRB2024AbouelkomsanBand,
  title = {Band Mixing in the Quantum Anomalous {{Hall}} Regime of Twisted Semiconductor Bilayers},
  author = {Abouelkomsan, Ahmed and Reddy, Aidan P. and Fu, Liang and Bergholtz, Emil J.},
  year = 2024,
  month = mar,
  journal = {Phys. Rev. B},
  volume = {109},
  number = {12},
  pages = {L121107},
  publisher = {American Physical Society},
  doi = {10.1103/PhysRevB.109.L121107},
  urldate = {2026-08-21}
}

@article{PRB2024AhnNonAbelian,
  title = {Non-{{Abelian}} Fractional Quantum Anomalous {{Hall}} States and First {{Landau}} Level Physics of the Second Moir\'e Band of Twisted Bilayer {{MoTe}}{$_{2}$}},
  author = {Ahn, Cheong-Eung and Lee, Wonjun and Yananose, Kunihiro and Kim, Youngwook and Cho, Gil Young},
  year = 2024,
  month = oct,
  journal = {Phys. Rev. B},
  volume = {110},
  number = {16},
  pages = {L161109},
  publisher = {American Physical Society},
  doi = {10.1103/PhysRevB.110.L161109},
  urldate = {2026-10-16}
}

@article{PRB2025KwanMoire,
  title = {Moir\'e Fractional {{Chern}} Insulators. {{III}}. {{Hartree-Fock}} Phase Diagram, Magic Angle Regime for {{Chern}} Insulator States, Role of Moir\'e Potential, and {{Goldstone}} Gaps in Rhombohedral Graphene Superlattices},
  author = {Kwan, Yves H. and Yu, Jiabin and {Herzog-Arbeitman}, Jonah and Efetov, Dmitri K. and Regnault, Nicolas and Bernevig, B. Andrei},
  year = 2025,
  month = aug,
  journal = {Phys. Rev. B},
  volume = {112},
  number = {7},
  pages = {075109},
  publisher = {American Physical Society},
  doi = {10.1103/PhysRevB.112.075109},
  urldate = {2026-10-06}
}

@article{PRB2025MitscherlingGaugeinvariant,
  title = {Gauge-Invariant Projector Calculus for Quantum State Geometry and Applications to Observables in Crystals},
  author = {Mitscherling, Johannes and Avdoshkin, Alexander and Moore, Joel E.},
  year = 2025,
  month = aug,
  journal = {Phys. Rev. B},
  volume = {112},
  number = {8},
  pages = {085104},
  publisher = {American Physical Society},
  doi = {10.1103/qscv-qxqt},
  urldate = {2026-10-12}
}

@article{PRB2025PichlerSingleparticle,
  title = {Single-Particle Spectral Function of Fractional Quantum Anomalous {{Hall}} States},
  author = {Pichler, Fabian and Kadow, Wilhelm and Kuhlenkamp, Clemens and Knap, Michael},
  year = 2025,
  month = feb,
  journal = {Phys. Rev. B},
  volume = {111},
  number = {7},
  pages = {075108},
  publisher = {American Physical Society},
  doi = {10.1103/PhysRevB.111.075108},
  urldate = {2026-10-08}
}

@article{PRB2025YuMoire,
  title = {Moir\'e Fractional {{Chern}} Insulators. {{IV}}. {{Fluctuation}}-Driven Collapse in Multiband Exact Diagonalization Calculations on Rhombohedral Graphene},
  author = {Yu, Jiabin and {Herzog-Arbeitman}, Jonah and Kwan, Yves H. and Regnault, Nicolas and Bernevig, B. Andrei},
  year = 2025,
  month = aug,
  journal = {Phys. Rev. B},
  volume = {112},
  number = {7},
  pages = {075110},
  publisher = {American Physical Society},
  doi = {10.1103/PhysRevB.112.075110},
  urldate = {2026-08-18}
}

@article{PRL1982TsuiTwoDimensional,
  title = {Two-{{Dimensional Magnetotransport}} in the {{Extreme Quantum Limit}}},
  author = {Tsui, D. C. and Stormer, H. L. and Gossard, A. C.},
  year = 1982,
  month = may,
  journal = {Phys. Rev. Lett.},
  volume = {48},
  number = {22},
  pages = {1559--1562},
  publisher = {American Physical Society},
  doi = {10.1103/PhysRevLett.48.1559},
  urldate = {2026-09-23}
}

@article{PRL1983LaughlinAnomalous,
  title = {Anomalous {{Quantum Hall Effect}}: {{An Incompressible Quantum Fluid}} with {{Fractionally Charged Excitations}}},
  shorttitle = {Anomalous {{Quantum Hall Effect}}},
  author = {Laughlin, R. B.},
  year = 1983,
  month = may,
  journal = {Phys. Rev. Lett.},
  volume = {50},
  number = {18},
  pages = {1395--1398},
  publisher = {American Physical Society},
  doi = {10.1103/PhysRevLett.50.1395},
  urldate = {2026-08-21}
}

@article{PRL1984ArovasFractional,
  title = {Fractional {{Statistics}} and the {{Quantum Hall Effect}}},
  author = {Arovas, Daniel and Schrieffer, J. R. and Wilczek, Frank},
  year = 1984,
  month = aug,
  journal = {Phys. Rev. Lett.},
  volume = {53},
  number = {7},
  pages = {722--723},
  publisher = {American Physical Society},
  doi = {10.1103/PhysRevLett.53.722},
  urldate = {2026-09-23}
}

@article{PRL1984HalperinStatistics,
  title = {Statistics of {{Quasiparticles}} and the {{Hierarchy}} of {{Fractional Quantized Hall States}}},
  author = {Halperin, B. I.},
  year = 1984,
  month = apr,
  journal = {Phys. Rev. Lett.},
  volume = {52},
  number = {18},
  pages = {1583--1586},
  publisher = {American Physical Society},
  doi = {10.1103/PhysRevLett.52.1583},
  urldate = {2026-09-23}
}

@article{PRL1996PerdewGeneralized,
  title = {Generalized {{Gradient Approximation Made Simple}}},
  author = {Perdew, John P. and Burke, Kieron and Ernzerhof, Matthias},
  year = 1996,
  month = oct,
  journal = {Phys. Rev. Lett.},
  volume = {77},
  number = {18},
  pages = {3865--3868},
  publisher = {American Physical Society},
  doi = {10.1103/PhysRevLett.77.3865},
  urldate = {2026-09-23}
}

@article{PRL2001LichtensteinFiniteTemperature,
  title = {Finite-{{Temperature Magnetism}} of {{Transition Metals}}: {{An}} Ab Initio {{Dynamical Mean-Field Theory}}},
  shorttitle = {Finite-{{Temperature Magnetism}} of {{Transition Metals}}},
  author = {Lichtenstein, A. I. and Katsnelson, M. I. and Kotliar, G.},
  year = 2001,
  month = jul,
  journal = {Phys. Rev. Lett.},
  volume = {87},
  number = {6},
  pages = {067205},
  publisher = {American Physical Society},
  doi = {10.1103/PhysRevLett.87.067205},
  urldate = {2026-10-09}
}

@article{PRL2011NeupertFractional,
  title = {Fractional {{Quantum Hall States}} at {{Zero Magnetic Field}}},
  author = {Neupert, Titus and Santos, Luiz and Chamon, Claudio and Mudry, Christopher},
  year = 2011,
  month = jun,
  journal = {Phys. Rev. Lett.},
  volume = {106},
  number = {23},
  pages = {236804},
  publisher = {American Physical Society},
  doi = {10.1103/PhysRevLett.106.236804},
  urldate = {2026-08-21}
}

@article{PRL2011SunNearly,
  title = {Nearly {{Flatbands}} with {{Nontrivial Topology}}},
  author = {Sun, Kai and Gu, Zhengcheng and Katsura, Hosho and Das Sarma, S.},
  year = 2011,
  month = jun,
  journal = {Phys. Rev. Lett.},
  volume = {106},
  number = {23},
  pages = {236803},
  publisher = {American Physical Society},
  doi = {10.1103/PhysRevLett.106.236803},
  urldate = {2026-08-21}
}

@article{PRL2011TangHighTemperature,
  title = {High-{{Temperature Fractional Quantum Hall States}}},
  author = {Tang, Evelyn and Mei, Jia-Wei and Wen, Xiao-Gang},
  year = 2011,
  month = jun,
  journal = {Phys. Rev. Lett.},
  volume = {106},
  number = {23},
  pages = {236802},
  publisher = {American Physical Society},
  doi = {10.1103/PhysRevLett.106.236802},
  urldate = {2026-08-21}
}

@article{PRL2013SchulerOptimal,
  title = {Optimal {{Hubbard Models}} for {{Materials}} with {{Nonlocal Coulomb Interactions}}: {{Graphene}}, {{Silicene}}, and {{Benzene}}},
  shorttitle = {Optimal {{Hubbard Models}} for {{Materials}} with {{Nonlocal Coulomb Interactions}}},
  author = {Sch{\"u}ler, M. and R{\"o}sner, M. and Wehling, T. O. and Lichtenstein, A. I. and Katsnelson, M. I.},
  year = 2013,
  month = jul,
  journal = {Phys. Rev. Lett.},
  volume = {111},
  number = {3},
  pages = {036601},
  publisher = {American Physical Society},
  doi = {10.1103/PhysRevLett.111.036601},
  urldate = {2026-10-08}
}

@article{PRL2020AbouelkomsanParticleHole,
  title = {Particle-{{Hole Duality}}, {{Emergent Fermi Liquids}}, and {{Fractional Chern Insulators}} in {{Moir\'e Flatbands}}},
  author = {Abouelkomsan, Ahmed and Liu, Zhao and Bergholtz, Emil J.},
  year = 2020,
  month = mar,
  journal = {Phys. Rev. Lett.},
  volume = {124},
  number = {10},
  pages = {106803},
  publisher = {American Physical Society},
  doi = {10.1103/PhysRevLett.124.106803},
  urldate = {2026-10-06}
}

@article{PRL2021LiuGateTunable,
  title = {Gate-{{Tunable Fractional Chern Insulators}} in {{Twisted Double Bilayer Graphene}}},
  author = {Liu, Zhao and Abouelkomsan, Ahmed and Bergholtz, Emil J.},
  year = 2021,
  month = jan,
  journal = {Phys. Rev. Lett.},
  volume = {126},
  number = {2},
  pages = {026801},
  publisher = {American Physical Society},
  doi = {10.1103/PhysRevLett.126.026801},
  urldate = {2026-10-08}
}

@article{PRL2023DongComposite,
  title = {Composite {{Fermi Liquid}} at {{Zero Magnetic Field}} in {{Twisted MoTe}}{$_{2}$}},
  author = {Dong, Junkai and Wang, Jie and Ledwith, Patrick J. and Vishwanath, Ashvin and Parker, Daniel E.},
  year = 2023,
  month = sep,
  journal = {Phys. Rev. Lett.},
  volume = {131},
  number = {13},
  pages = {136502},
  publisher = {American Physical Society},
  doi = {10.1103/PhysRevLett.131.136502},
  urldate = {2026-08-18}
}

@article{PRL2023GoldmanZeroField,
  title = {Zero-{{Field Composite Fermi Liquid}} in {{Twisted Semiconductor Bilayers}}},
  author = {Goldman, Hart and Reddy, Aidan P. and Paul, Nisarga and Fu, Liang},
  year = 2023,
  month = sep,
  journal = {Phys. Rev. Lett.},
  volume = {131},
  number = {13},
  pages = {136501},
  publisher = {American Physical Society},
  doi = {10.1103/PhysRevLett.131.136501},
  urldate = {2026-08-21}
}

@article{PRL2024DongAnomalous,
  title = {Anomalous {{Hall Crystals}} in {{Rhombohedral Multilayer Graphene}}. {{I}}. {{Interaction-Driven Chern Bands}} and {{Fractional Quantum Hall States}} at {{Zero Magnetic Field}}},
  author = {Dong, Junkai and Wang, Taige and Wang, Tianle and Soejima, Tomohiro and Zaletel, Michael P. and Vishwanath, Ashvin and Parker, Daniel E.},
  year = 2024,
  month = nov,
  journal = {Phys. Rev. Lett.},
  volume = {133},
  number = {20},
  pages = {206503},
  publisher = {American Physical Society},
  doi = {10.1103/PhysRevLett.133.206503},
  urldate = {2026-08-21}
}

@article{PRL2024DongTheory,
  title = {Theory of {{Quantum Anomalous Hall Phases}} in {{Pentalayer Rhombohedral Graphene Moir\'e Structures}}},
  author = {Dong, Zhihuan and Patri, Adarsh S. and Senthil, T.},
  year = 2024,
  month = nov,
  journal = {Phys. Rev. Lett.},
  volume = {133},
  number = {20},
  pages = {206502},
  publisher = {American Physical Society},
  doi = {10.1103/PhysRevLett.133.206502},
  urldate = {2026-08-21}
}

@article{PRL2024ShavitQuantum,
  title = {Quantum {{Geometry}} and {{Stabilization}} of {{Fractional Chern Insulators Far}} from the {{Ideal Limit}}},
  author = {Shavit, Gal and Oreg, Yuval},
  year = 2024,
  month = oct,
  journal = {Phys. Rev. Lett.},
  volume = {133},
  number = {15},
  pages = {156504},
  publisher = {American Physical Society},
  doi = {10.1103/PhysRevLett.133.156504},
  urldate = {2026-10-09}
}

@article{PRL2024WangFractional,
  title = {Fractional {{Chern Insulator}} in {{Twisted Bilayer MoTe}}{$_{2}$}},
  author = {Wang, Chong and Zhang, Xiao-Wei and Liu, Xiaoyu and He, Yuchi and Xu, Xiaodong and Ran, Ying and Cao, Ting and Xiao, Di},
  year = 2024,
  month = jan,
  journal = {Phys. Rev. Lett.},
  volume = {132},
  number = {3},
  pages = {036501},
  publisher = {American Physical Society},
  doi = {10.1103/PhysRevLett.132.036501},
  urldate = {2026-08-18}
}

@article{PRL2025ShiFractional,
  title = {Fractional {{Topological States}} in {{Rhombohedral Multilayer Graphene Modulated}} by {{Kagome Superlattice}}},
  author = {Shi, Yanran and Xie, Bo and Ren, Fengfan and Cai, Xinyu and Guo, Zhongqing and Li, Qiao and Lu, Xin and Regnault, Nicolas and Liu, Zhongkai and Liu, Jianpeng},
  year = 2025,
  month = dec,
  journal = {Phys. Rev. Lett.},
  volume = {135},
  number = {25},
  pages = {256603},
  publisher = {American Physical Society},
  doi = {10.1103/21vs-prpj},
  urldate = {2026-10-11}
}

@article{PRL2026GoncalvesSpinless,
  title = {Spinless and {{Spinful Charge Excitations}} in {{Moir\'e Fractional Chern Insulators}}},
  author = {Gon{\c c}alves, Miguel and {Mendez-Valderrama}, Juan Felipe and {Herzog-Arbeitman}, Jonah and Yu, Jiabin and Xu, Xiaodong and Xiao, Di and Bernevig, B. Andrei and Regnault, Nicolas},
  year = 2026,
  month = may,
  journal = {Phys. Rev. Lett.},
  volume = {136},
  number = {19},
  pages = {196503},
  publisher = {American Physical Society},
  doi = {10.1103/gmkp-bxf2},
  urldate = {2026-10-09}
}

@article{PRL2026PanOptical,
  title = {Optical {{Signatures}} of {$-\frac{1}{3}$} {{Fractional Quantum Anomalous Hall State}} in {{Twisted MoTe}}{$_{2}$}},
  author = {Pan, Haiyang and Yang, Shunshun and Wang, Yuzhu and Cai, Xiangbin and Wang, Wei and Zhao, Yan and Watanabe, Kenji and Taniguchi, Takashi and Zhang, Linlong and Liu, Youwen and Yang, Bo and Gao, Weibo},
  year = 2026,
  month = feb,
  journal = {Phys. Rev. Lett.},
  volume = {136},
  number = {5},
  pages = {056601},
  publisher = {American Physical Society},
  doi = {10.1103/f4dj-7sts},
  urldate = {2026-08-21}
}

@article{PRMater2022TasInitio,
  title = {Ab Initio Calculation of the {{Hubbard}} {{$U$}} and {{Hund}} Exchange {{$J$}} in Local Moment Magnets: {{The}} Case of {{Mn}}-Based Full {{Heusler}} Compounds},
  author = {Tas, M. and {\c S}a{\c s}{\i}o{\u g}lu, E. and Bl{\"u}gel, S. and Mertig, I. and Galanakis, I.},
  year = 2022,
  month = nov,
  journal = {Phys. Rev. Mater.},
  volume = {6},
  number = {11},
  pages = {114401},
  publisher = {American Physical Society},
  doi = {10.1103/PhysRevMaterials.6.114401},
  urldate = {2026-10-09}
}

@article{PRResearch2020LedwithFractional,
  title = {Fractional {{Chern}} Insulator States in Twisted Bilayer Graphene: {{An}} Analytical Approach},
  shorttitle = {Fractional {{Chern}} Insulator States in Twisted Bilayer Graphene},
  author = {Ledwith, Patrick J. and Tarnopolsky, Grigory and Khalaf, Eslam and Vishwanath, Ashvin},
  year = 2020,
  month = may,
  journal = {Phys. Rev. Res.},
  volume = {2},
  number = {2},
  pages = {023237},
  publisher = {American Physical Society},
  doi = {10.1103/PhysRevResearch.2.023237},
  urldate = {2026-10-09}
}

@article{PRResearch2020RepellinChern,
  title = {Chern Bands of Twisted Bilayer Graphene: {{Fractional Chern}} Insulators and Spin Phase Transition},
  shorttitle = {Chern Bands of Twisted Bilayer Graphene},
  author = {Repellin, C{\'e}cile and Senthil, T.},
  year = 2020,
  month = may,
  journal = {Phys. Rev. Res.},
  volume = {2},
  number = {2},
  pages = {023238},
  publisher = {American Physical Society},
  doi = {10.1103/PhysRevResearch.2.023238},
  urldate = {2026-10-06}
}

@article{PRResearch2021LiSpontaneous,
  title = {Spontaneous Fractional {{Chern}} Insulators in Transition Metal Dichalcogenide Moir\'e Superlattices},
  author = {Li, Heqiu and Kumar, Umesh and Sun, Kai and Lin, Shi-Zeng},
  year = 2021,
  month = sep,
  journal = {Phys. Rev. Res.},
  volume = {3},
  number = {3},
  pages = {L032070},
  publisher = {American Physical Society},
  doi = {10.1103/PhysRevResearch.3.L032070},
  urldate = {2026-08-21}
}

@article{PRResearch2023DiSanteElectronic,
  title = {Electronic Correlations and Universal Long-Range Scaling in Kagome Metals},
  author = {Di Sante, Domenico and Kim, Bongjae and Hanke, Werner and Wehling, Tim and Franchini, Cesare and Thomale, Ronny and Sangiovanni, Giorgio},
  year = 2023,
  month = jan,
  journal = {Phys. Rev. Res.},
  volume = {5},
  number = {1},
  pages = {L012008},
  publisher = {American Physical Society},
  doi = {10.1103/PhysRevResearch.5.L012008},
  urldate = {2026-10-09}
}

@article{PRResearch2023Morales-DuranPressureenhanced,
  title = {Pressure-Enhanced Fractional {{Chern}} Insulators along a Magic Line in Moir\'e Transition Metal Dichalcogenides},
  author = {{Morales-Dur{\'a}n}, Nicol{\'a}s and Wang, Jie and Schleder, Gabriel R. and Angeli, Mattia and Zhu, Ziyan and Kaxiras, Efthimios and Repellin, C{\'e}cile and Cano, Jennifer},
  year = 2023,
  month = aug,
  journal = {Phys. Rev. Res.},
  volume = {5},
  number = {3},
  pages = {L032022},
  publisher = {American Physical Society},
  doi = {10.1103/PhysRevResearch.5.L032022},
  urldate = {2026-10-08}
}

@article{PRResearch2025AbouelkomsanNonAbelian,
  title = {Non-{{Abelian}} Spin {{Hall}} Insulator},
  author = {Abouelkomsan, Ahmed and Fu, Liang},
  year = 2025,
  month = apr,
  journal = {Phys. Rev. Res.},
  volume = {7},
  number = {2},
  pages = {023083},
  publisher = {American Physical Society},
  doi = {10.1103/PhysRevResearch.7.023083},
  urldate = {2026-10-08}
}

@article{PRX2011RegnaultFractional,
  title = {Fractional {{Chern Insulator}}},
  author = {Regnault, N. and Bernevig, B. Andrei},
  year = 2011,
  month = dec,
  journal = {Phys. Rev. X},
  volume = {1},
  number = {2},
  pages = {021014},
  publisher = {American Physical Society},
  doi = {10.1103/PhysRevX.1.021014},
  urldate = {2026-08-21}
}

@article{PRX2023XuObservation,
  title = {Observation of {{Integer}} and {{Fractional Quantum Anomalous Hall Effects}} in {{Twisted Bilayer MoTe}}{$_{2}$}},
  author = {Xu, Fan and Sun, Zheng and Jia, Tongtong and Liu, Chang and Xu, Cheng and Li, Chushan and Gu, Yu and Watanabe, Kenji and Taniguchi, Takashi and Tong, Bingbing and Jia, Jinfeng and Shi, Zhiwen and Jiang, Shengwei and Zhang, Yang and Liu, Xiaoxue and Li, Tingxin},
  year = 2023,
  month = sep,
  journal = {Phys. Rev. X},
  volume = {13},
  number = {3},
  pages = {031037},
  publisher = {American Physical Society},
  doi = {10.1103/PhysRevX.13.031037},
  urldate = {2026-08-21}
}

@article{RMP2008NayakNonAbelian,
  title = {Non-{{Abelian}} Anyons and Topological Quantum Computation},
  author = {Nayak, Chetan and Simon, Steven H. and Stern, Ady and Freedman, Michael and Das Sarma, Sankar},
  year = 2008,
  month = sep,
  journal = {Rev. Mod. Phys.},
  volume = {80},
  number = {3},
  pages = {1083--1159},
  publisher = {American Physical Society},
  doi = {10.1103/RevModPhys.80.1083},
  urldate = {2026-08-21}
}

@article{TheJournalofChemicalPhysics2003HeydHybrid,
  title = {Hybrid Functionals Based on a Screened {{Coulomb}} Potential},
  author = {Heyd, Jochen and Scuseria, Gustavo E. and Ernzerhof, Matthias},
  year = 2003,
  month = may,
  journal = {J. Chem. Phys.},
  volume = {118},
  number = {18},
  pages = {8207--8215},
  issn = {0021-9606},
  doi = {10.1063/1.1564060},
  urldate = {2026-09-23}
}

@article{Theoreticachimicaacta1964OhnoRemarks,
  title = {Some Remarks on the {{Pariser-Parr-Pople}} Method},
  author = {Ohno, Kimio},
  year = 1964,
  month = jan,
  journal = {Theoret. Chim. Acta},
  volume = {2},
  number = {3},
  pages = {219--227},
  issn = {1432-2234},
  doi = {10.1007/BF00528281},
  urldate = {2026-08-18},
  langid = {english}
}

\end{document}


\title{Supplementary Information for ``Theory of High-Temperature Fractional Chern Phases in Monolayer van der Waals Kagome Materials A$_2$T$_3$X$_4$'' }
 
\author{Soomin Oh}
\affiliation{Department of Physics, Korea Advanced Institute of Science and Technology, Daejeon 34141, Republic of Korea}

\author{Jiyong Han}
\affiliation{Department of Physics and Astronomy, Seoul National University, Seoul 08826, Republic of Korea}

\author{Jinhyeok Choi}
\affiliation{Department of Physics and Astronomy, Seoul National University, Seoul 08826, Republic of Korea}

\author{Donghae Seo}
\affiliation{Natural Science Research Institute, Korea Advanced Institute of Science and Technology, Daejeon 34141, Republic of Korea}

\author{Mingu Kang}
\thanks{iordia@snu.ac.kr}
\affiliation{Department of Physics and Astronomy, Seoul National University, Seoul 08826, Republic of Korea}

\author{Gil Young Cho}
\thanks{gilyoungcho@kaist.ac.kr}
\affiliation{Department of Physics, Korea Advanced Institute of Science and Technology, Daejeon 34141, Republic of Korea}
\affiliation{Center for Artificial Low Dimensional Electronic Systems, Institute for Basic Science, Pohang 37673, Korea}

\maketitle

\tableofcontents

\section{Density Functional Theory Calculation}
In this supplementary section, we describe the computational methods employed to obtain the band structures of A$_2$T$_3$X$_4$. The electronic structures of A$_2$T$_3$X$_4$ were calculated independently using Vienna Ab initio Simulation Package (VASP)~\cite{ComputationalMaterialsScience1996KresseEfficiency, PRB1996KresseEfficient} and Quantum ESPRESSO package (QE)~\cite{JPCM2009GiannozziQUANTUM, JPCM2017GiannozziAdvanced}, yielding consistent results. To ensure experimental relevance of our calculations, the density-functional-theory parameters were refined through comparison with the experimentally measured band structures of A$_2$T$_3$X$_4$~\cite{NatPhys2026BongMonolayer}. In particular, we found that the Heyd-Scuseria-Ernzerhof (HSE)~\cite{TheJournalofChemicalPhysics2003HeydHybrid} hybrid exchange-correlation functional, which incorporates a fraction ($a_{\rm HF}$) of Hartree-Fock exchange, was necessary to accurately reproduce the experimental dispersion of A$_2$T$_3$X$_4$. We used $a_{\rm HF}=5\%$, $5\%$, $15\%$, $15\%$, and $15\%$ for Cs$_2$Ni$_3$S$_4$, Cs$_2$Ni$_3$Se$_4$, (Rb,Cs)$_2$Pd$_3$S$_4$, Rb$_2$Pd$_3$Se$_4$, and Cs$_2$Pt$_3$S$_4$, respectively. Monolayer calculations were performed after relaxing the lattice structure with the Perdew-Burke-Ernzerhof (PBE)~\cite{PRL1996PerdewGeneralized} functional using a $\Gamma$-centered $12\times12\times1$ mesh, and a vacuum spacing of up to $20~\text{\AA}$ was introduced along the out-of-plane direction. For the electronic structure calculations, VASP used projector augmented-wave datasets~\cite{PRB1994BlochlProjector} with a plane-wave energy cutoff of 450~eV for $\Gamma$-centered $5\times5\times1$ $k$-point mesh, whereas QE used optimized norm-conserving Vanderbilt (ONCV) pseudopotentials~\cite{ComputerPhysicsCommunications2018vanSettenPseudoDojo} with a wave cutoff of 100 Ry for $\Gamma$-centered $6\times6\times1$ $k$-point mesh. Spin-orbit coupling effects were included for all calculations. Following the self-consistent HSE calculations, maximally localized Wannier functions were constructed using the Wannier90 package~\cite{ComputerPhysicsCommunications2014MostofiUpdated}. The resulting Wannier Hamiltonians of monolayer A$_2$T$_3$X$_4$, constructed from transition metal (T) $d$ orbitals and chalcogen (X) $p$ orbitals, were used to interpolate the HSE band structures along high-symmetry paths.

\section{Tight-Binding Model and Method} \label{sec:TB Hamiltonian}
In this supplementary section, we write the tight-binding Hamiltonian on a kagome lattice and methods we used for fitting the tight-binding parameters to density-functional-theory band structures. The unit cell of a kagome lattice is spanned by two lattice vectors $\bm{a}_1=a(1, 0)$, $\bm{a}_2=a(1/2, \sqrt{3}/2)$, where we set $a=1$. We write the single-electron Hamiltonian in a 6-component spinor basis $$C_{\bm{k}}= \begin{pmatrix}
c_{\bm{k}A\uparrow} & c_{\bm{k}B\uparrow} & c_{\bm{k}C\uparrow} & c_{\bm{k}A\downarrow} & c_{\bm{k}B\downarrow} & c_{\bm{k}C\downarrow}
\end{pmatrix}^{T},$$
where $A, B, C$ denote the three sublattices of a kagome lattice. The Hamiltonian in momentum space is 
$$H_0=\sum_{\bm{k}} C^\dag_{\bm{k}} \begin{pmatrix}
    h_+(\bm{k}) & 0 \\ 0 & h_-(\bm{k})
\end{pmatrix} C_{\bm{k}}.$$ 

\begin{figure}
    \centering
    \includegraphics[width=0.9\linewidth]{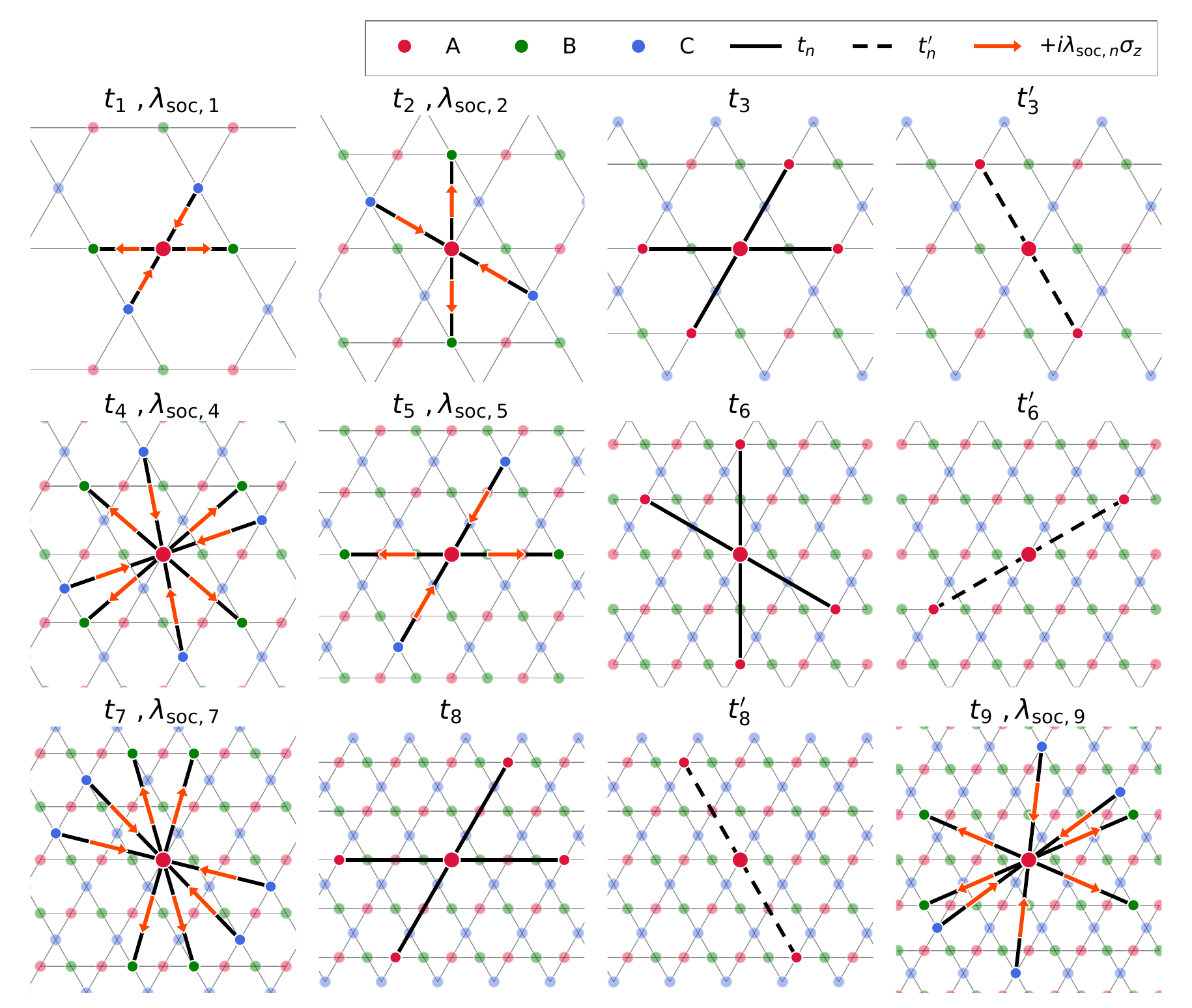}
    \caption{\textbf{Kagome Lattice Configurations.} Schematic illustration of the three sublattices $A$, $B$, and $C$, hopping amplitudes $t_n$ and $t_n'$, and spin-orbit couplings $\lambda_n$. All hopping paths are illustrated with reference to the $A$ sublattice.}
    \label{Fig_kagome_hopping}
\end{figure}

The diagonal components are
\begin{widetext}
\begin{align}
h_{AA\pm}(\bm{k})
= &-t_3 \left(e^{ik_1} + e^{-ik_1}+e^{ik_2} + e^{-ik_2}\right)-t_3'\left(e^{i(k_1-k_2)} + e^{i(k_2-k_1)}\right) \nonumber \\
&-t_6 \left(e^{i(k_1-2k_2)} + e^{i(2k_2-k_1)} + e^{i(2k_1-k_2)} + e^{i(k_2-2k_1)}\right) -t_6' \left(e^{i(k_1+k_2)} + e^{-i(k_1+k_2)}\right) \nonumber \\
&-t_8 \left(e^{i2k_1} + e^{-i2k_1} + e^{i2k_2} + e^{-i2k_2}\right) - t_8' \left(e^{i(2k_1-2k_2)} + e^{i(2k_2-2k_1)}\right), \\
h_{BB\pm}(\bm{k})
= &-t_3 \left(e^{ik_1} + e^{-ik_1}+ e^{i(k_1-k_2)} + e^{i(k_2-k_1)}\right)-t_3' \left(e^{ik_2} + e^{-ik_2}\right) \nonumber \\
&-t_6 \left(e^{i(k_1+k_2)} + e^{-i(k_1+k_2)} + e^{i(k_1-2k_2)} + e^{i(2k_2-k_1)}\right) -t_6' \left(e^{i(2k_1-k_2)} + e^{i(k_2-2k_1)}\right) \nonumber \\
&-t_8 \left(e^{i2k_1} + e^{-i2k_1} + e^{i(2k_1-2k_2)} + e^{i(2k_2-2k_1)}\right) - t_8'\left(e^{i2k_2} + e^{-i2k_2}\right), \\
h_{CC\pm}(\bm{k}) 
= &-t_3 \left(e^{ik_2} + e^{-ik_2} + e^{i(k_1-k_2)} + e^{i(k_2-k_1)}\right)-t_3' \left(e^{ik_1} + e^{-ik_1}\right) \nonumber \\
&-t_6 \left(e^{i(k_1+k_2)} + e^{-i(k_1+k_2)} + e^{i(2k_1-k_2)} + e^{i(k_2-2k_1)}\right)-t_6' \left( e^{i(k_1-2k_2)} + e^{i(2k_2-k_1)}\right) \nonumber \\
&-t_8 \left(e^{i2k_2} + e^{-i2k_2} + e^{i(2k_1-2k_2)} + e^{i(2k_2-2k_1)}\right) - t_8' \left(e^{i2k_1} + e^{-i2k_1}\right),
\end{align}
and the off-diagonal components are
\begin{align}
h_{AB\pm}(\bm{k})=h_{BA\pm}^*(\bm{k})=  &-\left(t_1 \pm i\lambda_1\right)\left(1+e^{-ik_1}\right) -\left(t_2\pm i\lambda_2 \right)\left(e^{-ik_2}+e^{i(k_2-k_1)}\right) \nonumber \\
&-\left(t_4\pm i\lambda_4 \right) \left(e^{ik_2}+e^{-i(k_1+k_2)}+e^{i(k_2-2k_1)}+e^{i(k_1-k_2)}\right) \nonumber \\
&-\left(t_5\pm i\lambda_5 \right) \left(e^{ik_1}+e^{-i2k_1}\right) -\left(t_7\pm i\lambda_7 \right) \left(e^{i(2k_2-k_1)}+e^{-i2k_2}+e^{i(2k_2-2k_1)}+e^{i(k_1-2k_2)}\right) \nonumber \\
&-\left(t_9\pm i\lambda_9 \right)\left(e^{-i(2k_1+k_2)}+e^{i(k_2-3k_1)}+e^{i(k_1+k_2)}+e^{i(2k_2-k_1)}\right), \\
h_{AC\pm}(\bm{k})=h_{CA\pm}^*(\bm{k})=  &-\left(t_1 \mp i\lambda_1\right)\left(1+e^{-ik_2}\right)-\left(t_2\mp i\lambda_2 \right)\left(e^{-ik_1}+e^{i(k_1-k_2)}\right) \nonumber \\
&-\left(t_4\mp i\lambda_4 \right) \left(e^{ik_1}+e^{-i(k_1+k_2)}+e^{i(k_1-2k_2)}+e^{i(k_2-k_1)}\right) \nonumber \\ 
&-\left(t_5\mp i\lambda_5 \right) \left(e^{ik_2}+e^{-i2k_2}\right) -\left(t_7\mp i\lambda_7 \right) \left(e^{i(2k_1-k_2)}+e^{-i2k_1}+e^{i(2k_1-2k_2)}+e^{i(k_2-2k_1)}\right) \nonumber \\
&-\left(t_9\mp i\lambda_9 \right)\left(e^{i(2k_2-k_1)} + e^{i(k_1+k_2)} + e^{i(k_1-3k_2)} + e^{-i(k_1+2k_2)}\right), \\
h_{BC\pm}(\bm{k})=h_{CB\pm}^*(\bm{k})=  &-\left(t_1\pm i\lambda_1 \right)\left(1+e^{i(k_1-k_2)}\right)-\left(t_2\pm i\lambda_2 \right)\left(e^{ik_1} + e^{-ik_2}\right) \nonumber \\
&-\left(t_4\pm i\lambda_4 \right) \left(e^{-ik_1} + e^{ik_2} + e^{i(k_1-2k_2)} + e^{i(2k_1-k_2)}\right) \nonumber \\
&-\left(t_5\pm i\lambda_5 \right) \left(e^{i(k_2-k_1)}+e^{i(2k_1-2k_2)}\right) -\left(t_7\pm i\lambda_7 \right) \left(e^{i2k_1}+e^{-i2k_2}+e^{i(k_1+k_2)}+e^{-i(k_1+k_2)}\right) \nonumber \\
&-\left(t_9\pm i\lambda_9 \right)\left(e^{i(2k_1-3k_2)} + e^{i(3k_1-2k_2)} + e^{i(2k_1-k_2)} + e^{i(k_2-2k_1)}\right).
\end{align}
\end{widetext}
Here, $k_i=\bm{k}\cdot \bm{a}_i$, and $t_n$, $\lambda_n$ represent the $n$th-nearest-neighbor hopping and spin-orbit coupling parameters, respectively. Note that there are two symmetry-inequivalent contributions of third-, sixth- and eighth-nearest-neighbor hoppings, which are labeled $t_3$, $t_6$, $t_8$ and $t_3'$, $t_6'$, $t_8'$. This is because there are two distinct real-space paths through different intermediate atoms. The real space configurations of sublattices, hoppings, and spin-orbit couplings are illustrated in Fig.~\ref{Fig_kagome_hopping}.

We next diagonalize the Hamiltonian as $$\sum_{\beta}h_{\alpha\beta\sigma}(\bm{k})u_{n\beta\sigma}(\bm{k})=\varepsilon_{n\sigma}(\bm{k})u_{n\alpha\sigma}(\bm{k}),$$ where $\alpha, \beta=A,B,C$ denote the sublattices. We define the operators in the band basis $c^\dag_{n\bm{k}\sigma}=\sum_\alpha u_{n\alpha\sigma}(\bm{k})c^\dag_{\bm{k}\alpha\sigma}$. With this, $H_0$ becomes
\begin{equation}
H_0=\sum_{\bm{k}n\sigma}\varepsilon_n(\bm{k})c^\dag_{n\bm{k}\sigma}c_{n\bm{k}\sigma}.
\end{equation}
The three kagome bands are spin-degenerate, i.e. $\varepsilon_{n\sigma}(\bm{k})\equiv \varepsilon_n(\bm{k})$, before introducing ferromagnetism.

The parameters of the kagome materials A\textsubscript{2}T\textsubscript{3}X\textsubscript{4} are listed in Tables \ref{tab:tightbindingparams} and \ref{tab:socparams}. For Cs$_2$Pd$_3$S$_4$ and Rb$_2$Pd$_3$S$_4$, we have adopted the parameters used in \cite{NatPhys2026BongMonolayer}. For the remaining materials in the A$_2$T$_3$X$_4$ family, we performed a neural-network-assisted optimization of the tight-binding Hamiltonian to obtain an unbiased fit to the density-functional-theory band structures. Three independently trainable neural-network modules were assigned to distinct subsets of the tight-binding parameters. The first module (N1) generated the on-site energy and nearest-neighbor hopping integrals; the second module (N2) generated the remaining hopping integrals; and the third module (N3) generated the spin–orbit-coupling parameters. The network parameters were optimized by minimizing the weighted mean squared error between the tight-binding and density-functional-theory band structures. Within each optimization cycle, N1 and N2 were optimized sequentially using the density-functional-theory bands without spin–orbit coupling, followed by optimization of N3 using the density-functional-theory bands with spin–orbit coupling. For each material, we compared two initialization protocols, random initialization and initialization from tight-binding parameters extracted from the Wannier Hamiltonians, yielding consistent results. The optimized tight-binding parameter sets closely reproduce the density-functional-theory bands within the target energy window, as shown in Fig. \ref{Fig2_DFT_TB_comparison}.

\begin{table*}[h!]
\centering
\begin{tabular}{c|cccccccccccc}
A$_2$T$_3$X$_4$ & $t_1$ & $t_2$ & $t_3$ & $t_3'$ & $t_4$ & $t_5$ & $t_6$ & $t_6'$ & $t_7$ & $t_8$ & $t_8'$ & $t_9$ \\
\hline
Cs\textsubscript{2}Pd\textsubscript{3}S\textsubscript{4} & 0.1134 & 0.0164 & 0.0255 & 0.0255 & 0.0067 & -0.0064 & -0.0024 & 0.0009 & 0.008 & 0 & 0 & 0 \\
Rb\textsubscript{2}Pd\textsubscript{3}S\textsubscript{4} & 0.1146 & 0.0154 & 0.0195 & 0.0176 & 0.0105 & -0.0095 & 0.0014 & -0.0037 & 0.0063 & 0 & 0 & 0 \\
Cs\textsubscript{2}Ni\textsubscript{3}S\textsubscript{4} & 0.0323 & 0.0013 & 0.0078 & 0.0047 & 0.0088 & -0.0149 & -0.0039 & 0.0033 & 0.0034 & -0.0024 & 0.0018 & 0 \\
Cs\textsubscript{2}Ni\textsubscript{3}Se\textsubscript{4} & 0.0351 & -0.0146 & 0.0193 & 0.0039 & 0.0129 & -0.0154 & -0.0060 & 0.0048 & 0.0034 & -0.0039 & 0.0046 & -0.0004 \\
Rb\textsubscript{2}Pd\textsubscript{3}Se\textsubscript{4} & 0.1103 & -0.0255 & 0.0484 & 0.0019 & 0.0118 & -0.009 & 0.0055 & -0.0096 & 0.0049 & 0 & 0 & 0 \\
Cs\textsubscript{2}Pt\textsubscript{3}S\textsubscript{4} & 0.0757 & 0.0403 & -0.0025 & 0.0389 & -0.0067 & 0.0002 & -0.0061 & 0.0042 & 0.0052 & -0.0026 & -0.002 & 0.0002 \\
\end{tabular}
\caption{\textbf{Tight-Binding Parameters.} The parameters are in eV units.}
\label{tab:tightbindingparams}
\end{table*}

\begin{table*}[h!]
\centering
\begin{tabular}{c|cccccc}
A$_2$T$_3$X$_4$ &  $\lambda_{1}$ & $\lambda_2$ & $\lambda_4$ & $\lambda_5$ & $\lambda_7$ & $\lambda_9$\\
\hline
Cs\textsubscript{2}Pd\textsubscript{3}S\textsubscript{4} & 0.0167 & 0 & 0 & 0 & 0 & 0 \\
Rb\textsubscript{2}Pd\textsubscript{3}S\textsubscript{4} & 0.0165 & 0 & 0 & 0 & 0 & 0 \\
Cs\textsubscript{2}Ni\textsubscript{3}S\textsubscript{4} & 0.011 & 0.0001 & 0.0014 & -0.0005 & -0.0002 & 0 \\
Cs\textsubscript{2}Ni\textsubscript{3}Se\textsubscript{4} & 0.0117 & -0.0004 & 0.0005 & -0.0010 & 0.0001 & 0 \\
Rb\textsubscript{2}Pd\textsubscript{3}Se\textsubscript{4} & 0.0115 & -0.0092 & 0.0028 & -0.0023 & 0.0007 & 0 \\
Cs\textsubscript{2}Pt\textsubscript{3}S\textsubscript{4} & 0.0046 & 0.0335 & -0.003 & 0.0036 & 0.0007 & 0 \\
\end{tabular}
\caption{\textbf{Spin-Orbit Coupling Parameters.} The parameters are in eV units.}
\label{tab:socparams}
\end{table*}

\begin{table*}[h!]
\centering
\begin{tabular}{c|cccccc}
A$_2$T$_3$X$_4$ & Cs$_2$Pd$_3$S$_4$ & Rb$_2$Pd$_3$S$_4$ & Cs$_2$Ni$_3$S$_4$ & Cs$_2$Ni$_3$Se$_4$ & Rb$_2$Pd$_3$Se$_4$ & Cs$_2$Pt$_3$S$_4$ \\
\hline
Full Bandwidth & 1.086 & 1.059 & 0.392 & 0.519 & 1.158 & 0.810 \\
SOC gap at Dirac point & 0.058 & 0.057 & 0.049 & 0.054 & 0.144 & 0.105 \\
Flat Bandwidth & 0.151 & 0.149 & 0.154 & 0.264 & 0.480 & 0.171 \\
\end{tabular}
\caption{\textbf{Characteristic Energy Scales.} Bandwidths and spin-orbit coupling (SOC)-induced gaps of monolayer kagome A$_2$T$_3$X$_4$ compounds, obtained from the tight-binding band structures. All values are given in eV.}
\label{tab:bandparams}
\end{table*}

\section{Quantum Band Geometry}
\begin{figure}
    \centering
    \includegraphics[width=0.9\linewidth]{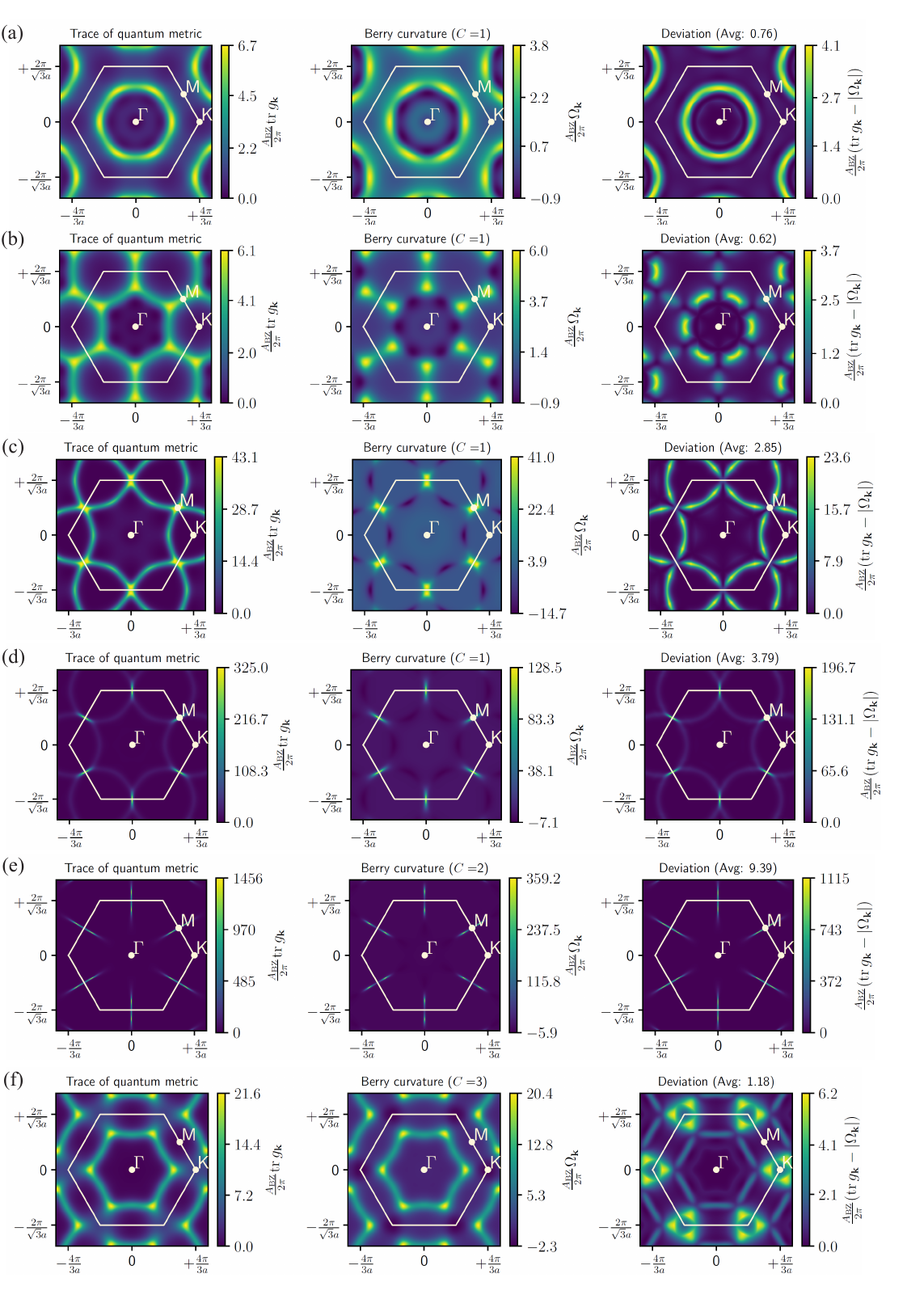}
    \caption{\textbf{Flat-Band Quantum Geometry.} $\mathrm{Tr}\,g(\bm{k})$ (left), $\Omega(\bm{k})$ (middle), and $\mathrm{Tr}\,g(\bm{k})-|\Omega(\bm{k})|$ (right) for \textbf{(a)} Cs\textsubscript{2}Pd\textsubscript{3}S\textsubscript{4}, \textbf{(b)} Rb\textsubscript{2}Pd\textsubscript{3}S\textsubscript{4}, \textbf{(c)} Cs\textsubscript{2}Ni\textsubscript{3}S\textsubscript{4}, \textbf{(d)} Cs\textsubscript{2}Ni\textsubscript{3}Se\textsubscript{4}, \textbf{(e)} Rb\textsubscript{2}Pd\textsubscript{3}Se\textsubscript{4}, and \textbf{(f)} Cs\textsubscript{2}Pt\textsubscript{3}S\textsubscript{4} are presented. The trace-condition violations $\mathcal{T} = \frac{1}{2\pi}\int_{\bm{k}}\,[\operatorname{Tr}g(\bm{k})-|\Omega(\bm{k})|]$ are \textbf{(a)} 0.76, \textbf{(b)} 0.62, \textbf{(c)} 2.85, \textbf{(d)} 3.79, \textbf{(e)} 9.39, and \textbf{(f)} 1.18.}
    \label{Fig_Quantum_Geometry}
\end{figure}

{In this supplementary section, we compare the quantum geometry of the flat kagome bands across the six compounds to characterize their relative suitability for fractional Chern insulators. We evaluate the quantum metric, $\mathrm{Tr}\,g(\bm{k})$, the Berry curvature, $\Omega(\bm{k})$, and their difference, $\mathrm{Tr}\,g(\bm{k})-|\Omega(\bm{k})|$, which are displayed as momentum-space maps in Fig.~\ref{Fig_Quantum_Geometry}.} These quantities are evaluated using the gauge-invariant projector expressions~\cite{PRB2025MitscherlingGaugeinvariant}
\begin{align}
    \mathrm{Tr}\,g(\bm{k})
    &= \frac{1}{2}\mathrm{tr}\left[
    \bigl(\partial_x\hat{P}(\bm{k})\bigr)^2
    + \bigl(\partial_y\hat{P}(\bm{k})\bigr)^2
    \right], \\
    \Omega(\bm{k})
    &= i\,\mathrm{tr}\left[
    \hat{P}(\bm{k})
    \left[\partial_x\hat{P}(\bm{k}),\partial_y\hat{P}(\bm{k})\right]
    \right],
\end{align}
where $\hat{P}(\bm{k})=|u_{n\bm{k}}\rangle\langle u_{n\bm{k}}|$ projects onto the normalized cell-periodic Bloch eigenstate of the isolated band $n$. Here, $\mathrm{tr}$ denotes the trace over the orbital and spin degrees of freedom of the Bloch Hamiltonian, whereas $\mathrm{Tr}\,g=g_{xx}+g_{yy}$ denotes the trace over Cartesian momentum indices. All three quantities plotted [Fig.~\ref{Fig_Quantum_Geometry}] are multiplied by $A_{\mathrm{BZ}}/2\pi$, where $A_{\mathrm{BZ}}$ is the area of the first Brillouin zone.

To model ferromagnetic spin polarization, we introduce a sufficiently large Zeeman splitting to fully separate the spin-up and spin-down bands in energy. We then evaluate the quantum geometry of the flat band with a positive Chern number. The results are presented in Fig.~\ref{Fig_Quantum_Geometry}. $\mathrm{Cs}_2\mathrm{Pd}_3\mathrm{S}_4$ and $\mathrm{Rb}_2\mathrm{Pd}_3\mathrm{S}_4$ offer small trace-condition violations of about 0.6--0.8, where the rest of the materials range from 1.1 to 9.3.

\section{Many-Body Hamiltonian}
{In this supplementary section, we specify the interacting Hamiltonians used in the monolayer and bilayer calculations.} Rewriting the full monolayer many-body Hamiltonian including the interacting term, we have
\begin{align}
    H &= H_0 + H_{\rm int} \nonumber \\
    &= \sum_{\bm{k}n\sigma}\varepsilon_n(\bm{k})c^\dag_{n\bm{k}\sigma}c_{n\bm{k}\sigma}
    + 
    \frac{1}{2A}\sum_{\bm{kk'q}}\sum_{\substack{nn'mm' \\ \sigma\sigma'}}\hat{V}^{nn'mm'\sigma\sigma'}_{\bm{k},\bm{k'} ,\bm{q}}c^\dag_{n'\bm{k+q}\sigma}c^\dag_{m'\bm{k'-q}\sigma'}c_{m\bm{k'}\sigma'}c_{n\bm{k}\sigma}, \label{manybodyHamiltonian}
\end{align}
where $A$ is the sample area. $\hat{V}^{nn'mm'\sigma\sigma'}_{\bm{k},\bm{k'},\bm{q}}$ is the two-body interaction matrix 
\begin{equation}
    \hat{V}^{nn'mm'\sigma\sigma'}_{\bm{k},\bm{k'},\bm{q}} = \sum_{\bm Q} V(\bm q+ \bm Q)\Lambda^{n'n\sigma}_{\bm{k},\bm{q}+\bm{Q}} \left(\Lambda^{mm'\sigma'}_{\bm{k}'-\bm q,\bm{q}+\bm{Q}}\right)^*, \label{eq:twobodyinteractionMatrix}
\end{equation}
where $\bm{Q}$ are the reciprocal lattice vectors, and the band form factor $\Lambda^{n'n\sigma}_{\bm{k},\bm{q}+\bm{Q}}$ is
\begin{equation}
\Lambda^{n'n\sigma}_{\bm{k},\bm{q}+\bm{Q}}=\sum_\alpha
    u^*_{ n'\alpha\sigma}(\bm{k}+\bm{q})
    u_{ n\alpha\sigma}(\bm{k})
    e^{i(\bm{q}+\bm{Q})\cdot \tau_\alpha}.
    \label{eq:form_factor}
\end{equation}
Here, $\tau_{\alpha=A,B,C}$ represents the sublattice coordinates, and the interaction $V(\bm q+ \bm Q)$ is introduced in the main text, \begin{equation}
    V(\mathbf{p})=\frac{e^2}{2 \epsilon p}e^{-pU_C/U_0}, \label{FTOhnoInteraction}
\end{equation}
where $p=|\bm{p}|$, $U_0$ is the onsite Hubbard repulsion, and $U_C=e^2/4\pi\epsilon$ is the Coulomb prefactor. We use $U_0=3~\mathrm{eV}$ and $\epsilon=5\epsilon_0$ as in the main text. A uniform neutralizing background removes the spatially uniform component of the interaction, and we set $V(\bm{p}=0)=0$. The sum over reciprocal vectors $\bm Q$ is an infinite sum, but in practice, we truncate it at an appropriate cutoff as the interaction \eqref{FTOhnoInteraction} exponentially decays.

{To extend the model to a bilayer, we include the interlayer Coulomb interaction in addition to the intralayer Ohno interaction. We neglect interlayer tunneling, so the particle number in each layer is separately conserved.} Further assuming that the ferromagnetism is robust, we drop the spin index $\sigma$ and the Hamiltonian is of the form
\begin{align}
    H_{\rm bilayer} &= H_0 + H_{\rm int} \nonumber \\
    &= \sum_{\bm{k}n\ell}\varepsilon_n^{(\ell)}(\bm{k})c^\dag_{\ell n\bm{k}}c_{\ell n\bm{k}}+ 
    \frac{1}{2A}\sum_{\bm{kk'q}}\sum_{\substack{\ell\ell' \\ nn'mm'}}\hat{V}^{\ell\ell'nn'mm'}_{\bm{k},\bm{k'},\bm{q}}c^\dag_{\ell n'\bm{k+q}}c^\dag_{\ell'm'\bm{k'-q}}c_{\ell' m\bm{k'}}c_{\ell n\bm{k}}, \label{BilayerMBHamiltonian}
\end{align}
where $\ell, \ell'=1, 2$ are layer pseudospin indices, and we restrict ourselves to homogeneous bilayers, $\varepsilon_n^{(1)}(\bm{k})=\varepsilon_n^{(2)}(\bm{k})$. The interaction matrix is now
\begin{equation}
    \hat{V}^{\ell\ell'nn'mm'}_{\bm{k},\bm{k'},\bm{q}} = \sum_{\bm Q} V^{\ell\ell'}(\bm q+ \bm Q)\Lambda^{\ell n'n}_{\bm{k},\bm{q}+\bm{Q}} \left(\Lambda^{\ell'mm'}_{\bm{k}'-\bm q,\bm{q}+\bm{Q}}\right)^*, \label{eq:bilayertwobodyinteractionMatrix}
\end{equation}
where we introduce different types of potentials for intralayer and interlayer interactions,
\begin{equation}
    V^{\ell\ell'}(\mathbf{p})=\frac{e^2}{2\epsilon p}
    \begin{cases} 
e^{-pU_C/U_0}, \quad \ell = \ell' \\ 
e^{-pd}, \quad \ell\neq\ell' 
\end{cases}, \label{eq:bilayerinteraction}
\end{equation}
with the interlayer distance $d$. The form factor $\Lambda$ essentially has the same expression as \eqref{eq:form_factor}, replacing the spin index $\sigma$ with the layer index $\ell$. The uniform neutralizing background leaves a finite layer-antisymmetric contribution by the capacitive correction,
\begin{align}
    E_{\rm cap} &=\frac{(N_1-N_2)^2}{4A}\lim_{p \to 0} \left[V^{\ell\ell}(\mathbf p) - V^{\ell\bar{\ell}}(\mathbf p)\right] \nonumber \\ 
    &= \frac{e^2}{8\epsilon A}\left(d-\frac{U_C}{U_0}\right)(N_1 - N_2)^2. \label{eq:capacitiveE}
\end{align}

\section{Numerical Details}
In this supplementary section, we discuss the details of the Hamiltonian and band mixing process that we apply in our exact diagonalization calculations.

\subsection{Spinful Exact Diagonalization}
\begin{figure}
    \centering
    \includegraphics[width=0.5\linewidth]{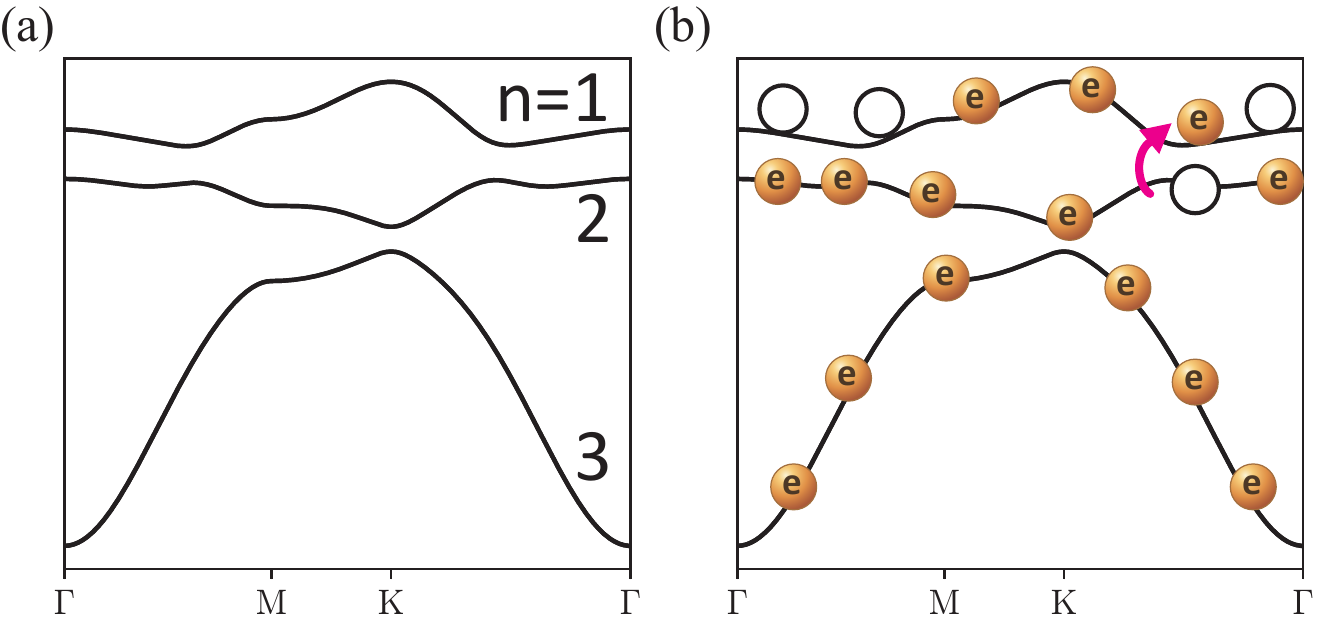}
    \caption{\textbf{Band Labels and Band-Mixing Scheme.} \textbf{(a)} The three kagome bands labeled $n=1,2,3$ in descending order of energy. \textbf{(b)} Schematic illustration of the band-mixing scheme employed in our calculations. We keep band 3 fully occupied and allow at most one interband excitation from band 2 to band 1. The tight-binding band structure of Cs$_2$Pd$_3$S$_4$ is used for illustration.}
    \label{Fig_Bandmixing_schematic}
\end{figure}

We describe the effective Hamiltonian used to investigate the possible ferromagnetism. We start from the many-body Hamiltonian \eqref{manybodyHamiltonian}. This Hamiltonian contains a total of 6 bands, 3 for each spin species, which is impractical to handle in exact diagonalization. Therefore, as we perform $S_z$-resolved calculations, we retain only the topmost flat band of each spin species as an active degree of freedom. Labeling the three kagome bands by $n=1,2,3$ in descending order of energy [Fig.~\ref{Fig_Bandmixing_schematic}(a)], we assume that bands $n=2, 3$ are fully occupied and inert for both spins. Then the spin number $S_z=\left(N_{\uparrow}-N_{\downarrow}\right)/2$ is determined by the difference of electrons filling band 1 for each spin component.

The effective Hamiltonian is obtained by projecting onto this restricted Hilbert space,
$$
H_{\mathrm{eff}}=PHP,
\quad
P=\prod_{\bm k}\prod_{\sigma=\uparrow,\downarrow}
\prod_{a=2,3}\hat n_{a\bm k\sigma},
$$
where $\hat n_{a\bm k\sigma} =c^\dagger_{a\bm k\sigma}c_{a\bm k\sigma}.$
Here, $H_{\mathrm{eff}}$ is understood as an operator acting within the projected subspace. This projection retains the static Hartree and Fock contributions from the filled bands while excluding excitations that create holes in them. Up to a constant, the resulting Hamiltonian reads
\begin{align}
&H_{\mathrm{eff}}
=
\sum_{\bm k\sigma}
\left[\varepsilon_1(\bm k)+\Sigma_\sigma(\bm k)\right]
c^\dagger_{1\bm k\sigma}c_{1\bm k\sigma} +\frac{1}{2A}
\sum_{\bm k\bm k'\bm q}
\sum_{\sigma\sigma'}
\hat V^{1111\sigma\sigma'}_{\bm k,\bm k',\bm q}
c^\dagger_{1\bm k+\bm q\sigma}
c^\dagger_{1\bm k'-\bm q\sigma'}
c_{1\bm k'\sigma'}c_{1\bm k\sigma}, \label{eq:effspinfulHamiltonian}
\end{align}

where the filled-band self-energy is
$$
\Sigma_\sigma(\bm k)
=
\frac{1}{A}\sum_{\bm k'}\sum_{a=2,3}
\left[
\sum_{\tau=\uparrow,\downarrow}
\hat V^{11aa\sigma\tau}_{\bm k,\bm k',0}
-
\hat V^{1aa1\sigma\sigma}_{\bm k,\bm k',\bm k'-\bm k}
\right].
$$
The Hartree contribution includes both spin species in the filled bands, whereas the Fock contribution involves only filled-band electrons with the same spin as the electron in band $n=1$.

For the bilayer systems, the strategy we use is essentially the same, as we drop the spin index $\sigma$ and replace it with the layer pseudospin index $\ell$. The interaction matrix $\hat{V}^{1111\sigma\sigma'}_{\bm k,\bm k',\bm q}$ in \eqref{eq:effspinfulHamiltonian} is replaced with $\hat{V}^{\ell\ell'1111}_{\bm k,\bm k',\bm q}$, introduced in \eqref{eq:bilayertwobodyinteractionMatrix}. Here, the potential $V^{\ell\ell'}(\mathbf{p})$ differs from the monolayer spinful case.

\subsection{Band Mixing in Ferromagnetic Regime}
We next incorporate interband mixing within the maximally spin-polarized regime, provided that the spin gap is sufficiently large to restrict the low-energy analysis to the maximally spin-polarized sector. For this, we first show that the fully occupied majority-spin bands contribute only an additive constant within the spin-$S_z$ sector. We then retain the two upper minority-spin bands and restrict their configurations to at most one interband electron–hole excitation.

Taking $\downarrow$ to be the majority spin, we introduce a projection operator for the ferromagnetic sector $$P_{\rm FM}=\prod_{\bm k}\prod_{m=1}^{3}\hat n_{m\bm k\downarrow},$$ where the three majority-spin bands are fully filled. Note that the many-body Hamiltonian \eqref{manybodyHamiltonian} can be decomposed into intraspin and interspin components, $$H= H_{\uparrow} + H_{\downarrow} + H_{\rm mix}.$$ Here, the intraspin component, $H_{\sigma=\uparrow, \downarrow}$, is defined as
\begin{align}
    &H_{\sigma} = \sum_{\bm{k}n}\varepsilon_n(\bm{k})c^\dag_{n\bm{k}\sigma}c_{n\bm{k}\sigma} + 
    \frac{1}{2A}\sum_{\bm{kk'q}}\sum_{nn'mm'}\hat{V}^{nn'mm'\sigma\sigma}_{\bm{k},\bm{k'} ,\bm{q}}c^\dag_{n'\bm{k+q}\sigma}c^\dag_{m'\bm{k'-q}\sigma}c_{m\bm{k'}\sigma}c_{n\bm{k}\sigma}, 
\end{align}
and $H_{\rm mix}$ denotes the interspin component which is a sum of $(\sigma,\sigma')=(\uparrow,\downarrow)$ and
$(\downarrow,\uparrow)$ terms in $H_{\rm int}$. {As we project onto the subspace of the maximally spin-polarized sector, the effective Hamiltonian becomes }
\begin{align}
P_{\rm FM}HP_{\rm FM} &= P_{\rm FM}(H_{\uparrow} + H_{\downarrow} + H_{\rm mix})P_{\rm FM} \nonumber \\
    &= (H_{\uparrow} + E_{\downarrow})P_{\rm FM} + P_{\rm FM}H_{\rm mix}P_{\rm FM}. \label{eq:effspinlessHamiltonian}
\end{align}
Here, $E_\downarrow$ is a constant energy since all spin-$\downarrow$ electrons occupy their orbitals. We also show that the interspin interaction $H_{\rm mix}$ only gives a constant shift as well. The last term is expressed as
$$P_{\rm FM}H_{\rm mix}P_{\rm FM}
=
\sum_{\bm{k},nn'}
\Sigma^{H,\downarrow}_{n'n}(\bm{k})
c^\dagger_{n'\bm{k}\uparrow}
c_{n\bm{k}\uparrow}P_{\rm FM},$$ where
\begin{equation}
    \Sigma^{H,\downarrow}_{n'n}(\bm{k})
=
\frac{1}{A}
\sum_{\bm{p}}\sum_{m=1}^{3}
\hat{V}^{nn'mm\uparrow\downarrow}_
{\bm{k},\bm{p},\bm{0}}
\end{equation}
is the Hartree self-energy. Calculating this term explicitly,
\begin{equation}
\begin{split}
    \Sigma^{H,\downarrow}_{n'n}(\bm{k})
&=
\frac{1}{A}
\sum_{\bm{p}}\sum_{m=1}^{3}\sum_{\bm{Q}}
V(\bm{Q})\,
\Lambda^{n'n\uparrow}_{\bm{k},\bm{Q}}
\left(
\Lambda^{mm\downarrow}_{\bm{p},\bm{Q}}
\right)^* = \frac{1}{A}\sum_{\bm{Q}\alpha\beta}
V(\bm{Q})
e^{i\bm{Q}\cdot(\tau_\alpha-\tau_\beta)}u^*_{ n'\alpha\uparrow}(\bm{k})
    u_{ n\alpha\uparrow}(\bm{k})
\sum_{\bm{p}, m}
|u_{m\beta\downarrow}(\bm{p})|^2 \\
&= \frac{N}{A}\sum_{\alpha} u^*_{ n'\alpha\uparrow}(\bm{k})
    u_{ n\alpha\uparrow}(\bm{k})\sum_{\bm{Q}\beta} V(\bm{Q})
e^{i\bm{Q}\cdot(\tau_\alpha-\tau_\beta)}.
\end{split}
\end{equation}
We have used the completeness of the Bloch eigenvectors, $\sum_{m=1}^{3}
u_{m\alpha\downarrow}(\bm{p})
u^*_{m\beta\downarrow}(\bm{p})
=
\delta_{\alpha\beta}.$
Due to the threefold rotational symmetry of the three sublattices $\alpha, \beta = A, B, C$, and provided that $\bm{Q}$ is summed properly, the term $\sum_{\bm{Q} \beta} V(\bm{Q})
e^{i\bm{Q}\cdot(\tau_\alpha-\tau_\beta)}$ is independent of $\alpha$. As a result, together with $N/A$, this term can be treated as a constant $C_H$ while summing over $\alpha$,
\begin{equation}
\begin{split}
\Sigma^{H,\downarrow}_{n'n}(\bm{k})
&=
C_H\sum_\alpha
u^*_{n'\alpha\uparrow}(\bm{k})
u_{n\alpha\uparrow}(\bm{k})
\\
&=
C_H\delta_{n'n},
\end{split}
\end{equation}
using orthonormality again. Therefore,
\begin{equation}
    \sum_{\bm{k}nn'}
\Sigma^{H,\downarrow}_{n'n}(\bm{k})
c^\dagger_{n'\bm{k}\uparrow}
c_{n\bm{k}\uparrow}=C_H\sum_{\bm{k}n}
c^\dagger_{n\bm{k}\uparrow}
c_{n\bm{k}\uparrow} = C_H\hat{N}_{\uparrow},
\end{equation}
which is a constant shift as long as the particle number $N_\uparrow$ is fixed. Consequently, the effective Hamiltonian \eqref{eq:effspinlessHamiltonian} becomes
\begin{align}
    P_{\rm FM}HP_{\rm FM} &= (H_{\uparrow} + E_{\downarrow}+C_H\hat{N}_{\uparrow})P_{\rm FM} \nonumber \\
    &= H_{\uparrow}P_{\rm FM} + \text{const.}
\end{align}
The remaining problem is therefore equivalent to a spinless three-band system, and we suppress the minority-spin index $\uparrow$ below.

To incorporate band mixing within a tractable Hilbert space, we keep band $3$ fully occupied and allow at most one electron to be promoted from band $2$ to band $1$ [Fig.~\ref{Fig_Bandmixing_schematic}(b)]. We define
\begin{align*}
    &P_3=\prod_{\bm k}\hat n_{3\bm k}, \quad
\hat N_{h,2} = \sum_{\bm k}\left(1-\hat n_{2\bm k}\right), \quad Q_{\leq1}= \Pi_{N_{h,2}=0}+\Pi_{N_{h,2}=1},
\end{align*}
where $\hat n_{n\bm k}=c^\dagger_{n\bm k}c_{n\bm k}$ and $\Pi_{N_{h,2}=r}$ projects onto states with exactly $r$ holes in band $2$. At fixed minority-spin particle number, the two retained sectors correspond to zero and one interband excitation, respectively. {The full projector and the corresponding truncated Hamiltonian are}
$$
P_{\rm full}=P_{\mathrm{FM}}P_3Q_{\leq1},
\qquad
H_{\mathrm{trunc}}=P_{\rm full}HP_{\rm full}.
$$
Here, $H_{\mathrm{trunc}}$ acts within the projected Hilbert space. After suppressing the frozen degrees of freedom and omitting additive constants, it can be written as
$$
H_{\mathrm{trunc}}=Q_{\leq1}H_{12}Q_{\leq1},
$$
where
\begin{align}
&H_{12}=
\sum_{\bm k}\sum_{n,n'=1,2}
\left[
\varepsilon_n(\bm k)\delta_{nn'}
+\Sigma^{(3)}_{nn'}(\bm k)
\right]
c^\dagger_{n\bm k}c_{n'\bm k} +
\frac{1}{2A}
\sum_{\bm k\bm k'\bm q}
\sum_{\substack{n,n',m,m'=1,2}}
\hat V^{nn'mm'}_{\bm k,\bm k',\bm q}
c^\dagger_{n'\bm k+\bm q}
c^\dagger_{m'\bm k'-\bm q}
c_{m\bm k'}c_{n\bm k}. 
\end{align}

Here, the fully occupied band $3$ contributes the static Hartree–Fock self-energy
$$
\Sigma^{(3)}_{nn'}(\bm k)
=
\frac{1}{A}\sum_{\bm p}
\left[
\hat V^{n'n33}_{\bm k,\bm p,\bm 0}
-
\hat V^{n'33n}_{\bm k,\bm p,\bm p-\bm k}
\right].
$$
Unlike the Hartree contribution from the complete majority-spin sector, this correction need not be proportional to the identity or independent of momentum. The final projection retains all matrix elements within and between the zero- and one-hole sectors of band $2$, while excluding states with two or more holes in band $2$ or any holes in band $3$. \\

In the calculations described above, the valence bands are fully occupied and inert, and their static Hartree–Fock contribution is retained. A common alternative is to formulate exact diagonalization by projecting onto a target band without an explicit self-energy contribution from the inert bands~\cite{PRL2020AbouelkomsanParticleHole, PRResearch2020RepellinChern, PRResearch2021LiSpontaneous, PRB2023ReddyFractional, PRL2024WangFractional}. These two treatments are analogous, respectively, to the average (AVE) and charge-neutrality (CN) interaction schemes discussed in Refs.~\cite{PRB2025KwanMoire, PRB2025YuMoire}, which differ in the reference used to define the interaction and the associated one-body terms. Unless otherwise specified, all results presented in this paper retain the inert-band self-energy, analogous to the AVE convention.

\section{Spontaneous Symmetry Breaking}
\begin{figure}
    \centering
    \includegraphics[width=0.95\linewidth]{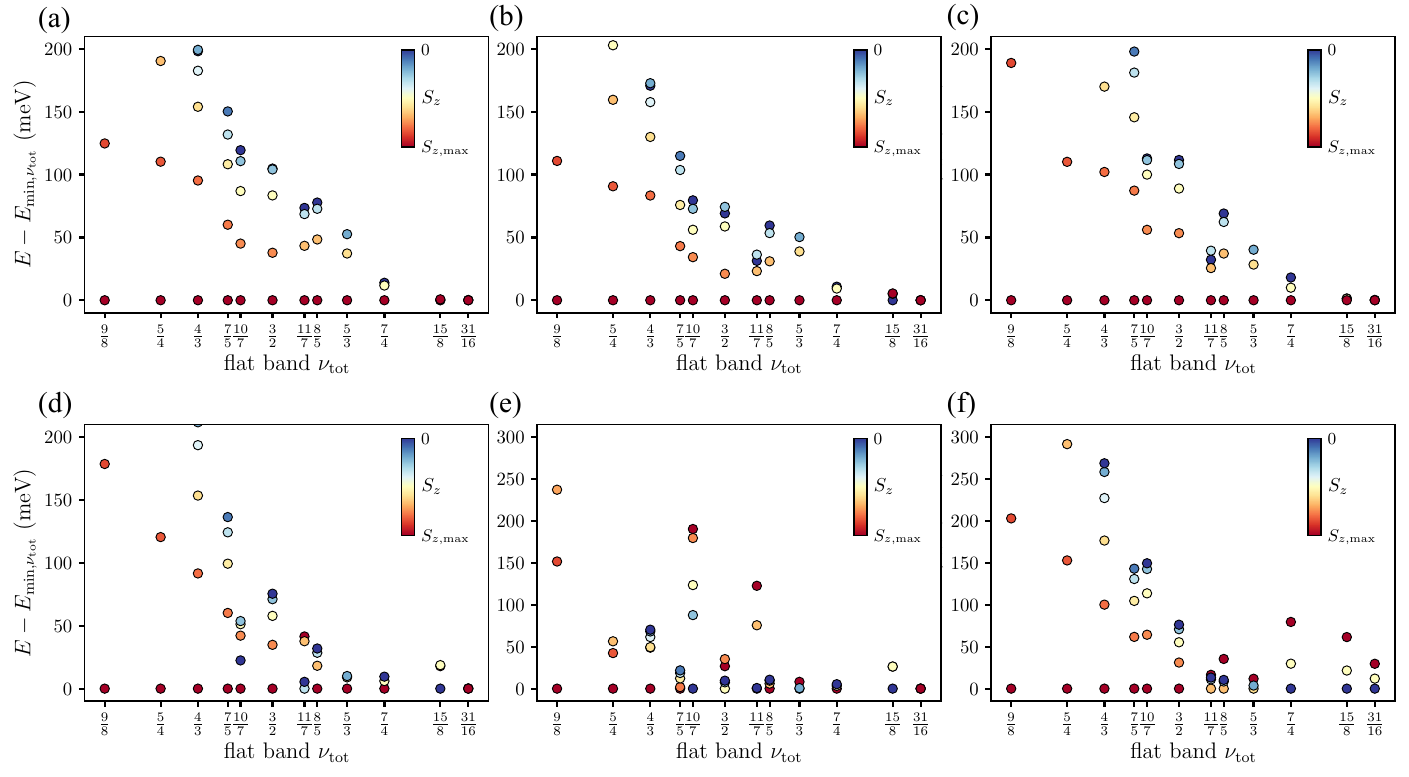}
    \caption{\textbf{Spinful Many-Body Spectra.} The spin-resolved many-body spectra at various fillings are presented for \textbf{(a)} Cs\textsubscript{2}Pd\textsubscript{3}S\textsubscript{4}, \textbf{(b)} Rb\textsubscript{2}Pd\textsubscript{3}S\textsubscript{4}, \textbf{(c)} Cs\textsubscript{2}Ni\textsubscript{3}S\textsubscript{4}, \textbf{(d)} Cs\textsubscript{2}Ni\textsubscript{3}Se\textsubscript{4}, \textbf{(e)} Rb\textsubscript{2}Pd\textsubscript{3}Se\textsubscript{4}, and \textbf{(f)} Cs\textsubscript{2}Pt\textsubscript{3}S\textsubscript{4}.}
    \label{Fig_spinful_ED_Ohno}
\end{figure}

In this supplementary section, we present the spin-resolved many-body spectra used to determine the ferromagnetic regimes displayed in Fig.~\ref{Fig1} of the main text. Fig.~\ref{Fig_spinful_ED_Ohno} shows the lowest energy in each $S_z$ sector at fixed total filling, measured relative to the overall minimum at that filling. For the momentum clusters, $N=14$ was used for $\nu_{\rm tot}=\frac{10}{7}, \frac{11}{7}$, $N=15$ was used for $\nu_{\rm tot}=\frac{4}{3}, \frac{7}{5}, \frac{8}{5}, \frac{5}{3}$, $N=16$ was used for $\nu_{\rm tot}=\frac{9}{8}, \frac{5}{4}, \frac{3}{2}, \frac{7}{4}$, and  $N=32$ and $64$ were used for $\nu_{\rm tot}=\frac{15}{8}, \frac{31}{16}$, respectively. $\nu_{\rm tot}=\frac{31}{16}$ is the largest accessible filling which can be separated into at least three $S_z$ sectors. We note that the spectra at $\nu_{\rm tot}=\frac{15}{8}, \frac{31}{16}$ either favor paramagnetic sector or have 0.1 meV or less order of separations (the spectra overlap and only the ferromagnetic sector is shown in the figure)  for all cases except Cs\textsubscript{2}Ni\textsubscript{3}S\textsubscript{4} at $\nu_{\rm tot}=\frac{15}{8}$, and therefore identify them as paramagnetic. Cs\textsubscript{2}Ni\textsubscript{3}S\textsubscript{4} has a ferromagnetic ground state with a spin gap of about 1 meV at $\nu_{\rm tot}=\frac{15}{8}$.

Comparing the flat bandwidth presented in Table~\ref{tab:bandparams} with the stability of ferromagnetism over different electron fillings, we observe that the robustness of ferromagnetism over a broad range of fillings is correlated with the flatness of the topmost kagome band. The ordering of the flat bandwidths is:
\begin{align*}
    \mathrm{Cs}_2\mathrm{Pd}_3\mathrm{S}_4 \sim \mathrm{Rb}_2\mathrm{Pd}_3\mathrm{S}_4 \sim \mathrm{Cs}_2\mathrm{Ni}_3\mathrm{S}_4 < \mathrm{Cs}_2\mathrm{Pt}_3\mathrm{S}_4 < \mathrm{Cs}_2\mathrm{Ni}_3\mathrm{Se}_4 < \mathrm{Rb}_2\mathrm{Pd}_3\mathrm{Se}_4. 
\end{align*}
Notably, the ferromagnetic filling range tends to decrease as the flat bandwidth increases. We organize the relation in Table~\ref{tab:FerroandFlatbandwidth}:
\begin{table*}[h!]
\centering
\begin{tabular}{c|c|c|c|c|c|c}
A$_2$T$_3$X$_4$ & Cs$_2$Pd$_3$S$_4$ & Rb$_2$Pd$_3$S$_4$ & Cs$_2$Ni$_3$S$_4$ & Cs$_2$Pt$_3$S$_4$ & Cs$_2$Ni$_3$Se$_4$ & Rb$_2$Pd$_3$Se$_4$  \\
\hline
Flat Bandwidth & 0.151 & 0.149 & 0.154 & 0.171 & 0.264 & 0.480  \\
\hline
Ferromagnetism & $1<\nu_{\rm tot}<15/8$ & $1<\nu_{\rm tot}<15/8$ & $1<\nu_{\rm tot}<15/8$ & $1<\nu_{\rm tot}<11/7$ & $1<\nu_{\rm tot}<3/2$ & $1<\nu_{\rm tot}<7/5$
\end{tabular}
\caption{\textbf{Correlation of Flat Bandwidth and Ferromagnetism.} A comparison of flat bandwidth and the range of ferromagnetic regions for the kagome materials A$_2$T$_3$X$_4$ is summarized. The flat bandwidth is written in eV units.}
\label{tab:FerroandFlatbandwidth}
\end{table*}

\begin{figure}
    \centering
    \includegraphics[width=0.95\linewidth]{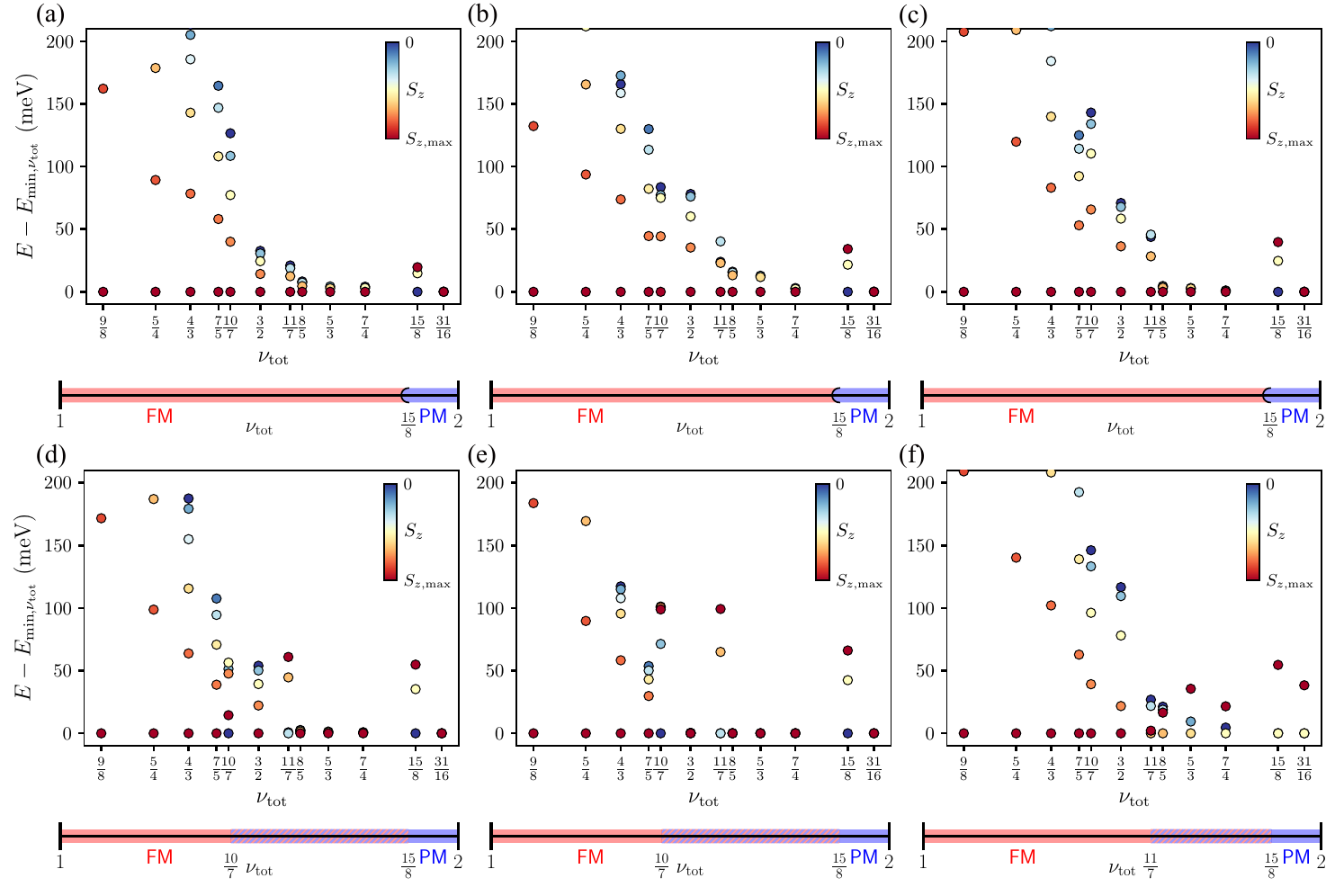}
    \caption{\textbf{Spinful Hubbard Many-Body Spectra and Phase Diagrams.} The spin-resolved many-body spectra at various fillings with onsite Hubbard repulsion are presented. The corresponding phase diagrams are shown below each energy spectrum, for \textbf{(a)} Cs\textsubscript{2}Pd\textsubscript{3}S\textsubscript{4}, \textbf{(b)} Rb\textsubscript{2}Pd\textsubscript{3}S\textsubscript{4}, \textbf{(c)} Cs\textsubscript{2}Ni\textsubscript{3}S\textsubscript{4}, \textbf{(d)} Cs\textsubscript{2}Ni\textsubscript{3}Se\textsubscript{4}, \textbf{(e)} Rb\textsubscript{2}Pd\textsubscript{3}Se\textsubscript{4}, and \textbf{(f)} Cs\textsubscript{2}Pt\textsubscript{3}S\textsubscript{4}.}
    \label{Fig_spinful_ED_Hubbard}
\end{figure}

We further find that our phase diagram remains robust under varying the interaction type. For onsite Hubbard interaction, the many-body spectra [Fig.~\ref{Fig_spinful_ED_Hubbard}] reveal a similar trend, and consequently the symmetry breaking remains largely unchanged.

\section{Momentum Grids}
\begin{figure}
    \centering
    \includegraphics[width=1\linewidth]{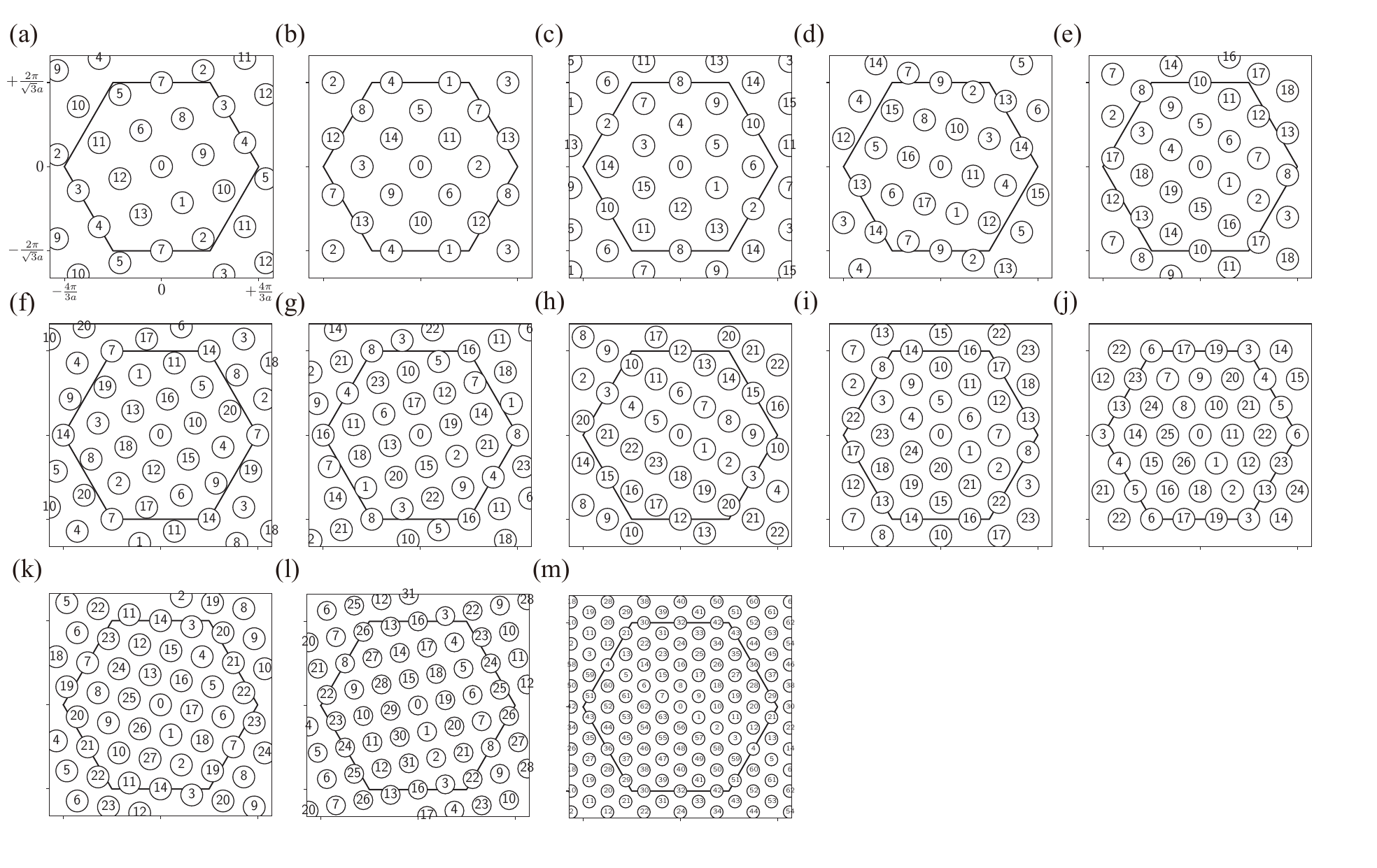}
    \caption{\textbf{Finite-Size Brillouin Zones.} Discrete momentum grids used in our exact diagonalization calculations. The parameters $(N,u_{11},u_{12},u_{21},u_{22})$ are given by \textbf{(a)} $(14, 7, 0, 2, 2)$, \textbf{(b)} $(15, 5, 0, 6, 3)$, \textbf{(c)} $(16, 4, 0, 0, 4)$, \textbf{(d)} $(18, 9, 0, 4, 2)$, \textbf{(e)} $(20, 5, 0, 0, 4)$, \textbf{(f)} $(21, 7, 7, -6, -3)$, \textbf{(g)} $(24, 8, 8, -12, -9)$, \textbf{(h)} $(24, 6, 0, 0, 4)$, \textbf{(i)} $(25, 5, 0, 0, 5)$, \textbf{(j)} $(27, 9, 0, 3, 3)$, \textbf{(k)} $(28, 14, 0, 4, 2)$, \textbf{(l)} $(32, 16, 0, 4, 2)$, \textbf{(m)} $(64, 8, 0, 0, 8)$.}
    \label{Fig_BZgrid}
\end{figure}

For our exact diagonalization calculations, we used finite-size periodic lattices of various system sizes $N$. To diagonalize on an $N$-unit cell lattice with periodic boundary conditions, we define two boundary vectors $\mathbf{L}_1, \mathbf{L}_2$ such that there exists $u_{ij}\in\mathbb{Z}$ with $\mathbf{L}_i=\sum_{j=1,2}u_{ij}\bm{a}_j$. Then the number of unit cells corresponds to $$N=\frac{||\mathbf{L}_1\times \mathbf{L}_2||}{||\bm{a}_1\times \bm{a}_2||}=\det u,$$ and the two vectors $\bm{g}_1, \bm{g}_2$ that satisfy $\mathbf{L}_i\cdot \bm{g}_j=2\pi\delta_{ij}$ span the discrete Brillouin zone. We further introduce $N_1, N_2$ so that our $u$ satisfies $u_{ij}=N_i\Bar{u}_{ij}$ where $N_i, \Bar{u}_{ij}\in\mathbb{Z}$ and $\det \Bar{u}=1$. Then $N=N_1N_2$, and we have a parametrization $K=k_1 + N_1k_2$ in our calculation. The grids that we used are plotted in Fig.~\ref{Fig_BZgrid}. 

\section{Additional Results at Fractional Fillings}
In this supplementary section, we present additional numerical studies at fractional fillings for different kagome monolayers. All energy spectra presented and discussed in this section adopt band mixing with at most one interband electron--hole excitation, as mentioned in the main text.
\paragraph{Many-Body Chern Number}
We calculate the many-body Berry curvature on the torus of boundary twists \(\boldsymbol{\Phi}=(\Phi_1,\Phi_2)\), with \(\Phi_\mu\in[0,2\pi)\) using the gauge-invariant discretization~\cite{JPSJ2005FukuiChern}. For an $N_{\rm GS}$-state ground-state multiplet, we define the link variables \(U_\mu(\boldsymbol{\Phi})=\det W_\mu(\boldsymbol{\Phi})/|\det W_\mu(\boldsymbol{\Phi})|\), where \(W_\mu\) is the overlap matrix between neighboring multiplets, including the evolution of the single-particle basis. The discretized multiplet Berry curvature is
\begin{equation}
    F(\Phi_1,\Phi_2)=\frac{\operatorname{Arg}\!\left[U_1(\boldsymbol{\Phi})
U_2(\boldsymbol{\Phi}+\boldsymbol{\delta}_1)U_1^*(\boldsymbol{\Phi}+\boldsymbol{\delta}_2)
U_2^*(\boldsymbol{\Phi})\right]}{\Delta\Phi_1\Delta\Phi_2},
\end{equation}
where \(\boldsymbol{\delta}_\mu=\Delta\Phi_\mu\hat{\boldsymbol e}_\mu\). Its integral divided by the ground-state degeneracy $N_{\rm GS}$ yields the many-body Chern number, $$C_{\rm MB}=\frac{1}{2\pi N_{\rm GS}}\sum_{\boldsymbol{\Phi}}F(\Phi_1,\Phi_2)\Delta\Phi_1\Delta\Phi_2 = \frac{2\pi}{N_\Phi N_{\rm GS}} \sum_{\boldsymbol{\Phi}}F(\Phi_1,\Phi_2),$$ where $N_\Phi$ is the number of grid meshes. Our many-body Chern number calculation was conducted on at least $N_\Phi=12\times 12$ grid. In the following subsection, we plot the normalized average Berry curvature $2\pi F(\Phi_1, \Phi_2)/N_{\rm GS}$, where its average over the boundary-twist torus equals $C_{\rm MB}$. \\

\begin{figure}[]
    \centering
    \includegraphics[width=1\linewidth]{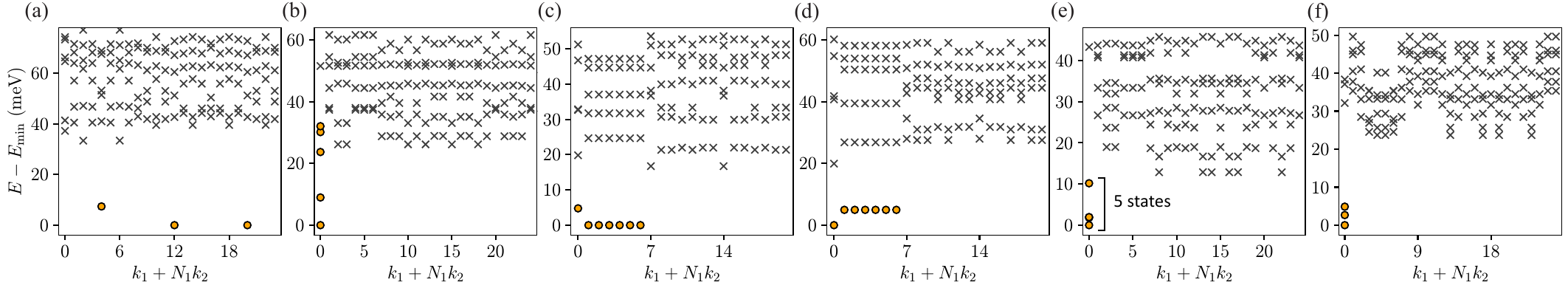}
    \caption{\textbf{Many-Body Spectra of Cs\textsubscript{2}Pd\textsubscript{3}S\textsubscript{4}.} Exact diagonalization energy spectra at \textbf{(a)} $\nu_{\rm tot}=4/3$, $N=24$, \textbf{(b)} $\nu_{\rm tot}=7/5$, $N=25$, \textbf{(c)} $\nu_{\rm tot}=10/7$, $N=21$, \textbf{(d)} $\nu_{\rm tot}=11/7$, $N=21$, \textbf{(e)} $\nu_{\rm tot}=8/5$, $N=25$, and \textbf{(f)} $\nu_{\rm tot}=5/3$, $N=27$.}
    \label{Fig_CsPdS_Jain}
\end{figure}

\begin{figure}
    \centering
    \includegraphics[width=0.85\linewidth]{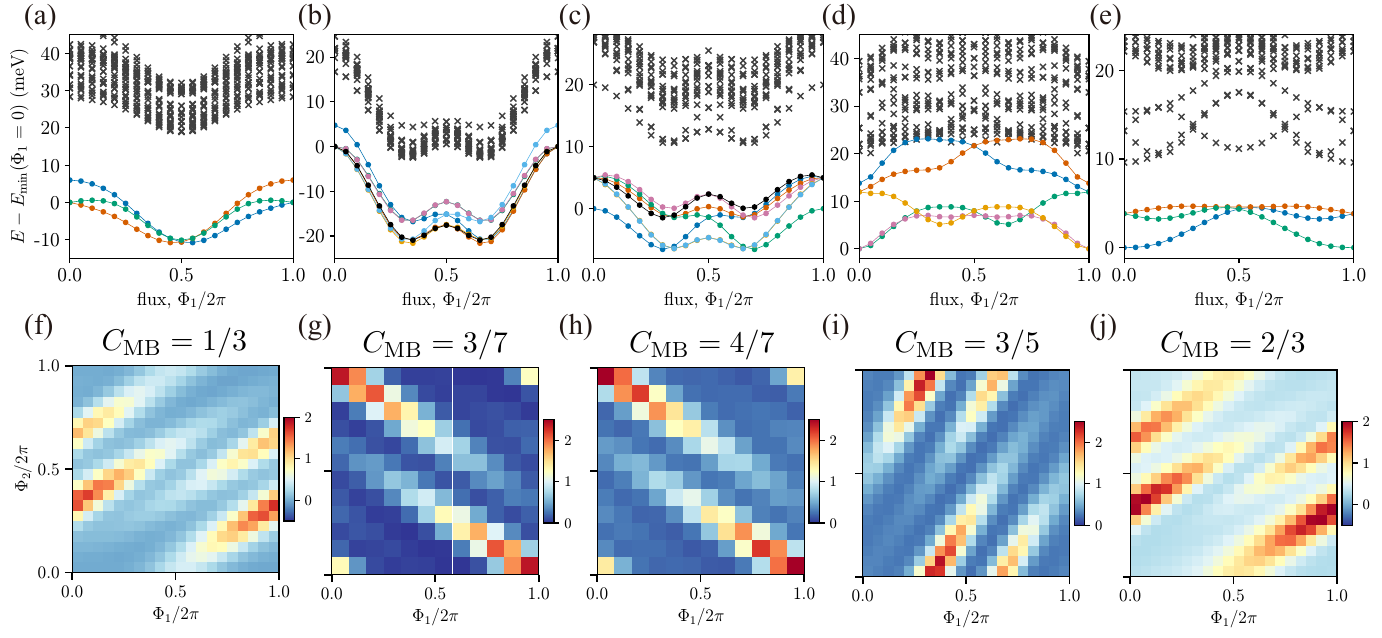}
    \caption{\textbf{Spectral Flow and Many-Body Berry Curvature of Cs\textsubscript{2}Pd\textsubscript{3}S\textsubscript{4}.} \textbf{(a-e)} Spectral flow under one flux insertion and \textbf{(f-j)} normalized average Berry curvature $2\pi F(\Phi_1, \Phi_2)/N_{\rm GS}$ of Cs\textsubscript{2}Pd\textsubscript{3}S\textsubscript{4} at \textbf{(a,f)} $\nu_{\rm tot}=4/3$, $N=18$, \textbf{(b,g)} $\nu_{\rm tot}=10/7$, $N=21$, \textbf{(c,h)} $\nu_{\rm tot}=11/7$, $N=21$, \textbf{(d,i)} $\nu_{\rm tot}=8/5$, $N=15$, and \textbf{(e,j)} $\nu_{\rm tot}=5/3$, $N=18$.}
    \label{Fig_CsPdS_Jain_spectral_mbChern}
\end{figure}

\paragraph{Fractional Chern Insulators}
We report additional details of fractional Chern insulators that we discovered in our exact diagonalization study. We first summarize the many-body spectra of $\mathrm{Cs}_2\mathrm{Pd}_3\mathrm{S}_4$ at Jain sequence fillings $\nu_{\rm tot}=4/3,\,7/5,\,10/7,\,11/7,\,8/5,\,5/3$ at the largest accessible system size [Fig.~\ref{Fig_CsPdS_Jain}]. The fractional Chern insulator at $\nu_{\rm tot}=4/3$ [Fig.~\ref{Fig_CsPdS_Jain}(a)] is also provided in the main text [Fig.~\ref{Fig3_EDspectra}(a)]. The result for $\nu_{\rm tot}=7/5$ [Fig.~\ref{Fig_CsPdS_Jain}(b)] is shown as an example for an undetermined phase. The fivefold states are not ground states, and hence the corresponding fractional Chern insulator phase is absent from the phase diagram [Fig.~\ref{Fig1}(d)]. The calculations of spectral flow under flux insertion and many-body Chern number were done for the five identified Jain sequence fractional Chern insulators [Fig.~\ref{Fig_CsPdS_Jain_spectral_mbChern}], where we used a smaller system size if possible to reduce computational cost. In the spectral flow plot, at most three eigenstates per momentum sector are presented. The spectral flow for $\nu_{\rm tot}=4/3$ [Fig.~\ref{Fig_CsPdS_Jain_spectral_mbChern}(a)] is also provided in the main text [Fig.~\ref{Fig3_EDspectra}(b)].

\begin{figure}
    \centering
    \includegraphics[width=1\linewidth]{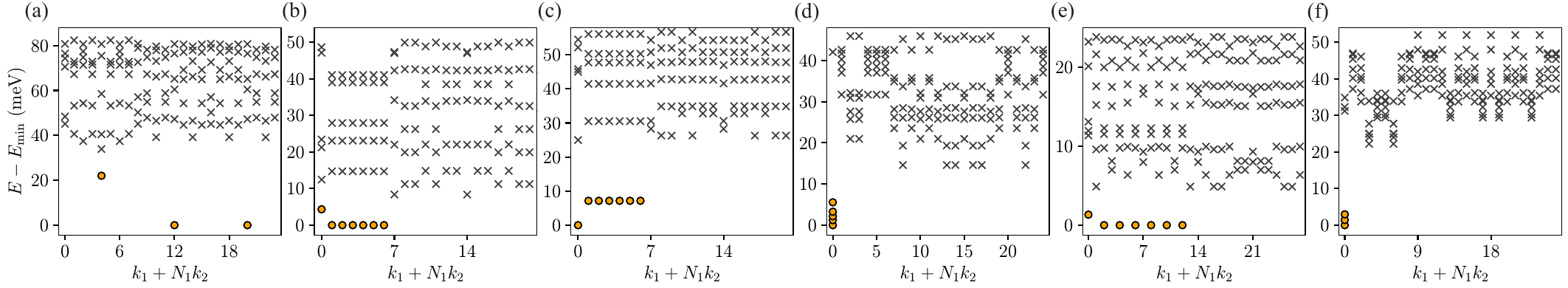}
    \caption{\textbf{Many-Body Spectra of Rb\textsubscript{2}Pd\textsubscript{3}S\textsubscript{4}.} Exact diagonalization energy spectra at \textbf{(a)} $\nu_{\rm tot}=4/3$, $N=24$, \textbf{(b)} $\nu_{\rm tot}=10/7$, $N=21$, \textbf{(c)} $\nu_{\rm tot}=11/7$, $N=21$, \textbf{(d)} $\nu_{\rm tot}=8/5$, $N=25$, \textbf{(e)} $\nu_{\rm tot}=12/7$, $N=28$, and \textbf{(f)} $\nu_{\rm tot}=5/3$, $N=27$.}
    \label{Fig_RbPdS_Jain}
\end{figure}

\begin{figure}
    \centering
    \includegraphics[width=1\linewidth]{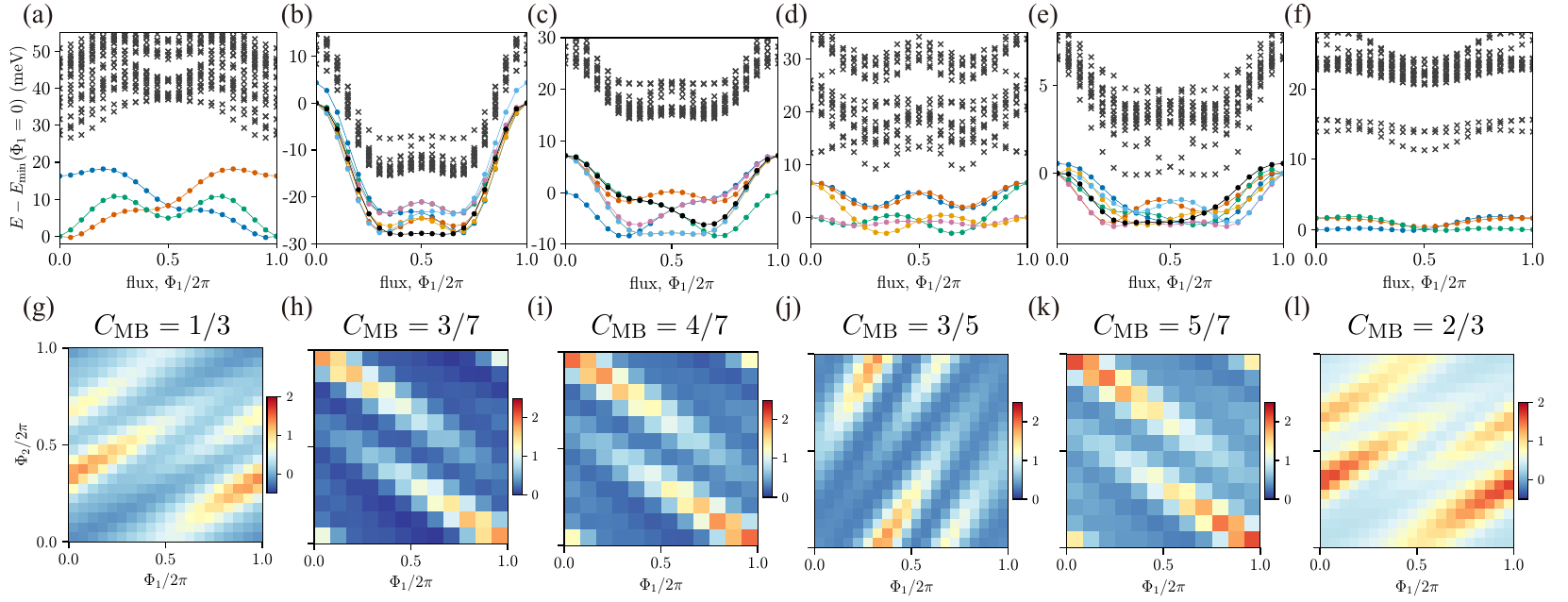}
    \caption{\textbf{Spectral Flow and Many-Body Berry Curvature of Rb\textsubscript{2}Pd\textsubscript{3}S\textsubscript{4}.} \textbf{(a-f)} Spectral flow under one flux insertion and \textbf{(g-l)} normalized average Berry curvature $2\pi F(\Phi_1, \Phi_2)/N_{\rm GS}$ of Rb\textsubscript{2}Pd\textsubscript{3}S\textsubscript{4} at \textbf{(a,g)} $\nu_{\rm tot}=4/3$, $N=18$, \textbf{(b,h)} $\nu_{\rm tot}=10/7$, $N=21$, \textbf{(c,i)} $\nu_{\rm tot}=11/7$, $N=21$, \textbf{(d,j)} $\nu_{\rm tot}=8/5$, $N=15$, \textbf{(e,k)} $\nu_{\rm tot}=12/7$, $N=21$, and \textbf{(f,l)} $\nu_{\rm tot}=5/3$, $N=18$.}
    \label{Fig_RbPdS_Jain_spectral_mbChern}
\end{figure}

We next turn to $\mathrm{Rb}_2\mathrm{Pd}_3\mathrm{S}_4$, where a similar set of fractional Chern insulators occurs [Fig.~\ref{Fig_RbPdS_Jain}]. In addition to the five fillings that were also identified in $\mathrm{Cs}_2\mathrm{Pd}_3\mathrm{S}_4$, we also report a fractional Chern insulator at $\nu_{\mathrm{tot}}=12/7$ [Fig.~\ref{Fig_RbPdS_Jain}(e)]. The spectral flow under flux insertion and many-body Chern numbers are further confirmed [Fig.~\ref{Fig_RbPdS_Jain_spectral_mbChern}]. Finally, we identify additional fractional Chern insulator phase{s} in $\mathrm{Cs}_2\mathrm{Ni}_3\mathrm{S}_4$. The quantum geometry of this material is less favorable than that of the preceding two materials (see Fig.~\ref{Fig_Quantum_Geometry}), and we identify a fractional Chern insulator only at $\nu_{\rm tot}=5/3$ {and $11/7$}. The spectral flow exhibits sharp variations {and overlaps between excited states}, and there were insufficient overlaps at a twist grid of $20\times20$ points for a reliable many-body Chern number calculation. We therefore adaptively refine the grid in the regions and confirm convergence to $C_{\rm MB}=2/3$ and $4/7$. The many-body gaps of fractional Chern insulators are reported in Table~\ref{tab:manybodygap}. \\


\begin{figure}
    \centering
    \includegraphics[width=0.8\linewidth]{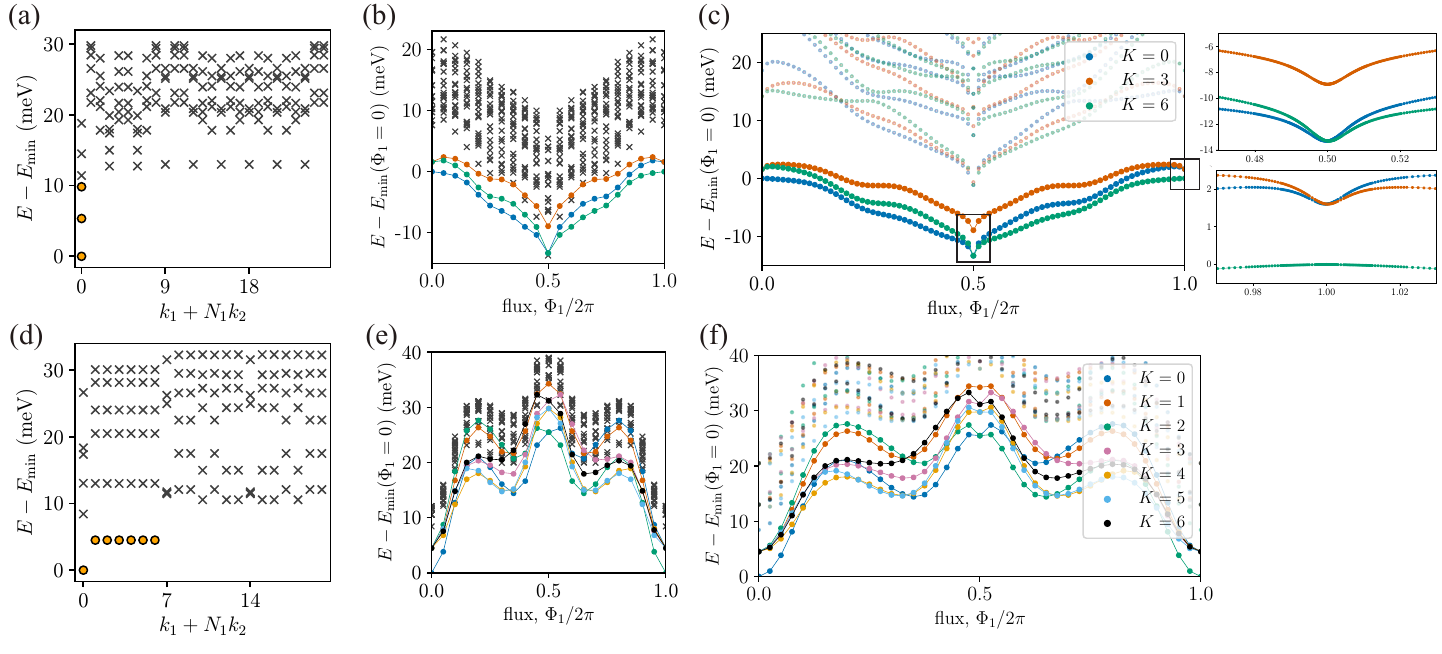}
    \caption{\textbf{Fractional Chern Insulator in Cs\textsubscript{2}Ni\textsubscript{3}S\textsubscript{4}.} \textbf{(a,d)} Many-body spectra of Cs\textsubscript{2}Ni\textsubscript{3}S\textsubscript{4} at \textbf{(a)} $\nu_{\rm tot}=5/3$, $N=27$ and \textbf{(d)} $\nu_{\rm tot}=11/7$, $N=21$. \textbf{(b,c,e,f)} Spectral flow under one flux insertion at \textbf{(b,c)} $\nu_{\rm tot}=5/3$, $N=18$ and \textbf{(e,f)} $\nu_{\rm tot}=11/7$, $N=21$, where \textbf{(b,e)} all momentum sectors are plotted, and \textbf{(c,f)} only the momentum sectors that contain the ground states are plotted to resolve the gaps within these momentum sectors. For (c), enlarged views of sharp variations near $\Phi_1/2\pi=0.5$ and $1$ (boxed in rectangles) are shown as two small plots on the right side.}
    \label{Fig_CsNiS_FCI}
\end{figure}

\begin{table*}[h]
\centering
\begin{tabular}{c|c|c|c|c|c|c|c|c}
$\nu_{\rm tot}$ &  $4/3$ & $7/5$  & $10/7$ & $11/7$ & $8/5$ & $5/3$ & $12/7$ \\ \hline
 system size &  $N=24$ & $N=25$ & $N=21$ & $N=21$ & $N=25$ &$N=27$ & $N=28$  \\
\hline
Cs\textsubscript{2}Pd\textsubscript{3}S\textsubscript{4} & 26 & - & 12  & 15 & 3 & 19 & - \\
Rb\textsubscript{2}Pd\textsubscript{3}S\textsubscript{4} & 12 & - & 4 & 18 & 9 & 19 & 4 \\
Cs\textsubscript{2}Ni\textsubscript{3}S\textsubscript{4} & - & - & - & 4 & - & 2 & - \\
\end{tabular}
\caption{\textbf{Many-Body Gaps of Fractional Chern Insulators}. The energy gap between the highest-energy state in the ground-state multiplet and the lowest excited state is presented for the largest accessible system size at each filling. The gaps are given in meV and rounded to the nearest meV. Note that the accessible size gets larger as the system gets closer to full filling $\nu_{\rm tot}=2$, because higher electron filling leads to fewer holes and hence a smaller Hilbert-space dimension at fixed $N$.}
\label{tab:manybodygap}
\end{table*}

\begin{figure}
    \centering
    \includegraphics[width=0.9\linewidth]{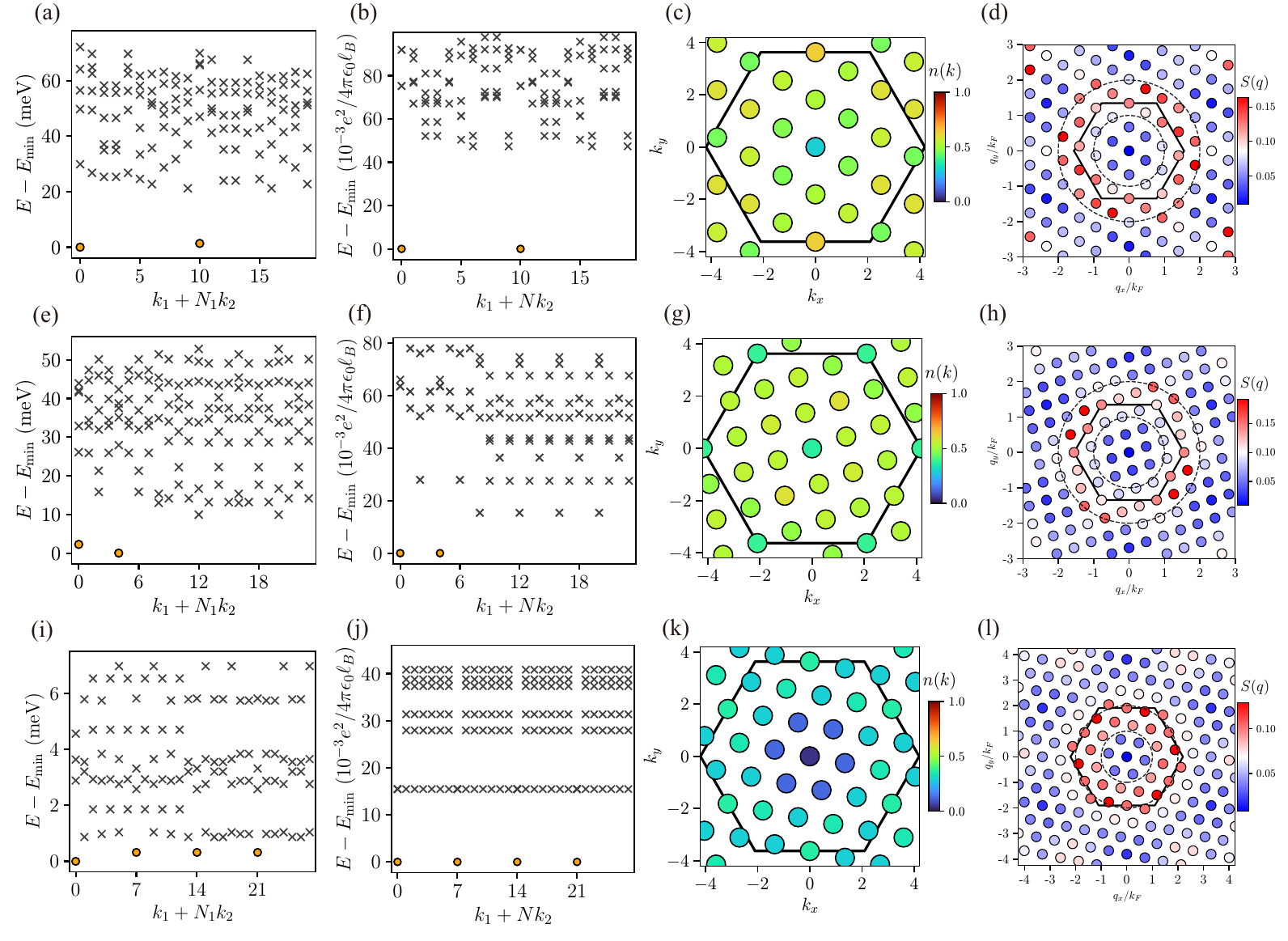}
    \caption{\textbf{Signatures of Anomalous Composite Fermi Liquids in Cs\textsubscript{2}Pd\textsubscript{3}S\textsubscript{4}.} \textbf{(a)} Many-body spectra of Cs\textsubscript{2}Pd\textsubscript{3}S\textsubscript{4} at  $\nu_{\rm tot}=3/2$, $N=20$. \textbf{(b)} Many-body spectra of half-filled lowest Landau level at $N=20$. \textbf{(c)} Momentum-space occupation number $n(\bm{k})$ and \textbf{(d)} structure factor $S(\bm{q})$ averaged over the two low-lying states in (a). \textbf{(e)} Many-body spectra of Cs\textsubscript{2}Pd\textsubscript{3}S\textsubscript{4} at  $\nu_{\rm tot}=3/2$, $N=24$. \textbf{(f)} Many-body spectra of half-filled lowest Landau level at $N=24$. \textbf{(g)} Momentum-space occupation number $n(\bm{k})$ and \textbf{(h)} structure factor $S(\bm{q})$ averaged over the two low-lying states in (e). \textbf{(i)} Many-body spectra of Cs\textsubscript{2}Pd\textsubscript{3}S\textsubscript{4} at  $\nu_{\rm tot}=5/4$, $N=28$. \textbf{(j)} Many-body spectra of quarter-filled lowest Landau level at $N=28$. \textbf{(k)} Momentum-space occupation number $n(\bm{k})$ and \textbf{(l)} structure factor $S(\bm{q})$ averaged over the four low-lying states in (i). The two dashed circles in (d,h,l) are $|\bm q|=k_F$ and $|\bm q|=2k_F$, respectively, where $k_F=\sqrt{A_{\rm BZ}/2\pi}$ in (d,h) and $k_F=\sqrt{A_{\rm BZ}/4\pi}$ in (l). }
    \label{Fig_CsPdS_ACFL}
\end{figure}

\paragraph{Anomalous Composite Fermi Liquids}
We next provide results for anomalous composite Fermi liquids in $\mathrm{Cs}_2\mathrm{Pd}_3\mathrm{S}_4$. In addition to the many-body spectra, ground state occupation number, and structure factor displayed in the main text [Fig.~\ref{Fig3_EDspectra}(c-f)], we present the same set of results at $N=20$ [Fig.~\ref{Fig_CsPdS_ACFL}(a-d)], and also $N=24$ [Fig.~\ref{Fig_CsPdS_ACFL}(e-h)] with a more symmetric grid [Fig.~\ref{Fig_BZgrid}(g)]. The calculations in the main text use an $N=6\times 4$ grid [Fig.~\ref{Fig_BZgrid}(h)] to explore the twofold degeneracies of low-lying states. Whereas the structure factor in main text [Fig.~\ref{Fig3_EDspectra}(f)] largely exhibits a Fermi-liquid-like behavior, there are also peaks at large $|\bm q|$ points. In our additional calculations, we observe that the peaks in the structure factor diminish with increasing system size [Fig.~\ref{Fig_CsPdS_ACFL}(d,h)]. We further note that the fluctuations in the occupation number decrease with increasing system size, from $|n(\bm k)-1/2|<0.22$ at $N=20$ [Fig.~\ref{Fig_CsPdS_ACFL}(c)] to $|n(\bm k)-1/2|<0.14$ at $N=24$ [Figs.~\ref{Fig_CsPdS_ACFL}(g) and \ref{Fig3_EDspectra}(e)].

We also report a signature consistent with a quarter-filled anomalous composite Fermi liquid at $\nu_{\rm tot}=5/4, ~N=28$ [Fig.~\ref{Fig_CsPdS_ACFL}(i-l)]. The low-lying momentum sectors [Fig.~\ref{Fig_CsPdS_ACFL}(i)] match those of quarter-filled lowest Landau level [Fig.~\ref{Fig_CsPdS_ACFL}(j)], and the momentum occupation number [Fig.~\ref{Fig_CsPdS_ACFL}(k)] has no sharp discontinuity, although there exists some modulation ($|n(\bm k)-1/4|<0.21$). The structure factor [Fig.~\ref{Fig_CsPdS_ACFL}(l)] shows peaks near $|\bm{q}|=2k_F$ and nearly vanishes for $|\bm{q}|>2k_F$. \\

\begin{figure}
    \centering
    \includegraphics[width=0.95\linewidth]{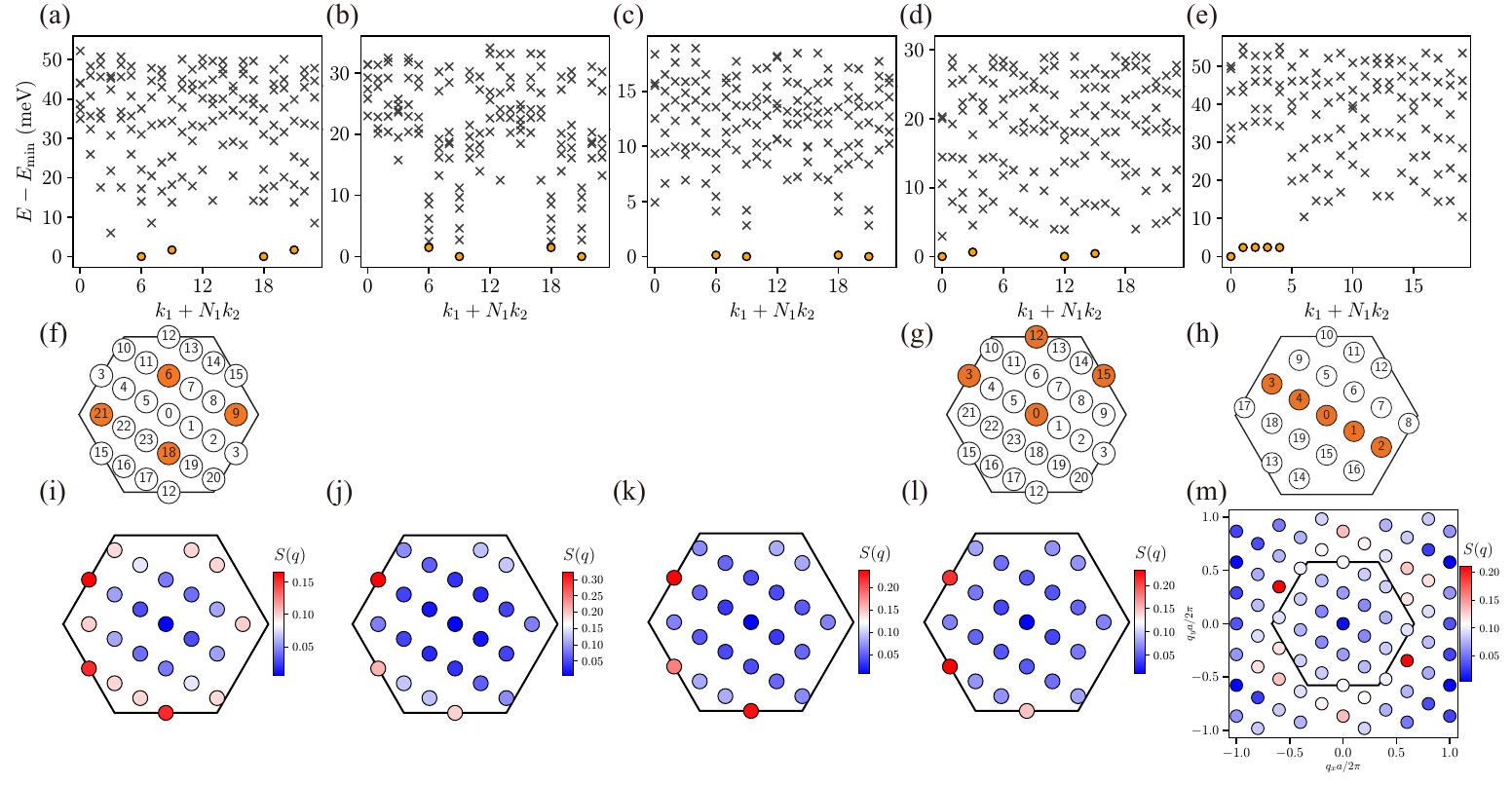}
    \caption{\textbf{Signatures of Charge Density Waves.} \textbf{(a-e)} Many-body spectra of \textbf{(a)} Cs\textsubscript{2}Pd\textsubscript{3}S\textsubscript{4}, \textbf{(b)} Rb\textsubscript{2}Pd\textsubscript{3}S\textsubscript{4}, and \textbf{(c)} Cs\textsubscript{2}Ni\textsubscript{3}S\textsubscript{4} at  $\nu_{\rm tot}=7/4$, $N=24$, Cs\textsubscript{2}Ni\textsubscript{3}S\textsubscript{4} at \textbf{(d)} $\nu_{\rm tot}=3/2$, $N=24$ and \textbf{(e)} $\nu_{\rm tot}=7/5$, $N=20$. \textbf{(f-h)} Momentum grids used where ground-state momentum sectors are marked in orange for \textbf{(f)} (a-c), \textbf{(g)} (d), \textbf{(h)} (e). \textbf{(i-m)} Structure factor $S(\bm{q})$ averaged over the multifold low-lying states in \textbf{(i)} (a), \textbf{(j)} (b), \textbf{(k)} (c), \textbf{(l)} (d), and \textbf{(m)} (e). Unlike the other panels, (m) includes an extended momentum range to reveal the peaks outside the first Brillouin zone.}
    \label{Fig_CDW}
\end{figure}

\paragraph{Charge Density Waves}
We summarize the evidence for charge density waves found at several fillings for different materials. The many-body spectra and ground state structure factor are presented [Fig.~\ref{Fig_CDW}]. We first observe charge density waves at $\nu_{\rm tot}=7/4$ for three different compounds,  $\mathrm{Cs}_2\mathrm{Pd}_3\mathrm{S}_4$, $\mathrm{Rb}_2\mathrm{Pd}_3\mathrm{S}_4$, and $\mathrm{Cs}_2\mathrm{Ni}_3\mathrm{S}_4$ [Fig.~\ref{Fig_CDW}(a-c)]. The fourfold ground-state momentum sectors [Fig.~\ref{Fig_CDW}(f)] are related to each other with $M$-point vectors, consistent with the Bragg peaks in $S(\bm{q})$ at $M$-points [Fig.~\ref{Fig_CDW}(i-k)]. For $\mathrm{Cs}_2\mathrm{Ni}_3\mathrm{S}_4$, we find two additional charge density waves at $\nu_{\rm tot}=3/2$ and $7/5$ [Fig.~\ref{Fig_CDW}(d,e)], where the ground-state momenta are again consistent with Bragg peaks [Fig.~\ref{Fig_CDW}(g,h,l,m)].

\section{Higher Chern Bands}
\label{sec:higher_chern_flat_bands}

In this supplementary section, we discuss the higher Chern numbers of
the flat bands in Cs$_2$Pt$_3$S$_4$ and Rb$_2$Pd$_3$Se$_4$. We adopt
the hopping and spin-orbit coupling conventions introduced in
Sec.~\ref{sec:TB Hamiltonian} and assume a sufficiently large Stoner
splitting, $H=H_0+\Delta\sigma_z$ with $\Delta>0$, to separate the two
spin sectors. Throughout this section, we focus on the upper spin
sector and list the three Chern numbers in descending energy order,
$\boldsymbol{C}=(C_{\mathrm{top}},C_{\mathrm{mid}},C_{\mathrm{bot}})$.

\begin{figure}
    \centering
    \includegraphics[width=1\linewidth]{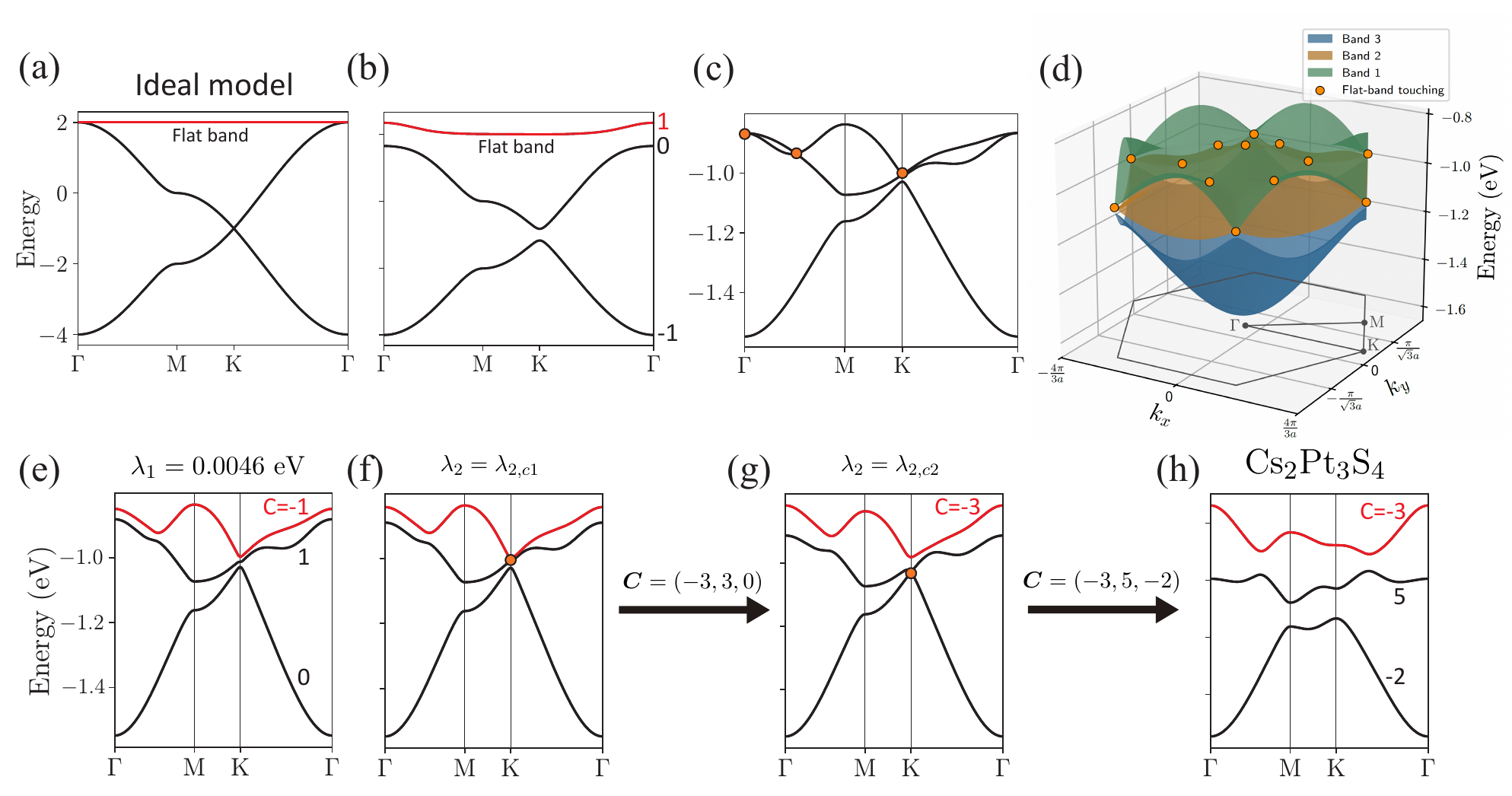}
    \caption{\textbf{Higher Chern Band Analysis.} \textbf{(a)} The ideal kagome model with $t_1=1$ and all other parameters zero. \textbf{(b)} Turning on $\lambda_1=0.1$ from (a). \textbf{(c)} A toy kagome model by retaining the hopping parameters of
Cs$_2$Pt$_3$S$_4$ with spin-orbit coupling turned off. The orange dots mark the Dirac touchings at the topmost flat band. \textbf{(d)} Band structure of (c) plotted in 3D. We observe a $\Gamma$ touching, six $K, K'$ touchings, and six additional Dirac touchings sitting on the $\Gamma-M$ paths. \textbf{(e)} Turning on $\lambda_1=0.0046$ eV from (c). Within consistent sign conventions, the flat band acquires $C=-1$, in contrast to the ideal model. \textbf{(f)} Turning on $\lambda_2=\lambda_{2, c1}=0.0023$ eV from (e). The upper two bands touch at $K$ and $K'$. \textbf{(g)} Turning on $\lambda_2=\lambda_{2, c2}=0.0091$ eV from (e). The lower two bands touch at $K$ and $K'$. \textbf{(h)} The resulting band structure of Cs$_2$Pt$_3$S$_4$ with $\lambda_2=0.0335$ eV and all remaining parameters set to their fitted values.}
    \label{fig:Higher Chern number}
\end{figure}

First, we consider the ideal nearest-neighbor
kagome model as a reference, choosing $t_1=1$ and setting all
further-neighbor hopping and spin-orbit coupling amplitudes to zero (Fig. \ref{fig:Higher Chern number}(a)). As we turn on a
positive nearest-neighbor spin-orbit coupling, $\lambda_1\to0^+$, the quadratic band touching
at $\Gamma$ and the Dirac crossings at $K$ and $K'$ are gapped, producing
$\boldsymbol{C}=(+1,0,-1)$ (Fig. \ref{fig:Higher Chern number}(b)).

Compared to this ideal model, the Chern numbers for the three bands of Cs$_2$Pt$_3$S$_4$ and Rb$_2$Pd$_3$Se$_4$ with the same sign conventions are $\boldsymbol{C}=(-3,5,-2)$, and $\boldsymbol{C}=(-2,3,-1)$, respectively, which are quite different. Using our tight-binding model, we examine the origin of this peculiarity, with emphasis on the Cs$_2$Pt$_3$S$_4$ case. \\

We introduce a kagome model by retaining the fitted hopping parameters of
Cs$_2$Pt$_3$S$_4$ (see Table \ref{tab:tightbindingparams}) while initially setting all spin-orbit coupling amplitudes to zero (Fig. \ref{fig:Higher Chern number}(c),(d)). We first notice that the lowest band is directly separated from
the other two. This is because the symmetry-enforced doublet at $K$ and
$K'$ now belongs to the upper two bands. This reordering can be seen
from the spin-orbit-coupling-free eigenvalues at $K$,
\begin{align*}
    &E_s=d_K+2f_K,\quad E_d=d_K-f_K,\\
    &f_K=t_1-2t_2+2t_4-2t_5+2t_7+2t_9,
\end{align*}
where $E_s$ and $E_d$ denote the singlet and doublet energies,
respectively, and $d_K$ is their common diagonal contribution.
Whereas $f_K=t_1>0$ in the nearest-neighbor model places the doublet below the
singlet, the fitted Cs$_2$Pt$_3$S$_4$ parameters give
$f_K=-0.0079~\mathrm{eV}$, reversing their order, mainly due to large
$t_2$. Since each spin sector retains an
effective spinless time-reversal symmetry in the absence of spin-orbit coupling, the isolated lowest band has $C_{\mathrm{bot}}=0$.

The upper two bands, meanwhile, exhibit additional crossings beyond
those of the ideal kagome model. Besides the quadratic touching at
$\Gamma$ and the Dirac touchings at $K$ and $K'$, there are six
symmetry-related Dirac points along the $\Gamma$--$M$ directions (marked with orange dots in Fig. \ref{fig:Higher Chern number}(c),(d)). As a weak positive nearest-neighbor spin-orbit coupling $\lambda_1\to0^+$ gaps all these crossings, the local Berry fluxes of the top
band, in units of $2\pi$, approach $+1$ near $\Gamma$, $+1/2$ near
each of $K$ and $K'$, and $-1/2$ near each of the six additional
Dirac points. Consequently,
\begin{equation}
    C_{\mathrm{top}} = 1+2\times \frac{1}{2} - 6\times \frac{1}{2}=-1.
    \label{eq:highC_weak_SOC}
\end{equation}
The contribution from $\Gamma$ thus retains the sign found in the
ideal model, while the additional Dirac points reverse the sign of
the total upper-band Chern number. The middle band receives the
opposite contribution, giving $\boldsymbol{C}=(-1,1,0)$. No further
gap closing occurs as $\lambda_1$ increases to the fitted value
$0.0046~\mathrm{eV}$ with all other spin-orbit coupling terms held at zero
(Fig. \ref{fig:Higher Chern number}(e)).

We now fix $\lambda_1=0.0046~\mathrm{eV}$ and increase the
next-nearest-neighbor spin-orbit coupling, $\lambda_2$, from zero to its fitted
value $0.0335~\mathrm{eV}$, keeping the remaining spin-orbit coupling amplitudes zero. The first transition occurs at
$\lambda_{2,c1}=\lambda_1/2=0.0023~\mathrm{eV}$, where the upper two
bands touch simultaneously at $K$ and $K'$ (Fig.~\ref{fig:Higher Chern number}(f)). The reversal of the
Dirac mass transfers one unit of Chern number from the top band to
the middle band at each valley, changing the sequence from
$(-1,1,0)$ to $(-3,3,0)$. Upon further increasing $\lambda_2$,
the descending branch of the split doublet reaches the singlet when
$$\lambda_{2,c2}=\frac{\lambda_1-\sqrt{3}f_K}{2}
    \simeq0.0091~\mathrm{eV}.$$
This second transition involves the middle and lowest bands at both
valleys and changes their Chern numbers by $+2$ and $-2$,
respectively (Fig.~\ref{fig:Higher Chern number}(g)). The resulting evolution is
$$(-1,1,0) \xrightarrow{\lambda_{2,c1}} (-3,3,0) \xrightarrow{\lambda_{2,c2}} (-3,5,-2).$$
The $\Gamma$-point gap remains open along this path, and the final
sequence persists up to $\lambda_2=0.0335~\mathrm{eV}$. Restoring
the remaining fitted spin-orbit coupling terms continuously from zero also leaves
the direct gaps open, connecting this simplified interpolation to
the full Cs$_2$Pt$_3$S$_4$ model (Fig.~\ref{fig:Higher Chern number}(h)). \\

A related mechanism yields the $|C|=2$ upper band in
Rb$_2$Pd$_3$Se$_4$. Its spin-orbit-coupling-free hopping Hamiltonian also contains
six additional Dirac crossings between the upper two bands along
the symmetry-related $\Gamma$--$M$ directions. Positive nearest-neighbor spin-orbit coupling gaps
these crossings, each contributing $-1/2$ to the upper-band Chern
number in the weak-spin-orbit-coupling limit. Together with the $+1$ contribution
from the quadratic touching at $\Gamma$, this gives
$C_{\mathrm{top}}=1-6/2=-2$. In this material, the $K$ and $K'$
doublets remain in the lower two bands, and the complete Chern
sequence is $(-2,3,-1)$ in the same upper-spin convention.
This sequence is obtained with nearest-neighbor spin-orbit coupling alone and remains
unchanged upon continuously restoring the other fitted spin-orbit coupling amplitudes.

\clearpage
\bibliography{kagome_A2T3X4_theory}